\documentclass{aa}  

\usepackage{graphicx}
\usepackage{txfonts}
\usepackage{hyperref}
\usepackage{orcidlink}

\usepackage{comment}
\usepackage{booktabs} 
\usepackage{multirow}
\usepackage{multicol} 
\usepackage{float}

\usepackage[normalem]{ulem}

\def\kms{{\rm km/s}}
\def\feh{{\rm [Fe/H]}}

\def\tcc{t_{\rm cc} }
\def\trh{\tau_{\rm rh} }
\def\trhn{\tau_{\rm rh0} }
\def\tev{\tau_{\rm ev} }

\def\rh{r_{\rm h} }
\def\rhoh{\rho_{\rm h} }
\def\rhdot{\dot{r}_{\rm h} }
\def\rv{r_{\rm v} }
\def\rt{r_{\rm t} }
\def\RG{R_{\rm G} }
\def\Rht{R_{\rm ht} }
\def\Vc{V_{\rm c} }
\def\dr{{\rm d}}
\def\Mbh{M_{\rm BH}}
\def\Mst{M_{\rm *}}
\def\msun{{M}_\odot}
\def\pc{{\rm pc}}
\def\kpc{{\rm kpc}}

\def\Myrs{{\rm Myrs}}
\def\Gyr{{\rm Gyr}}
\def\Gyrs{{\rm Gyrs}}
\def\kms{{\rm km\,s}^{-1}}
\def\mbh{\langle m_{\rm BH}\rangle}
\def\mst{\langle m_*\rangle}
\def \m{\langle m\rangle}

\newcommand\cbh{\rm{\sc clusterBH}}
\newcommand\cbhbd{{\sc cBHBd}}
\newcommand\bhbd{{\sc BHBdynamics}}
\newcommand\nbodys{{\sc nbody7}}
\newcommand\cmc{{\rm {\sc cmc}}}
\newcommand\emacss{{\sc emacss}}
\newcommand\ssp{{\sc SSPtools}}
\newcommand\sse{{\sc sse}}
\newcommand\dynesty{{\sc dynesty}}
\newcommand\rapster{{\sc rapster}}
\newcommand\fcl{{\sc fastcluster}}
\newcommand\mvsq{\langle v^2\rangle}
\newcommand\vsqd{v_\Delta^2}

\newcommand{\eq}[2][]{\begin{align}
                          #2
\end{align}}
\newcommand{\BH}{\mathrm{BH}}

\newcommand{\lr}[1]{\left(#1\right)}

\renewcommand{\arraystretch}{1.5}

\begin{document} 

\title{cBHBd: A fast code for the evolution of tidally limited star clusters and their binary black hole mergers}
\titlerunning{A fast star cluster code}
\subtitle{}

\author{
Fotios Fronimos Pouliasis\orcidlink{0000-0003-4158-5044}\inst{1}
\and Nolan Dickson\orcidlink{0000-0002-6865-2369}\inst{2}
\and Daniel Marín Pina\orcidlink{0000-0001-6482-1842}\inst{1,3,4}
\and Mark Gieles\orcidlink{0000-0002-9716-1868}\inst{1,3,5}
\and Vincent Hénault-Brunet\orcidlink{0000-0003-2927-5465}\inst{2}
\and Fabio Antonini\inst{6}
}

\institute{Institut de Ciències del Cosmos (ICCUB), Universitat de Barcelona (UB), c. Martí i Franqués, 1, 08028 Barcelona, Spain
\and Department of Astronomy and Physics, Saint Mary’s University, 923 Robie Street, Halifax, NS B3H 3C3, Canada
\and Institut d'Estudis Espacials de Catalunya (IEEC), Edifici RDIT, Campus UPC, 08860 Castelldefels (Barcelona), Spain
\and Zentrum für Astronomie der Universität Heidelberg, Institut für Theoretische Astrophysik, Albert-Ueberle-Str. 2, 69120 Heidelberg
\and ICREA, Pg. Llu\'{i}s Companys 23, E08010 Barcelona, Spain
\and Gravity Exploration Institute, School of Physics and Astronomy, Cardiff University, Cardiff CF24 3AA, UK}

\authorrunning{Fronimos Pouliasis et al.}
\date{Received XXXXX; accepted YYYYY}

\abstract{
%Context
The evolution of star clusters is driven by internal and external mechanisms, including stellar mass loss, two-body relaxation, and evaporation in the Galactic tidal field. Fast modeling tools are crucial for exploring diverse initial conditions and predicting cluster population properties and their contribution to gravitational wave (GW) sources over cosmic timescales.  
}
{
%Aims
We present an improved version of the {\sc clusterBHBdynamics (\cbhbd)} code, designed to evolve star clusters consisting of stars and stellar-mass black holes (BHs). Regarding cluster evolution, we improve the description of evaporation in the Galactic tidal field and add the effect of different metallicities and different stellar mass functions. For the GW part, we add a new prescription for GW captures occurring in BBH-BBH
interactions and for GW captures in between resonant interactions due to distant encounters that increase the eccentricity of the BBH.
}
{
%Methods
The updated version of \cbhbd\ is validated against a suite of Cluster Monte Carlo (\cmc) models for different initial cluster masses, radii, metallicities, and Galactic orbits. We also compare to a suite of $N$-body simulations and the GW predictions are compared to the BBH mergers in the \cmc\ models and $N$-body simulations from literature.
}
{
%Results
Seven parameters of the \cbhbd\ model are fitted to the \cmc\ results with nested sampling.
With the best-fit values, the evolution of the total cluster mass, half-mass radius, and the mass of the BH population  over 13 \Gyr\ are reproduced to within $\sim10\%$. With the new GW capture recipes, \cbhbd\ now reproduces the BBH merger rate found in the \cmc\ models of massive clusters ($\gtrsim10^5\,\msun$) and in direct $N$-body models of lower mass clusters ($\lesssim10^5\,\msun$), to within $\sim 20\%$.
}
{
%Conclusions
The improved \cbhbd\ provides a fast and efficient tool for modeling cluster evolution, capturing essential dynamical effects while maintaining flexibility and scalability for large-scale star cluster population studies. With a typical runtime of one second per cluster, \cbhbd\ enables iterative applications such as the search for initial conditions of globular clusters (GCs), the modeling of stellar streams and GW population synthesis.
}


\maketitle

\section{Introduction}
Globular clusters (GCs) serve as unique laboratories for understanding stellar dynamics, black hole (BH) populations, and gravitational wave (GW) sources. Their dense stellar environments lead to complex dynamical interactions that shape their long-term evolution and influence the formation and retention of BHs \citep{2012ApJ...750L..27S,2018MNRAS.475L..15G}. While young massive clusters (YMCs) with similar masses as GCs are known \citep{2010ARA&A..48..431P,2019ARA&A..57..227K}, they are not as dense as proto-GCs recently found at high redshift  \citep{2024Natur.632..513A}. 
Recent observations with the James Webb Space Telescope
(JWST) may point to signatures of GC formation at high redshift \citep{2022ApJ...940L..53V,2023MNRAS.523.3516C,2023A&A...673L...7C} and connecting these observations to GCs in the Local Universe requires detailed dynamical modeling. Despite progress in simulations \citep{2020MNRAS.497..536W, 2023MNRAS.519.1366G,2024MNRAS.530..645L,2025MNRAS.538..639B}, fully self-consistent modeling of GC formation and evolution remains computationally challenging.

A fast evolution code is crucial for various astrophysical applications. The main advantage over more detailed and computationally expensive methods (such as Monte Carlo or direct $N$-body simulations) lies in enabling large-scale parameter space explorations. With runtimes of order seconds per cluster, one can systematically vary initial conditions—such as cluster mass, radius, metallicity, and galactic orbit—and study the impact of uncertain physical processes (e.g., stellar winds, supernovae, dynamical friction) on the resulting populations of GW sources. GCs may contribute significantly to GW sources involving massive BHs, because dynamical interactions in dense stellar environments can lead to BBH formation, hardening, and eventual coalescence \citep[e.g.,][]{2000ApJ...528L..17P,2016PhRvD..93h4029R,2018ApJ...855..124S}. Accurate and efficient models of cluster evolution are therefore essential for understanding the contribution of GCs to the observed GW event rate, particularly in the context of population synthesis studies. 

The evolution of clusters as the result of various processes can be captured by simple scaling relations  \citep{1977MNRAS.181P..37F, 1997ApJ...474..223G}.
\citet{2011MNRAS.413.2509G} presented analytic expressions for the evolution of the cluster mass and half-mass radius of tidally limited clusters. It describes the initial expansion by using H\'{e}non's model of an isolated cluster \citep{1965AnAp...28...62H} and the evaporation phase at constant density at late stages by H\'{e}non's model for a cluster filling its tidal radius \citep{1961AnAp...24..369H}. This simple model formed the base for \emacss, a fast model for cluster evolution, which efficiently models the evolution of tidally limited GCs \citep{2012MNRAS.422.3415A,2014MNRAS.437..916G,2014MNRAS.442.1265A}. \emacss\ does not consider the effect of BHs, which can greatly affect the dynamical evolution of clusters \citep{breen_heggie_2013} and this was the motivation to develop \cbh\ \citep[][hereafter AG20]{2020MNRAS.492.2936A}.
\cbh\ includes the additional effect of BHs on the dynamical evolution, based on the theory presented in \citet{breen_heggie_2013}. 
Combined with a description for the  dynamics of the BBHs (\textsc{BHBdynamics}, AG20), the combined \cbh+\textsc{BHBdynamics} code (\cbhbd\footnote{\cbhbd\, is available at \href{https://github.com/cBHBd/cBHBd}{https://github.com/cBHBd/cBHBd}.}) is a fast evolution code designed for population synthesis modeling of dynamically formed BBH mergers in evolving clusters. It has been used for population synthesis inference of the gravitational wave transient catalogs 2 and 3  
\citep[GWTC2, GWTC3,][respectively]{2020PhRvD.102l3016A,2023MNRAS.522..466A}, from which it was found that dynamics is particularly efficient in producing massive BBH mergers (with primary masses $\gtrsim30\,\msun$, see also \citealt{2025arXiv250904637A}). 
The differences between \cbh\ and \emacss\ are discussed in Appendix \ref{app:cbh_emacss_diff}.

In recent years, other fast codes have been developed and presented in literature.
The cluster evolution in the fast code \rapster\ \citep{2023PhRvD.108h3012K,2023PhRvD.108h3015N,2023PhRvD.108h4044A,2024PhRvD.110d3023K} is based on similar input physics as \cbh, and also follows the GW inspiral and BBH dynamics in a similar way as {\sc BHBdynamics}, as well as the growth of a central BH \citep[see also][]{2019MNRAS.486.5008A}. Other codes, such as \fcl\ \citep{2021MNRAS.505..339M,2022MNRAS.511.5797M}, {\sc b-pop} \citep{2023MNRAS.520.5259A} and {\sc qluster} \citep{Gerosa:2023rwu}
are  BBH population synthesis codes that focus on the formation and evolution of BBH mergers without, or simplified, evolution of the cluster. 

This paper examines the dynamical evolution of GCs from a macroscopic perspective. The main concept was introduced in AG20, where the \cbhbd\ model was developed to describe the evolution of isolated GCs containing a stellar-mass BHs. Although the framework allowed the initial BH masses to vary with metallicity the evolution of the cluster model was 
obtained for a single metallicity value ($\feh\simeq-1.5$) and the cluster was assumed to evolve in isolation (AG20) or have a constant mass loss rate due to the tidal field \citep{2023MNRAS.522..466A}. By focusing on macroscopic properties such as the total stellar mass ($\Mst$), total mass of the BH population ($\Mbh$), and the half-mass radius ($\rh$), \cbhbd\ efficiently captured the long-term evolution of such systems without resolving individual stars while enabling broader applications across different cluster environments and initial conditions. Building on this foundation, we extend the model to include the effects of an external tidal field  to improve the mass loss of the previous version and generalize it to different metallicities and initial mass functions (IMFs) for the cluster's evolution. 

Our goal is to develop a fast evolution code that delivers results with acceptable loss of accuracy compared to more computationally intensive methods \citep{1992PASJ...44..141M,1999PASP..111.1333A,2012MNRAS.424..545N,2015MNRAS.450.4070W,2016MNRAS.458.1450W,2020MNRAS.497..536W} in a computing time of the order of a second. By modeling two-component clusters consisting of stellar-mass BHs and stars, we aim to capture the key dynamical processes that drive cluster evolution. BHs significantly influence cluster expansion and mass loss by segregating rapidly to the cluster core and transferring energy to the outer regions through relaxation.

To test the capabilities of the updated \cbhbd, we compare its predictions to Cluster Monte Carlo (\cmc) models of massive GCs ($\gtrsim10^5\,\msun$) with different metallicities [Fe/H] \citep{2001A&A...375..711F,2002A&A...394..345F,2015PhRvL.115e1101R,rodriguez_2018,2018ApJ...852...29K,2019ApJ...871...91Z,2020ApJS..247...48K}. \cmc\, is a code that uses the Monte Carlo method \citep{1971Ap&SS..14..151H} to evolve a cluster.
This approach allows  to encompass a wide range of cluster masses and initial conditions within a tidal field. \cbhbd\ is also compared to lower mass $N$-body models from literature \citep{2021NatAs...5..957G,banerjee_2021,2025MNRAS.538..639B}.

\begin{figure*}[h!]
\includegraphics[width=1.8\columnwidth]{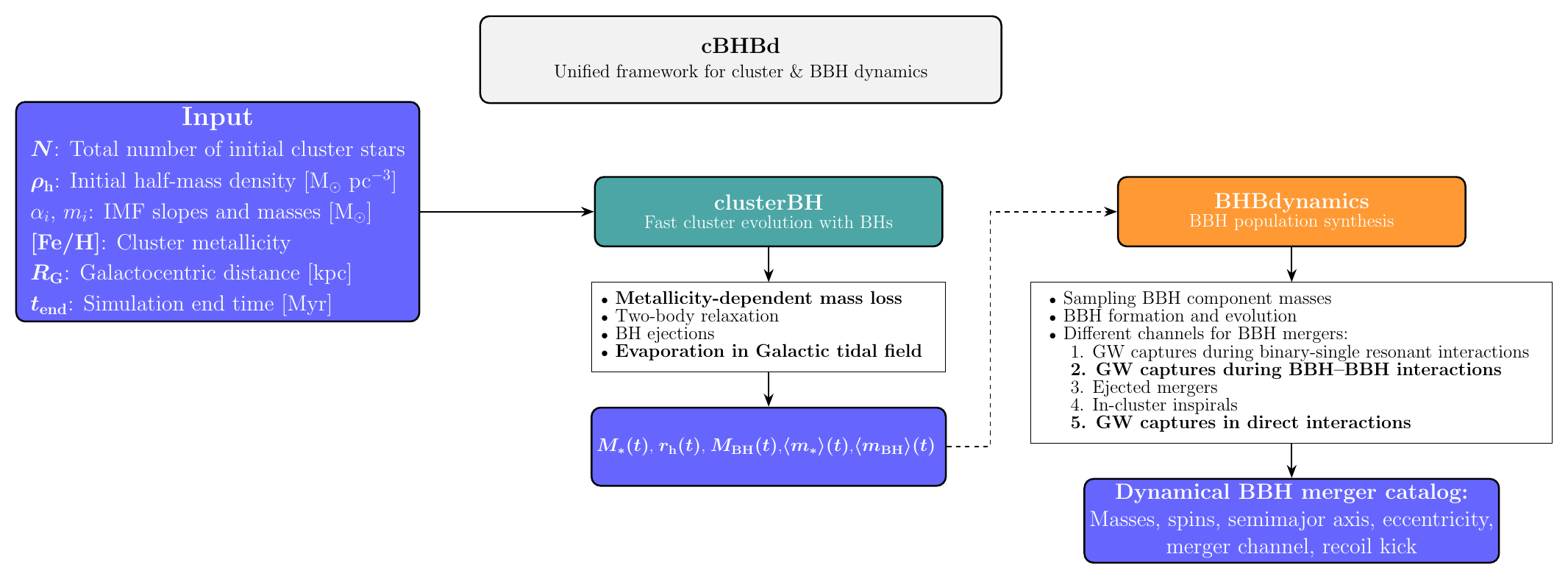}
\caption{Schematic diagram of \cbhbd. Bold mechanisms are the new inclusions to \cbh\ and \bhbd.}
\label{Fig:chart}
\end{figure*}

\cbhbd\ computes the dynamical formation and evolution of BBHs in stellar environments using semi-analytic prescriptions for BBH formation and hardening. It includes the formation of BBHs with arbitrary mass ratios $q=m_2/m_1$ with $m_2\leq m_1$, allowing for realistic modeling of mass ratio distributions. The binding energy and orbital eccentricity of each BBH evolves through repeated hardening encounters, with the code tracking whether mergers occur inside the cluster due to rapid dynamical hardening or outside after ejection, with post-ejection merger times computed via the \cite{peters_1964} formalism. Motivated by recent results \citep{marinpina_2023, marinpina_2025, randoforastier_2025}, we include two new GW capture mechanism in the new version of \cbhbd, as these can contribute significantly to the (eccentric) merger rate. The full dynamical history of each BBH, including component masses, mass ratios, and eccentricities at formation and ejection, is retained, enabling statistical predictions for the distributions of these properties, relevant for GW observations. A schematic diagram of the working of \cbhbd\ is shown in Figure~\ref{Fig:chart}.

This paper is structured as follows: In Section~\ref{sect_model},
we describe the two parts of the \cbhbd\, model: \cbh\ (subsections \ref{sect:BHMF}-\ref{sect:evap}) and \bhbd\ (subsection \ref{sec:bhbdyn}). 
Section~\ref{sect_fit}
focuses on fitting \cbh\ to \cmc\,models. In Section \ref{sect:bbhmergers}, we compare results for the production of BBH mergers in \cbhbd\ to those in \cmc\, models and $N$-body models. The paper concludes with a summary of the findings in Section \ref{sect:conclusion}. Finally, the Appendices discuss  differences between \cbh\ and \emacss\ (Appendix~\ref{app:cbh_emacss_diff}), mass segregation (Appendix~\ref{app:dEdM}), the effect BHs have on relaxation (Appendix~\ref{app:comp_h}) and finally 
a comparison to additional \cmc\ models that were not used to fit \cbhbd\ parameters (Appendix~\ref{app:LowN}).

\section{\cbhbd\  model description}
\label{sect_model}

\subsection{The initial BH population}
\label{sect:BHMF}
Here we describe how we obtain the initial BH mass function (BHMF). The initial BHMF (which includes all BHs that form in a cluster) is
determined using the mass function evolution algorithm implemented in the
\ssp\ library\footnote{Available at https://github.com/SMU-clusters/ssptools.
Version 3.1.0 is used throughout this work.},
which was first introduced by \citet{2018MNRAS.474.2479B} and further developed by
\citealt{2023MNRAS.522.5320D,2024MNRAS.529..331D}. We briefly summarize it here.

Within \ssp, the minimum initial stellar mass which is required to form a BH,
as well as the BH initial-final mass relation, are interpolated for a given
metallicity from a grid of Single Star Evolution (\sse) models, using the updated prescriptions for stellar winds and remnant formation described by \citet{2020A&A...639A..41B}. By default, we apply the rapid supernova scheme \citep{2012ApJ...749...91F}, and implement the effects of pair-instability supernova (PSN) and pulsation pair-instability supernova (PPSN) according to the prescriptions of \citet{Belczynski2016}. We adopt the same prescriptions for stellar winds, supernova mechanisms, and remnant formation as implemented in \cmc, ensuring consistency with widely used models. These prescriptions, however, remain subject to significant uncertainties. For instance, line-driven winds are uncertain, the physics of the red supergiant (RSG) phase and stars crossing the Humphreys–Davidson limit remains poorly understood, and different supernova models yield very different remnant mass distributions \citep{2025A&A...700A..20M,Willcox:2025cus}. These uncertainties propagate into the initial BH mass function and therefore into the predicted BBH population. The modular nature of \cbhbd, together with \ssp, allows users to readily replace the default prescriptions—such as the initial-final mass relation—and explore alternative assumptions as needed.
From a given IMF, this relation is used to construct
logarithmically-spaced BH mass bins.
Here we adopt a \citealt{2001MNRAS.322..231K} IMF over
the mass range \([0.08, 150]\,\msun\), however this, and other details of the stellar evolution prescriptions, can be changed by the user. For \bhbd, we explicitly sample BH masses in the aforementioned mass range by interpolating the same precomputed stellar-mass and [Fe/H] grid as used by \cbh. In Sect.~\ref{sec:bhbdyn}, we explain how we synchronise both codes.

If desired, the ejection of BHs through supernovae natal kicks can be
simulated by the inclusion of a retention fraction on newly formed BHs. We first assume that the kick velocity is drawn from a Maxwellian
distribution with a dispersion of $(1-f_{\rm fb})\, 265\, \kms$, where $f_{\rm fb}$ is the
fallback fraction, the fraction of the stellar ejecta that falls back onto the newly formed compact remnant \citep{2012ApJ...749...91F}, which depends on stellar progenitor mass and metallicity and is also interpolated from the \sse\ model grid. The
retention fraction is then computed by integrating the
Maxwellian distribution up to the initial cluster escape velocity, which is
given in terms of the initial half-mass density $\rhoh$ and total cluster mass $M$ as (AG20)

\begin{equation}
\label{eq:vesc}
v_{\rm esc}=50\,\kms\,\left(\frac{M}{10^5\,\msun}\right)^{1/3}\left(\frac{\rhoh}{10^5\,\msun\,\pc^{-3}}\right)^{1/6}\, ,
\end{equation}
where the constant of proportionality holds for a King model \citep{1966AJ.....71...64K} with dimensionless central potential of $W_0 = 7$ (AG20). From the BH mass bins representing the BHMF, the total initial number and mass
of BHs can be easily computed. For simplicity, the initial BH mass is included in the BH mass fraction $f_{\rm BH}={\Mbh/M}$ from $t=0$, although BHs have yet to be formed. This decision leads to an overestimation of the BH fraction the first few $\Myrs$, but it simplifies the modeling of the cluster's early relaxation. This decision does not affect the results since the influence of the BH fraction through relaxation appears after the core has been formed.

\subsection{Assumptions and definitions}
\label{sect:assum}
The general philosophy of \cbh\,is that it describes the evolution of the total energy of a cluster, $E$, as the result of various processes.  Specifically, $E$ refers to the `external energy' of the cluster, that is, the total cluster energy excluding the negative energy locked up in multiples \citep{1992MNRAS.257..513H}. Any change in the total mass, $M$, is accompanied by a change in $E$ and by assuming that all these changes are slow with respect to the crossing time, the evolution of $\rh$, can be obtained through the assumption of virial equilibrium \citep[see][]{2011MNRAS.413.2509G}.

In virial equilibrium, $E$ can be expressed in terms of $M$ and $\rh$ as  
\begin{equation}
E\simeq-\frac{GM^2}{5\rh},
\label{eq:E}
\end{equation}
where $G$ is the gravitational constant. The numerical constant follows from the ratio of the half mass radius ($\rh$) over the virial radius ($\rv$) of $\rh/\rv\simeq0.8$\footnote{This ratio depends on the cluster model. For a Plummer model \citep{1911MNRAS..71..460P} it is $0.78$ and for King models \citep{1966AJ.....71...64K} with dimensionless central potential in the range $2\le W_0\le8$ it is in the range $0.80-0.85$.}, where $\rv\equiv -GM^2/(2W)$ and $W$ is the total external gravitational energy, equaling $W=2E$ in virial equilibrium.
Then, from equation~(\ref{eq:E}) one finds that
\begin{equation}
\label{eq:rhdot}
\frac{\rhdot}{\rh} = \frac{\dot E}{|E|} + 2\frac{\dot M}{M}\, .
\end{equation}
This equation can be used to determine the radius evolution after contributions to $\dot{E}$ and $\dot{M}$ have been defined. As in AG20, we approximate the cluster by a two-component model, consisting of stars and BHs, such that $M=\Mst+\Mbh$ where $\Mst$ includes stars and also all remnants that are not BHs, that is, white dwarfs (WDs) and neutron stars (NSs).

Throughout the evolution of the cluster, the change of energy is driven by different mechanisms and we consider three dominant processes
\begin{equation}
\label{eq:Edot}
\dot{E} = \dot{E}_{\rm sev} +  \dot{E}_{\rm rlx} + \dot{E}_{\rm ev}\,,
\end{equation}
where subscripts `sev', `rlx' and `ev' stand for stellar evolution, relaxation  and evaporation, respectively. In the following subsections we discuss each of these three processes and the corresponding $\dot{M}$ and $\dot{E}$ terms in detail. 

\subsection{Stellar evolution}
\label{secsev}

Stellar evolution results in a change in energy following stellar mass loss, which becomes significant a few $\Myrs$ after cluster formation because of massive star winds and the first stars leaving the main sequence. As stars lose mass by winds and supernovae, mostly negative binding energy and a bit of positive kinetic energy is
carried away, such that $\dot{E}_{\rm{sev}}>0$,  leading to an expansion of the cluster.
Here, we track this stellar evolution mass loss through the \ssp\ library, using
similar prescriptions as in Section~\ref{sect:BHMF}, which are described below.

First, the main-sequence lifetime of stars of initial mass \(m\) is
approximated as
\begin{equation}
    t_{\rm{ms}}(m) = a_0\exp\left(a_1 m^{a_2}\right),
\end{equation}
where the value of the  \(a_i\) coefficients are computed from interpolated
Dartmouth Stellar Evolution Program models \citep{2007AJ....134..376D,2008ApJS..178...89D}, and
are dependent on metallicity.
Inverting this function gives the value of the turn-off mass \(m_{\rm{to}}\)
at any time \(t\).
The rate of change of main-sequence stars, at a given \(m_{\rm to}\) is then given
by:
\begin{equation}
    \dot{N}_{\rm{ms,sev}}(m_{\rm to}) = - \left.\frac{\dr N}{\dr m}\right|_{m_{\rm to}}\,
                            \left|%{\frac{dm_{\rm to}}{dt}}
                            \dot{m}_{\rm to}
                            \right|,
\end{equation}
where the number of stars of a given turn-off mass (\(\mathrm{d}N/\mathrm{d}m\))
is given by the stellar mass function at that mass. Here we make the approximation that this is given by the IMF, which should be valid for realistic clusters where the bulk of the mass lost through evaporation will be from stars below the turn-off mass, but which could break down in cases of extreme tidal fields leading to cluster dissolution on stellar evolution timescales. The rate of change of the turn-off mass itself is found by differentiating the function for \(m_{\rm{to}}(t)\):
\begin{equation}
   \dot{m}_{\rm to} 
    = \frac{1}{a_1 a_2 t} \left[
        \frac{1}{a_1}\ln\left(\frac{t}{a_0}\right)
    \right]^{\frac{1}{a_2} - 1}.
\end{equation}
The rate of change of the total mass in main-sequence stars is then simply
given by:
\begin{equation}
    \dot{M}_{\rm{ms, sev}}(m_{\rm to}) = \dot{N}_{\rm{ms, sev}}(m_{\rm to})\ m_{\rm to}
\end{equation}

Because we are considering here all non-BH objects, the actual rate of change of
these objects is given by:
\begin{equation}
    \label{eq:Nstdotsev}
    \dot{N}_{\ast,\rm{sev}}(m_{\rm to}) = \begin{cases}
        \dot{N}_{\rm{ms,sev}} & {\rm for}\ t < t_{\rm BH,max} \\
        \dot{N}_{\rm{ms, sev}}\ (1 - f_{\rm ret}) & {\rm for}\ t > t_{\rm BH,max}
    \end{cases}
\end{equation}
and the the total mass:
\begin{equation}
    \label{eq:Mstdotsev}
    \dot{M}_{\ast,\rm{sev}}(m_{\rm to}) = \begin{cases}
        \dot{N}_{\rm{ms, sev}}\, m_{\rm to} & {\rm for}\ t < t_{\rm BH,max} \\
        \dot{N}_{\rm{ms,sev}}\,(m_{\rm to} - m_{\rm rem}\,f_{\rm ret}) & \mathrm{for}\ t > t_{\rm BH,max}
    \end{cases}
\end{equation}
where $t_{\rm BH, max}$ corresponds to the lifetime of the lowest mass stars which
will form BHs, with the remaining remnants (WD and NS),
formed with masses \(m_{\rm rem}\), being included in \(M_\ast\). The parameter
\(f_{\rm ret}\) is the retention fraction of these remaining remnants, which we
take as 10 per cent for NS and 100 per cent for WD, by default.
The initial-final mass relation for WDs is interpolated from the MIST 2018
isochrones \citep{2016ApJS..222....8D,2016ApJ...823..102C}, while for NS we assume a constant
\(m_{\rm rem}=1.4\ M_\odot\).

Finally the rate of change of the average stellar mass is then given by:
\begin{equation}
    \label{eq:mst}
    %\frac{d}{dt}\mst 
    \dot{\mst}
    = \frac{\dr}{\dr t}\bigg(\frac{\Mst}{N_\ast}\bigg)
        = \frac{\dot{M}_\ast}{N_\ast} - \frac{M_\ast \dot{N}_\ast}{N_\ast^2}
%        = N_\ast^{-2} \left(N_\ast \dot{M}_\ast - M_\ast \dot{N}_\ast\right)
\end{equation}
As in \citet{2014MNRAS.442.1265A}, we introduce a parameter that relates the fractional energy change to the fractional mass loss\footnote{The authors denote the parameter $\eta$ as $\mathcal{M}$.} $\eta=\dr \ln |E|/\dr \ln M$ and is sensitive to the degree of mass segregation. For homologous mass loss (no preferred location), $\eta=3$ (Appendix~\ref{app:dEdM}). As a result of relaxation, massive stars sink toward the cluster center. Since these stars experience the most significant mass loss through stellar winds, this process occurs predominantly in the inner regions. The potential deepens, such that more energy is lost and $\eta>3$. The energy change can be written as
\begin{equation}
\label{eq:Edotsev}
\dot E_{ \rm sev}=\eta\dot{M}_{\rm sev}\frac{E}{M} .
\end{equation}
To account for different degrees of segregation, as clusters have different relaxation times, we postulate that before $t_{\rm cc}$, $\eta$ evolves linearly with time as

\begin{align}
\label{eq:eta}
\eta&=\eta_0 +(\eta_1-\eta_0)\frac{t}{\tcc},& t\leq\tcc,
\end{align}
where $\eta_0$ is the initial value and $\eta_1>\eta_0$ is the final value. In general, $\eta_0$ depends on the initial degree of segregation. Since all the clusters studied in this article are not mass segregated initially, the value is fixed to $\eta_0=3$. The second parameter $\eta_1$ is fitted on in subsequent sections.

\subsection{Two-body relaxation}
\label{sect:rlx}
\subsubsection{Timescale}
\label{sect:rlx2}
Relaxation starts to contribute to $E$ after the formation of the first dynamical (BH) binary at the time of core collapse\footnote{Historically, the term core collapse was used for  observed clusters that have a high stellar density \citep[for example,][]{1986ApJ...305L..61D}, but here we follow  \citet{breen_heggie_2013} who show that the core of BHs collapses well before that, and that the observable core collapse corresponds to the moment all BHs have been ejected.}.
We refer to the phase after core collapse as `balanced evolution' \citep{breen_heggie_2013},  when there exists a balance between the energy production in the core and the energy flux through $\rh$ as the result of two-body relaxation \citep{1961AnAp...24..369H}. Although the exact nature of the energy source is irrelevant for understanding the evolution of the bulk  cluster properties \citep{henon_1975}, we assume that the energy is provided by a BBH as long as the cluster contains BHs. As the binary interacts with its surroundings, it transfers energy outward, ensuring the cluster's energy demands are satisfied. The energy that flows to the outer layers prevents further collapse and allows the system to evolve. Over time, this continuous energy transfer reshapes the cluster’s $\rh$, driving its long-term evolution. To describe this process we write \citep{1961AnAp...24..369H} 
\begin{align}
\label{eq:dotErlx}
\dot E_{\rm rlx} &= \zeta \frac{|E|}{\trh}, &t\geq\tcc\, ,
\end{align}
where $\zeta\simeq 0.1$ \citep{1961AnAp...24..369H,henon_1975} describes the efficiency of energy `conduction' through the cluster by relaxation and is assumed to be constant. The corresponding timescale is the half-mass relaxation timescale, which is defined as \citep{1971ApJ...164..399S}
\begin{equation}
\trh=\frac{0.138}{\m\psi\ln(\gamma N)}\sqrt{\frac{M\rh^3}{G}},
\label{eq:trh}
\end{equation}
where $N$ is the total number of particles, $\m$ is the average mass of all stars and BHs, $\ln(\gamma N)$ is the Coulomb logarithm with $\gamma\simeq0.02$ \citep{1996MNRAS.279.1037G}, and $\psi$ is a parameter that depends on the mass spectrum \citep{1971ApJ...164..399S,1995ApJ...443..109L}. For equal-mass clusters $\psi=1$ and for a two-component cluster we can write $\psi$ as \citep{1971ApJ...164..399S}
\begin{equation}
\label{eq:psi}
\psi=\bigg(\frac{\mst}{\m}\bigg)^{1+\lambda}(1-f_{\rm BH}) + f_{\rm BH}\bigg(\frac{\mbh}{\m}\bigg)^{1+\lambda}\, ,
\end{equation}
where $f_{\rm BH}=\Mbh/M$ is the BH mass fraction and $\mbh$ and $\mst$ are the average BH and average stellar mass, respectively. The index $\lambda$ relates the velocity dispersion to mass as $\sigma\propto m^{-\lambda}$, where $\lambda=0$
corresponds to a mass-independent velocity dispersion and $\lambda = 0.5$ implies  equipartition  \citep{1971ApJ...164..399S}. Because the initial $\Mbh$ and $\mbh$ depend on the metallicity ($Z$), this formulation allows us to (indirectly) include the  effect of $Z$ on relaxation via $\psi$. To simplify the expression we can write equation~(\ref{eq:psi}) in the limit of small $f_{\rm BH}$ as
\begin{equation}
\label{eq:psi_approx}
\psi\simeq 1 + S\, ,
\end{equation}
where $S$ is similar to Spitzer's parameter \citep{1969ApJ...158L.139S} and is here  defined as
\begin{equation}
\label{eq:S}
S=f_{\rm BH}\bigg(\frac{\mbh}{\m}\bigg)^{1 + \lambda}\, .
\end{equation}
For typical values ($\mbh/\m\simeq20$, $f_{\rm BH}\simeq0.01$) we find $S\simeq1$ for $\lambda=0.5$, which is well above the maximum value for equipartition of $S=0.16$ found by Spitzer. Equipartition between stars and BHs is, therefore, not possible during most of the evolution \citep{breen_heggie_2013}.
From now on, we fix the exponent $\lambda=0.25$, in between the values for equipartition and mass-independent velocity dispersion. 
This value is a few percent off the value 
found from $N$-body models of clusters with $f_{\rm BH}\simeq0.06-0.07$ in \cite{2020MNRAS.491.2413W}. The quantities $\mbh$, $\m$, $\mst$, and $\psi$ are defined for properties within $\rh$, which  differ from the corresponding global values when the cluster is mass segregated (for example, all BHs may reside within $\rh$). However, for simplicity, we approximate these quantities using their global counterparts in our analysis. This approximation is motivated and justified in Appendix~\ref{app:comp_h}, where we show that using global values yields sufficiently accurate results. 
In reality the exponent $\lambda$ increases towards $\lambda=0.5$ when equipartition is reached.
This increases $S$, however it is neglected here since equipartition is achieved for small values of $S$ such that $\psi\simeq1$, so increasing $\lambda$ only mildly affects $\psi$.  The total number of cluster members at any time is,
\begin{equation}
\label{eq:N}
N=\frac{\Mbh}{\mbh}+\frac{\Mst}{\mst}\, .
\end{equation}

The energy change as the result of relaxation (equation~\ref{eq:dotErlx}) is only considered after the moment of core collapse, $\tcc$, which is a multiple of the initial $\trh$: $\tcc = N_{\trh}\trhn$, with $N_{\trh}$ of order unity. This expression applies to clusters that are not mass segregated at birth and the constant $N_{\trh}$ depends on the degree of initial mass segregation and the central concentration of the cluster, with $N_{\trh}=0$ signifying complete mass segregation or a very high initial central concentration. The time instance for core collapse for the \cmc\ models ranges from a few $\Myrs$ to $\Gyrs$ based on initial conditions such as $\rh$, $N$ and $Z$. The impact of metallicity is captured in $\psi$ in the expression for $\tcc$ at $t=0$ since $M_{\rm BH0}$ is accounted for. As a result, $\tcc$ is larger for metal-rich GCs because the lower values of $f_{\rm BH}$ and $\mbh$ reduce $\psi$, which increases $\trhn$.
The initial $\mbh$ and $\m$ needed for $S$ (equation~\ref{eq:S}) are found from the initial BHMF at $\tcc$ (described in Section~\ref{sect:BHMF}) and the stellar IMF at $t=0$ respectively. The dependence of the initial relaxation time $\trhn$ on the BH mass $M_{\rm BH0}$ through $\psi$ is an approximation used for modeling long relaxation clusters. This is because for such clusters, BHs are formed within the first few \tens of $\Myrs$ and therefore dominate in the core. The approximation in principle fails when short relaxation clusters are considered since core collapse may occur before BHs are formed. In Appendix~\ref{app:LowN} we compare the predictions of the above formula with actual time instances for core collapse in the \cmc\ models to motivate this decision.

\subsubsection{BH ejections}
\label{sect:bhejec}
A  side effect of energy generation by a BBH is BH ejections from the cluster. To describe this, we follow \cite{breen_heggie_2013} who showed that during balanced evolution, BHs are ejected on a timescale that depends on the cluster as a whole 
\begin{align}
\label{eq:Mbhdot}
\dot M_{\rm BH}&=-\beta\zeta\frac{M}{\trh}\, ,& t\geq\tcc\, ,
\end{align}
where $\beta$ is a dimensionless parameter. In the comparison to the \cmc\ models we noticed that this expression over-predicts the BH ejection rate near the depletion of the BH population. This is likely because near depletion the BHs have achieved equipartition with the stars, and the central BBH starts ejecting both stars and BH, reducing $|\dot M_{\rm BH}|$. To account for this, a dependence on $S$ is inserted in the definition of $\beta$ as, 
\begin{equation}
\label{eq:beta_form}
\beta = \beta_0 \bigg[1 - \exp(-S/S_0)\bigg]\, ,
\end{equation}
where $\beta_0$ describes BH ejection for $S\gg S_0$ and $S_0$ is a fitting parameter that sets when BH ejection becomes less efficient. The function is chosen so that $\beta\simeq\beta_0$  for $S\gg S_0$, and $\beta\simeq0$ for $S\ll S_0$. 
This inclusion was unnecessary in AG20 because their $N$-body models were clusters in isolation such that $\trh$ grows and the end of the simulation time ($12\,\Gyr$) was reached before the BHs were ejected.

Dynamics preferentially ejects the most massive BHs, reducing $\mbh$ \citep{2012MNRAS.422..841A}. We compute $\mbh$ by subtracting the cumulative ejected mass (equation~\ref{eq:Mbhdot}) from the initial BHMF, removing bins from highest to lowest mass; $\Mbh$ and $\mbh$ for $\psi$ are then obtained from the remaining population.

\subsection{Evaporation}
\label{sect:evap}
Evaporation refers to the gradual escape of stars over the tidal boundary and we assume that only stars are lost via this mechanism, a valid approximation as long as the BH fraction remains small. Since the cluster is within a tidal field from the start, the effect of evaporation is present at all times, as shown in equation~(\ref{eq:Edot}). Denoting $\xi_{\rm ev} $ as the fractional mass lost by evaporation per evaporation timescale $\tev$ \citep{1938ZaTsA..22...19A,1987degc.book.....S},

\begin{equation}
\label{eq:Mstdotev}
\dot \Mst_{, \rm ev}=-\xi_{\rm ev}\zeta \frac{\Mst}{\tev}\, ,
\end{equation}
where $\zeta$ is the same parameter as in equations~(\ref{eq:dotErlx}) and~(\ref{eq:Mbhdot}). The timescale $\tev$ is usually assumed to be $\trh$, but here we allow it to differ (see below). 
Because evaporation is more efficient in stronger tides, $\xi_{\rm ev} $ can be written as \citep{2008MNRAS.389L..28G}
\begin{equation}
\label{eq:xi}
\xi_{\rm ev} = \xi_0 \exp\left(\frac{\rh/\rt}{\Rht}\right)\,  ,
\end{equation}
where $\rt$ is the tidal radius and $\Rht$ and $\xi_0$ are free parameters.  
For a circular Galactic orbit at distance $\RG$ in a singular isothermal galaxy with circular velocity $\Vc$ we write \citep{1962AJ.....67..471K}
\begin{equation}
\label{eq:rt}
\rt=\bigg(\frac{GM}{2\Omega^2}\bigg)^{1/3}\, ,
\end{equation}
where $\Omega=\Vc/\RG$ is the angular frequency of the cluster orbit. 

Because evaporation occurs in the outer regions of the cluster where no BHs are present, $\tev$ in equation (\ref{eq:Mstdotev}) can be longer than $\trh$ which is defined for the average properties within $\rh$  (\ref{eq:trh}). We therefore postulate that
\begin{equation}
\tev=\trh\psi\, .
\end{equation}
Effectively, $\tev$ equals $\trh$ with $\psi=1$, so $\tev$ corresponds to $\trh$ of an equal-mass cluster. This step is necessary as $\trh$ captures the effect of BHs within $\rh$, which influences the cluster’s expansion but not its evaporation rate, because near the escape energy there are almost no BHs.

The nonlinear dependence of $\tev$ on $\trh$ as the result of potential escapers \citep{2000MNRAS.318..753F,2001MNRAS.325.1323B,2005A&A...429..173L,2011MNRAS.418..759R} is neglected for simplicity. In \cmc, stars are assumed to escape if they meet the energy criterion $\varepsilon>\varepsilon_{\rm crit}$
\citep{2010ApJ...719..915C,2022ApJS..258...22R}, where $\varepsilon$ denotes the specific energy and subscript `crit' the critical value for escape. The expression for $\varepsilon_{\rm crit}$ used in the \cmc\ models is $\varepsilon_{\rm crit}=c(N)\phi(\rt)$ where $c(N)$ is a parameter depending on the number of cluster components $N$. This criterion can be used to estimate $\dot E_{\rm ev}$ which then depends  on both $N$ and the $\rh/\rt$. Most of the models that we consider here have a small $\rh/\rt$, such that $\dot{E}_{\rm rlx}$ dominates over $\dot E_{\rm ev}$ at all ages. In this paper, we simplify the model by postulating $\varepsilon_{\rm crit}=0$, and therefore $\dot E_{\rm ev}=0$,  however in \cbh\ the option to include an alternative is available. We note that this does not mean that evaporation is not important, because it affects expansion  through the term $\dot M_{\rm \star, ev}$ (equation~\ref{eq:rhdot}).

In Section \ref{sect_fit}, we focus on determining the numerical values of the free parameters by performing a fit to several \cmc\ models. The list of parameters is available in Table \ref{tab:parameters}.

\begin{table}[H]
    \centering
    \renewcommand{\arraystretch}{1.2} 
    \begin{tabular}{l p{0.65\columnwidth}} 
        \toprule
        \textbf{Parameter} & \textbf{Description} \\ 
        \midrule
        $\zeta$ & Dimensionless energy change (Section~\ref{sect:rlx2}, equation~\ref{eq:dotErlx}). \\
        $\beta_0$ & Dimensionless BH ejection rate (Section~\ref{sect:bhejec}, equation~\ref{eq:beta_form}). \\
        $\Rht$ & Reference scale for ratio $\rh/\rt$ where the tidal mass loss rate is known (Section~\ref{sect:evap}, equation~\ref{eq:xi}). \\
        $\xi_0$ & Prefactor for tidal mass loss (Section~\ref{sect:evap}, equation~\ref{eq:xi}). \\
        $N_{\trh}$ & Number of initial relaxation times until $\tcc$ (Section~\ref{sect:rlx2}). \\
        $\eta_1$ & Relates  mass loss to energy change (Section~\ref{secsev}, ~\ref{eq:eta})). \\
        $S_0$ & Value for Spitzer's term used to slow down BH ejections for clusters with low $S$ (Section~\ref{sect:bhejec}, equation~\ref{eq:beta_form}). \\
        $\delta$ & Error tolerance for the fitting (Section~\ref{sect_fit}).  \\
        \midrule
        $\lambda=0.25$ & Exponent that connects velocity dispersion to average mass (Section~\ref{sect:rlx2}), equation~\ref{eq:psi}). \\
        $\gamma=0.02$ & Constant in argument of  Coulomb logarithm (Section~\ref{sect:rlx2}, equation~\ref{eq:trh}). \\
        \bottomrule
    \end{tabular}
    \caption{List of \cbh\ parameters and their descriptions. The first set of eight parameters are fitted to \cmc\ models while the second set of two parameters is assigned fixed values prior to fitting.}
    \label{tab:parameters}
\end{table}

%%%%%%%%%%%%%%%%%
\subsection{Few-body BH interactions and GW mergers}
\label{sec:bhbdyn}
Here we describe how we obtain the dynamically formed BBH mergers in the \bhbd\ part of \cbhbd. We follow the algorithm presented in \cite{2020MNRAS.492.2936A, antonini_gieles_2023} with two updates to the treatment of few-body dynamics and two new channels for BBH mergers, which we describe in Sections~\ref{sect:bbhbbh} and \ref{ssec:di}.

\subsubsection{Recap of \cbhbd\ GW recipes}
\label{ssec:recap}
In massive clusters, the majority of dynamical BBH mergers involve dynamically formed BBH, or `three-body' BBHs, even in the presence of a large number of BBHs that form from binary evolution \citep[referred to as primordial BBHs,][]{chattopadhyay_2022, torniamenti_2022,marinpina_2025}. Three-body BBHs are those formed in interactions involving at least three previously unbound BHs \citep{heggie_1975, tanikawa_hut_makino_2012}. Theory predicts that a cluster may host several three-body BBHs simultaneously \citep{1984ApJ...280..298G}, but in practice binary–binary interactions quickly ionize the wider binaries, leaving on average only one hard BBH at any given time \citep{marinpina_2023}.

At the start of the simulation, we sample a population of BHs from the initial BHMF (Section~\ref{sect:BHMF}), taking into account immediate ejections due to natal kicks with velocities that exceed the escape velocity of the cluster. Natal kicks are applied again to sample individual BH masses and populate the BHMF, since \bhbd\ needs discrete BH masses to track individual BH interactions, which can not be done with the mass bins of \cbh.  However, we ensure that $\Mbh(t)$ is the same in the \cbh\ and \bhbd\ parts (explained at the end of Section~\ref{ssec:di}). As in AG20, we assume that all BHs are initially single. After core collapse, we form a dynamical BBH from this population, with a semi-major axis at the hard-soft boundary $a_\mathrm{h}=G m_1 m_2/ ( 2\langle m\rangle \sigma^2)$, where $\sigma$ is the velocity dispersion within $\rh$, estimated from $\sigma^2\simeq 0.2 G M/r_\mathrm{h}$. The primary BH mass, $m_1$, is sampled from the BH population in the cluster, following a distribution $p(m_1)\propto m_1^{\alpha_1}$, while the secondary, $m_2$, is obtained by sampling the mass ratio, $q=m_2/m_1$, from $p(q)\propto q^{\alpha_2}$. Following \cite{antonini_gieles_2023}, we use the exponents $\alpha_1=8 + 2\alpha$ and $\alpha_2 = 3.5 + \alpha$, with $\alpha$ the slope of the BH mass function,  determined at each step from the remaining BH population using a maximum likelihood method. While the power-law BHMF is not fully realistic, it enables fast computations. The resulting differences relative to the BHMF from \cbh\ are at the level of a few percent.

Once the BBH is formed, it undergoes strong resonant interactions with single BHs. In such interactions, the three BHs engage in chaotic orbital motion, repeatedly exchanging energy and angular momentum during multiple close approaches. In scattering experiments, resonant interactions are defined as interactions for which the sum of the square distances between the three BHs has multiple minima \citep[see e.g.][]{mcmillan_hut_1996, randoforastier_2025}. \bhbd\ considers the few-body dynamics for an entire binary life cycle; the synchronization with \cbh\ is performed at the end of each binary cycle (see Section~\ref{ssec:di}). 

We now describe the three pathways to BBH coalescence that were considered in \citet{2023MNRAS.522..466A}, including improvements we introduce here:
\begin{itemize}
\item {\bf GW captures during binary-single resonant interactions:}
In each resonant interaction between a BBH and a single BH, we sample the mass of the interloper BH, $m_3$, from $p(m_3)\propto m_3^{\alpha_3}$, with $\alpha_3 = \alpha + 1/2$. To determine whether there are GW captures during these interactions, we follow the method of \cite{samsing_2014}. The probability of GW captures during resonant binary–single interactions can be estimated using the framework of \citet{2018MNRAS.481.5445S}. The interactions can be split into a number $N_\mathrm{IMS}$ of hierarchical metastable intermediate states, separated by chaotic democratic intermediate states. For a binary-single interaction with unequal masses, we compute $N_\mathrm{IMS}$ from equations~8 and 9 in \cite{randoforastier_2025}. This improves the method of \cite{antonini_gieles_2023}, which considered a fixed $N_\mathrm{IMS}$, independent of the mass ratio. For each of the $N_\mathrm{IMS}$ hierarchical states, we assume that the eccentricity of the BBH is sampled from a thermal distribution. At each intermediate state, there is a merger if the energy radiated in a single pericenter passage \citep{hansen_1972} is greater than the binding energy of the BBH. We refer to these mergers as `binary-single GW captures'. 

In this version of the code, we allow for the possibility that one of the components of the binary is exchanged with the interloper BH. For a resonant binary-single interaction involving three BHs of masses $m_\mathrm{a}$, $m_\mathrm{b}$, and $m_\mathrm{c}$, the probability that the ejected component is $m_\mathrm{c}$ is given by \citep[][equation 36]{ginat_perets_2021}
\begin{equation}
   p(m_\mathrm{c}) = \frac{m_\mathrm{a}^4 m_\mathrm{b}^4}{(m_\mathrm{a}+m_\mathrm{b})^ {5/2}\left[\frac{m_\mathrm{a}^4m_\mathrm{b}^4}{\left(m_\mathrm{a}+m_\mathrm{b}\right)^{5/2}} + \frac{m_\mathrm{a}^4m_\mathrm{c}^4}{\left(m_\mathrm{a}+m_\mathrm{c}\right)^{5/2}} 
    + \frac{m_\mathrm{b}^4m_\mathrm{c}^4}{\left(m_\mathrm{b}+m_\mathrm{c}\right)^{5/2}}\right]}.
\end{equation}
After the binary-single interaction, we eject one of the three BHs. We select the ejected BH via a weighted random choice, with the probability to eject the BH with mass $m_i$ given by $p(m_i)$. This means, e.g., that the binary will not experience an exchange in a fraction $p(m_3)$ of interactions. After a binary-single resonant interaction, on average, the binary will become more bound \citep{heggie_1975}. The fractional energy loss for unequal-mass BHs is given by equation~5 in \cite{randoforastier_2025}. This improves the method of \cite{antonini_gieles_2023}, which assumed a mass-independent fractional energy loss. Due to energy conservation, both the interloper and BBH receive a recoil kick. If their recoil kick velocity is larger than $v_\mathrm{esc}$, we assume that they are ejected from the cluster; otherwise, they reach a higher orbit around the cluster that decays to the core due to dynamical friction. 

\item {\bf Ejected mergers:} For an ejected BBH, we compute its merger timescale \citep{peters_1964} and, if it inspirals before the end of the simulation, we label it as `ejected merger'.  

After the interaction, the binary either (i) becomes more bound and remains in the cluster; (ii) becomes more bound and is ejected from the cluster due to the momentum recoil kick; (iii) merges and remains in the cluster, or (iv) merges and is ejected from the cluster as the result of a GW kick.  

\item {\bf In-cluster inspirals:}
Furthermore, we allow the possibility that a BBH is sufficiently tight and eccentric that it merges between resonant interactions. If the merger timescale \citep{peters_1964} is shorter than the time in between resonant interactions ($t_3$, see equation 20 in \citealt{2020MNRAS.492.2936A}), we label it as `in-cluster inspiral'. 

\end{itemize}

In the following subsections, we explain the numerical recipes for two new merger channels in \bhbd.

\subsubsection{GW captures during BBH-BBH interactions}
\label{sect:bbhbbh}
Until now, \bhbd\ considered only binary-single interactions. Here, we add GW captures during binary-binary interactions. Binary–binary encounters are  expected to play an important role in dense stellar environments, contributing to the dynamical formation and evolution of compact-object binaries \citep{2019ApJ...871...91Z}. We expect this merger channel to be more important in lower mass clusters  because of a high rate of BBH formation \citep{marinpina_2025}. The vast majority of binary-binary interactions among three-body BBHs lead to the ionisation of the wider binary and do not significantly change the SMA or the eccentricity of the tighter binary. For simplicity, we assume that they either trigger a merger or disrupt the wider binary without changing the properties of the smaller binary, as demonstrated by the scattering experiments in \cite{marinpina_2025}.
 
The mean number of binary-binary interactions between each binary-single interaction, $N_\mathrm{bb}$, is \citep{marinpina_2025}
\eq{N_\mathrm{bb}=\frac{\Gamma_\mathrm{bb}}{\Gamma_\mathrm{bs}}= 0.3 \, \lr{\frac{N_\BH}{10^2}}^{-1/3}.}
Here $\Gamma_{\rm bb}$ is the binary-binary interaction rate; $\Gamma_\mathrm{bs}=t_3^{-1}$ is the binary-single interaction rate and $N_\BH$ is the total number of BHs in the cluster. From post-Newtonian binary-binary scattering experiments \citet{marinpina_2025} found that the probability that a binary-binary interaction triggers a merger is given by
\eq{p_\mathrm{cap, bb} \simeq 0.034\lr{\frac{\langle m\rangle_4}{20\,\msun}}^{5/7}\lr{\frac{a_\mathrm{min}}{0.1\,\mathrm{AU}}}^{-5/7} \lr{1 + \lr{\frac{a_\mathrm{r}}{8.6}}^2}^{-0.83},}
where $\langle m\rangle_4$ is the mean mass of the four BHs, $a_\mathrm{min}$ is the SMA of the smallest binary, and $a_\mathrm{r} \ge 1$ is the SMA ratio. 
To compute $p_\mathrm{cap, bb}$, we create a new dynamical BBH with the same procedure as defined above: SMA at the hard-soft boundary, with masses sampled from $p(m_1)$ and $p(q)$. While a real BBH can undergo binary-binary interactions with multiple different BBHs, we treat them as effectively multiple interactions with this second BBH. We model GW captures in binary-binary interactions as a Poissonian process, so we assume them to occur before the next binary-single resonant interaction via a random choice with probability
\begin{equation}
p_\mathrm{bb}=1-e^{-p_\mathrm{cap, bb}N_\mathrm{bb}}.
\end{equation}
We assume that the binary with the smallest SMA merges, and label it as `binary-binary GW capture'. If there is no merger, we assume that the larger binary is ionised and the smaller binary remains unperturbed \citep{marinpina_2023}.

\subsubsection{GW captures in direct interactions}
\label{ssec:di}

In between resonant binary-single interactions, the BBH will undergo many distant non-resonant, or `direct', encounters with single BHs. The effect of these interactions is a small change in the eccentricity of the BBH while the binding energy is conserved. There is, however, the possibility that a binary is sufficiently tight and eccentric that a slight positive change in eccentricity triggers a runaway GW emission at periapsis that leads to a merger. This process, which we name `GW captures in direct interactions' \citep{trani_2024, randoforastier_2025}, increases the relative contribution of mergers inside the cluster versus ejected mergers. \cite{randoforastier_2025} show that the rate of GW captures in direct interactions is comparable to the in-cluster mergers between resonant encounters. To determine whether a BBH merges in this channel, we use the cross-section for eccentricity changes in wide fly-bys, $\Sigma(\delta e > \delta e_0)$, from \cite{heggie_rasio_1996}, where $\Sigma$ stands for the cross section of the interaction, $\delta e$ denotes the eccentricity change and $\delta e_0$ is the threshold eccentricity change required to trigger a merger. 

The timescale for such an interaction to happen is given by 
\begin{equation}
t_\mathrm{di}=\frac{1}{n_\BH\Sigma(\delta e > \delta e_0) v_\mathrm{bs}},
\end{equation}
where $n_\BH$ is the number density of BHs within the half-mass radius of the BH subcluster, $r_\mathrm{h, BH}$, computed as \citep{breen_heggie_2013}
\begin{equation}
    r_\mathrm{h, BH} = \rh \left(\frac{M_\BH}{M}\right)^{3 / 5} \left(\frac{\langle m_\BH\rangle}{\langle m \rangle}\frac{\ln (\gamma N_\BH)}{\ln (\gamma N)}\right) ^ {2 / 5},
\end{equation}
and $v_\mathrm{bs}$ is the mean relative velocity between binaries and singles within $r_\mathrm{h, \BH}$. Assuming equipartition, we get $v_\mathrm{bs}\simeq \sqrt{3/2} \sigma_\BH$, with $\sigma_\BH$ estimated as in \cite{breen_heggie_2013}, i.e.
\begin{equation}
    \sigma_\BH = \sqrt{0.2 \frac{G M_\BH}{r_\mathrm{h, BH}}}.
\end{equation}

We model GW captures in direct interactions as a Poissonian processes, so we assume them to occur before the next resonant interaction via a random choice with probability
\begin{equation}
p_\mathrm{di}=1-e^{-t_3/t_\mathrm{di}}.
\end{equation}
After a GW capture in a direct interaction, the merger remnant is retained (kicked out) if the GW recoil kick is smaller (larger) than $v_{\rm esc}$.

The binary cycle is concluded once the BBH merges or is ejected and then a new binary is created at time $m_{\rm ej}/\dot{M}_{\rm BH}$ after the formation of the previous binary (see equation~37 in \citealt{2020MNRAS.492.2936A}). Here $m_{\rm ej}<0$ is the total mass ejected by the previous binary (including itself) and $\dot{M}_{\rm BH}$ is provided by \cbh, evaluated at the time of the formation of the previous binary. We note that there could be a slight mismatch between this timestep and the internal time of \bhbd. We prioritize the time synchronization of $M_{\rm BH}$ between \cbh\ and \bhbd. 

The process is repeated until the end of the simulation or until all BHs are ejected from the cluster.

\section{Fitting {\cbh\ }to {\cmc\ }models}
\label{sect_fit}
\subsection{{\cmc\ }models}
\label{ssec:cmcmodels}
To calibrate 
the \cbh\ code, we use \cmc\ models, a set of Monte Carlo star cluster simulations \citep{2020ApJS..247...48K,rodriguez_2018,2018ApJ...852...29K,2015PhRvL.115e1101R}. 
\cmc\ serves as a convenient tool for evolving massive clusters, because the computational time scales as $\mathcal{O}(N\ln(N))$. The publicly available grid of \cmc\ models\footnote{The models used are available at https://cmc.ciera.northwestern.edu/} \citep{2020ApJS..247...48K} includes simulations with initial conditions for 
$N=[2\times10^5, 4\times10^5, 8\times10^5, 1.6\times10^6, 3.2\times10^6]$, $\rv=[0.5, 1, 2, 4]\,\pc$, $\RG=[2, 8, 20]\,\kpc$ and $Z=[0.0002, 0.002, 0.02]$, allowing for a broad exploration of cluster evolution across different environments. They sum to a total of 148 models. A few combinations are not available in the database, such as $N = 1.6 \times 10^6$ with $\rv = 0.5\,\pc$ at $Z = 0.01\,Z_\odot$, $N = 3.2 \times 10^6$ with $Z = 0.1\,Z_\odot$ and finally $N = 3.2 \times 10^6$ with $\rv = [0.5, 4]\,\pc$ at $Z = [0.01, 1]\,Z_\odot$. The relation between the $\rh$ and $\rv$ is $\rh=0.814\,\rv$ for the \cmc\ initial conditions (a King profile with dimensionless central potential $W_0=5$). As in Section~\ref{sect:BHMF} a Kroupa IMF over the mass range $[0.08,150]\,\msun$ is used with the initial average mass being $\mst_0=0.586\,\msun$. We determine our parameters from the models with $N=[8\times10^5, 1.6\times10^6, 3.2\times10^6]$ and $\rv=[1, 2, 4]\,\pc$. We limit the study to clusters with large number of particles, where the Monte Carlo method is more reliable, and three options for the virial radius by excluding the densest clusters with $\rv=0.5\,\pc$. This leaves us with 58 models to work with. The unused clusters experience stellar collisions which cannot be reproduced by the current version of \cbh. These stellar collisions do not necessarily occur at early times and the masses of colliding stars differ with age. While such collisions can affect the stellar population and BH formation, their impact on the number of merging BBHs is likely secondary compared to the dynamical interactions among BHs themselves. This subset defines the calibration sample used to determine the fitting parameters. We then validate the model by applying these fixed parameters to the remaining 90 \cmc\ models not included in the fit, as described in Appendix~\ref{app:LowN}.

\begin{figure*}[h!]
\includegraphics[width=2\columnwidth]{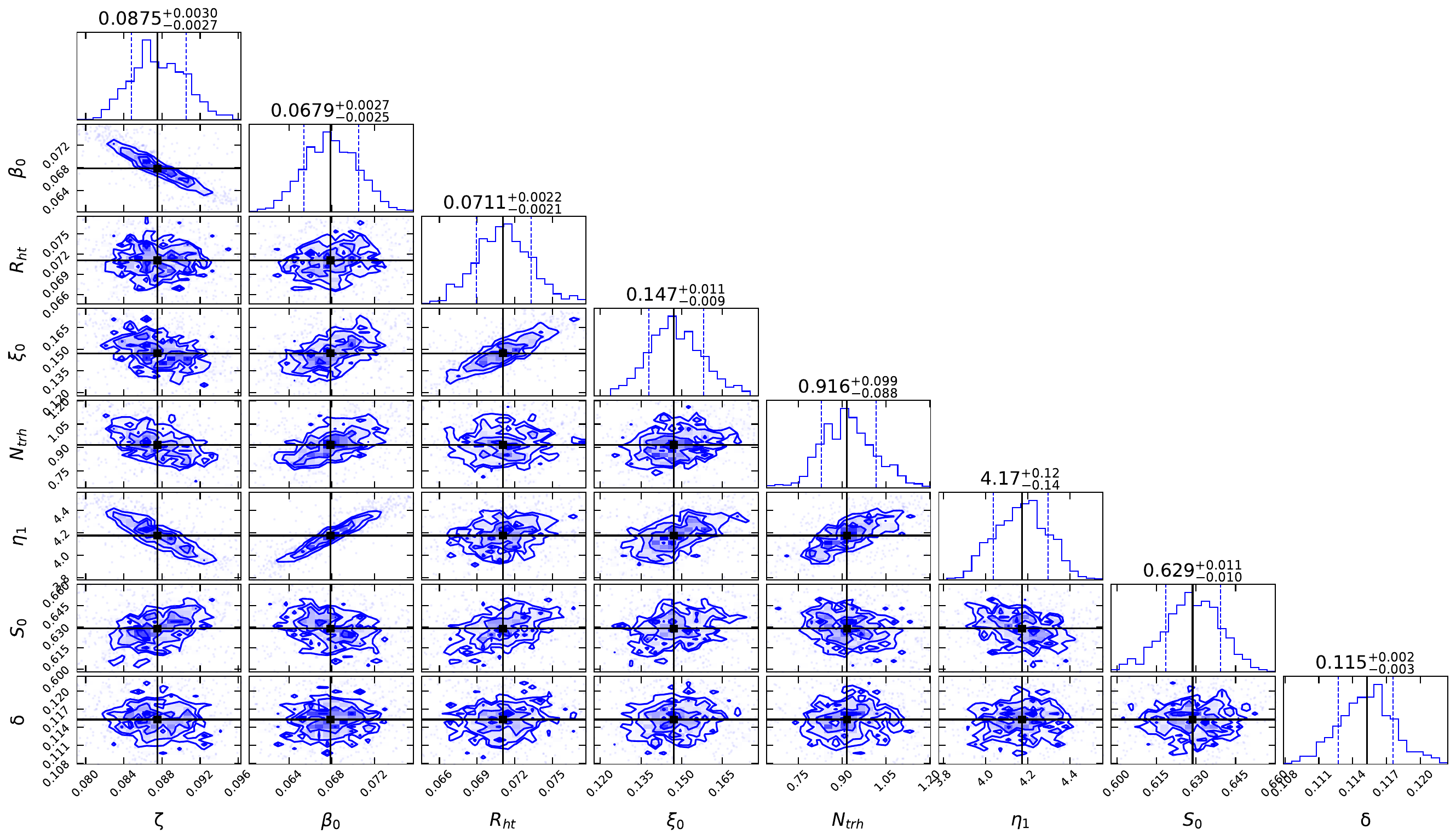}
\caption{Marginalized and 2D projections of the posterior probability distributions of the free parameters of \cbh. The plot displays contours up to the $3\sigma$ level, corresponding to the $68\%$, $95\%$, and $99.7\%$ credible regions in parameter space. 
Above each marginalized posterior distribution, the median and $1\sigma$ credibility interval for the corresponding parameter are shown.}
\label{Fig_paramplots}
\end{figure*}

\begin{figure*}
\includegraphics[width=\columnwidth]{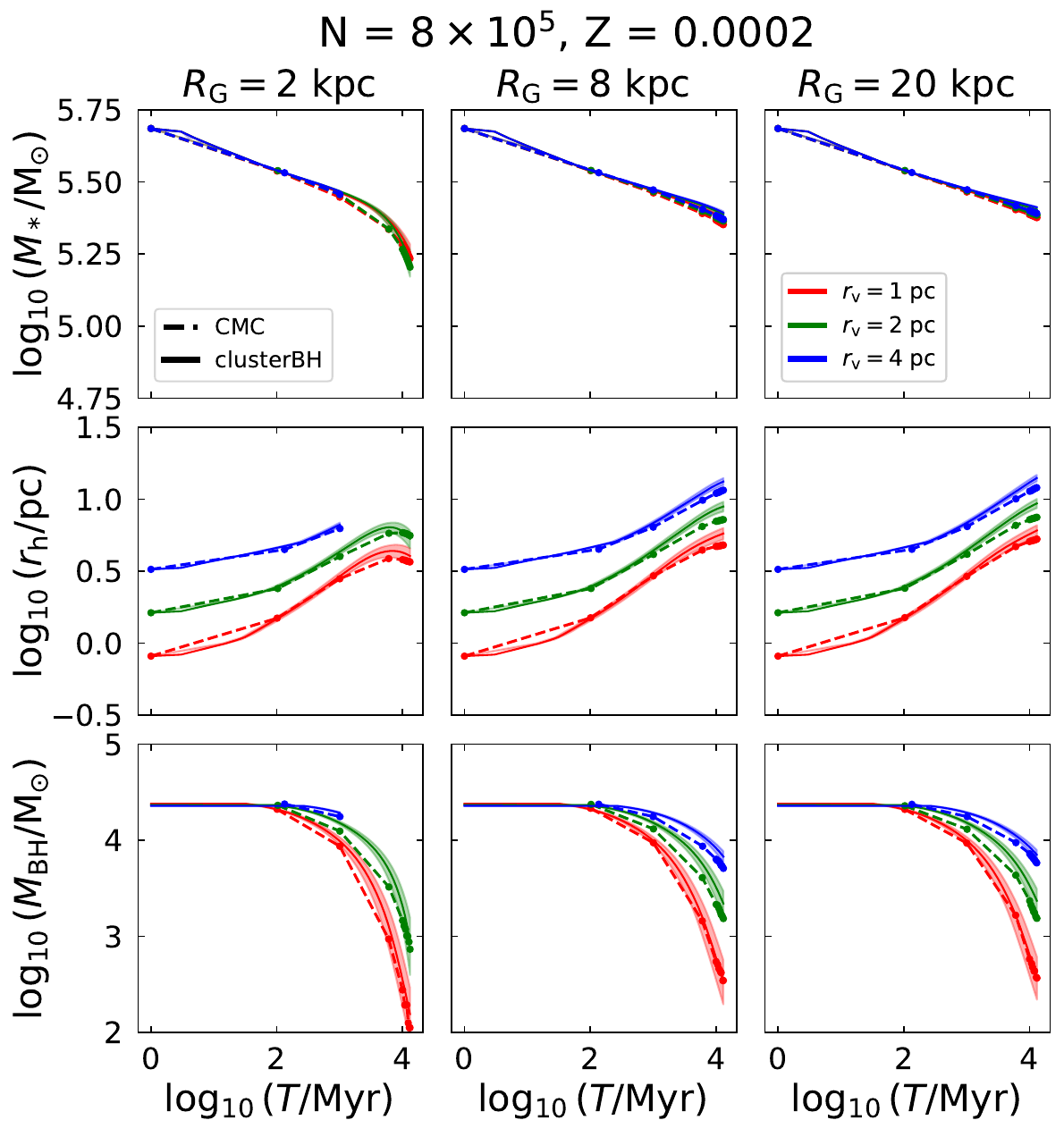}
\includegraphics[width=\columnwidth]{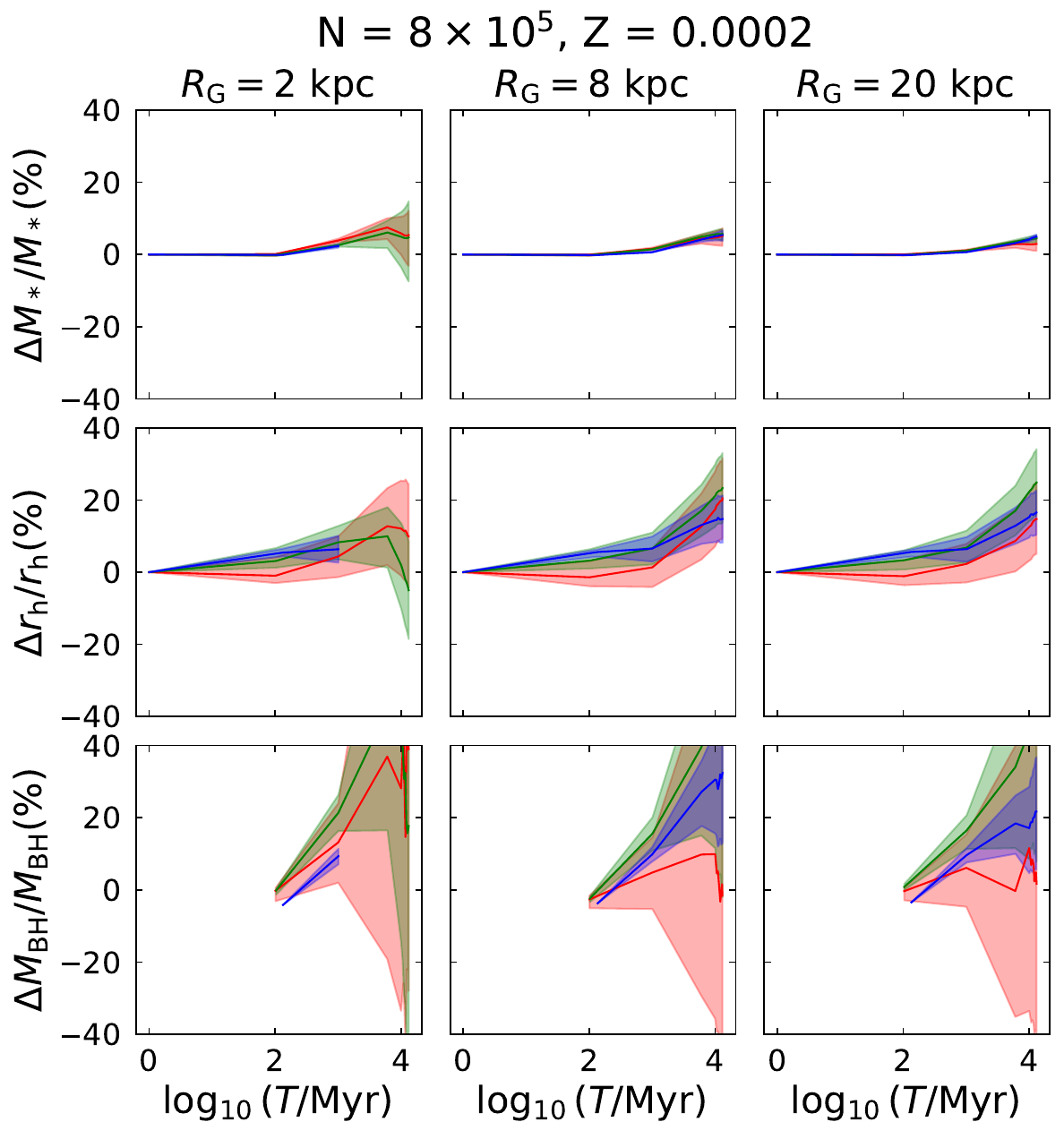}\\
\caption{Left: Evolution of $\Mst$, $\rh$ and $\Mbh$ for clusters with $N=8\times10^5$ and $Z=0.0002$. Right: Percentage difference between \cbh\,and \cmc, defined as $\Delta X/ X = (X^{\cbh}-X^{\cmc}) / X^{\cmc}$ for all three quantities. Dashed lines show the \cmc\ models while continuous lines the predictions of \cbh. Different colors signify different values of the initial $r_{\rm v}$, with red, green and blue corresponding to $1\,\pc$, $2\,\pc$ and $4\,\pc$ respectively. Shaded areas indicate the lowest-highest values of $\Mst$, $\rh$ or $\Mbh$ within $1\sigma$ at each time instance. They do not necessarily correspond to a single set of parameters. For the cluster with $\rv=4\,\pc$ at $\RG=2\,\kpc$, the available snapshots stop close to $1\,\Gyr$ because it dissolves fast.}
\label{Fig_N8e5_Z0.0002}
\end{figure*}

\begin{figure*}
\includegraphics[width=\columnwidth]{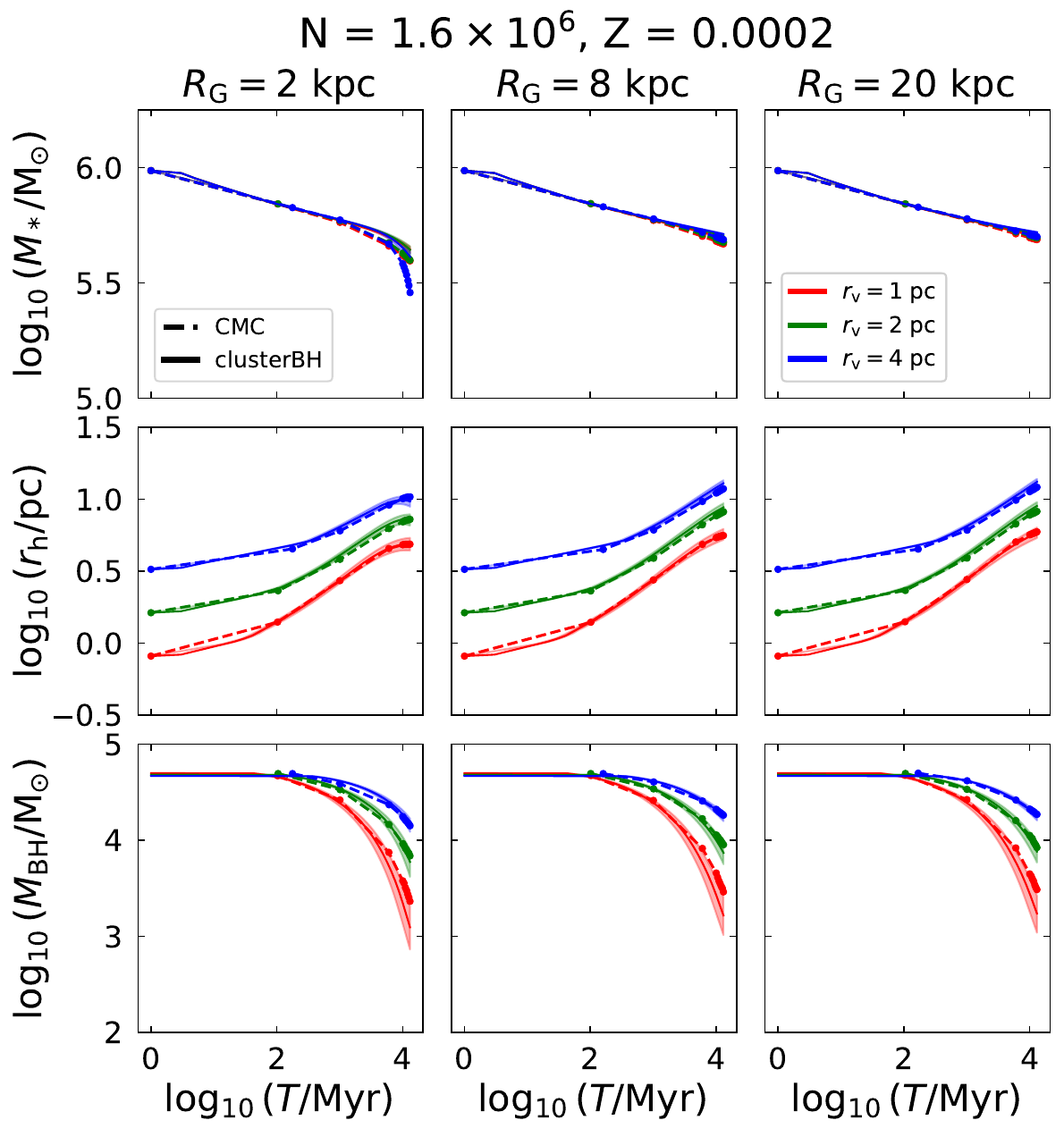}
\includegraphics[width=\columnwidth]{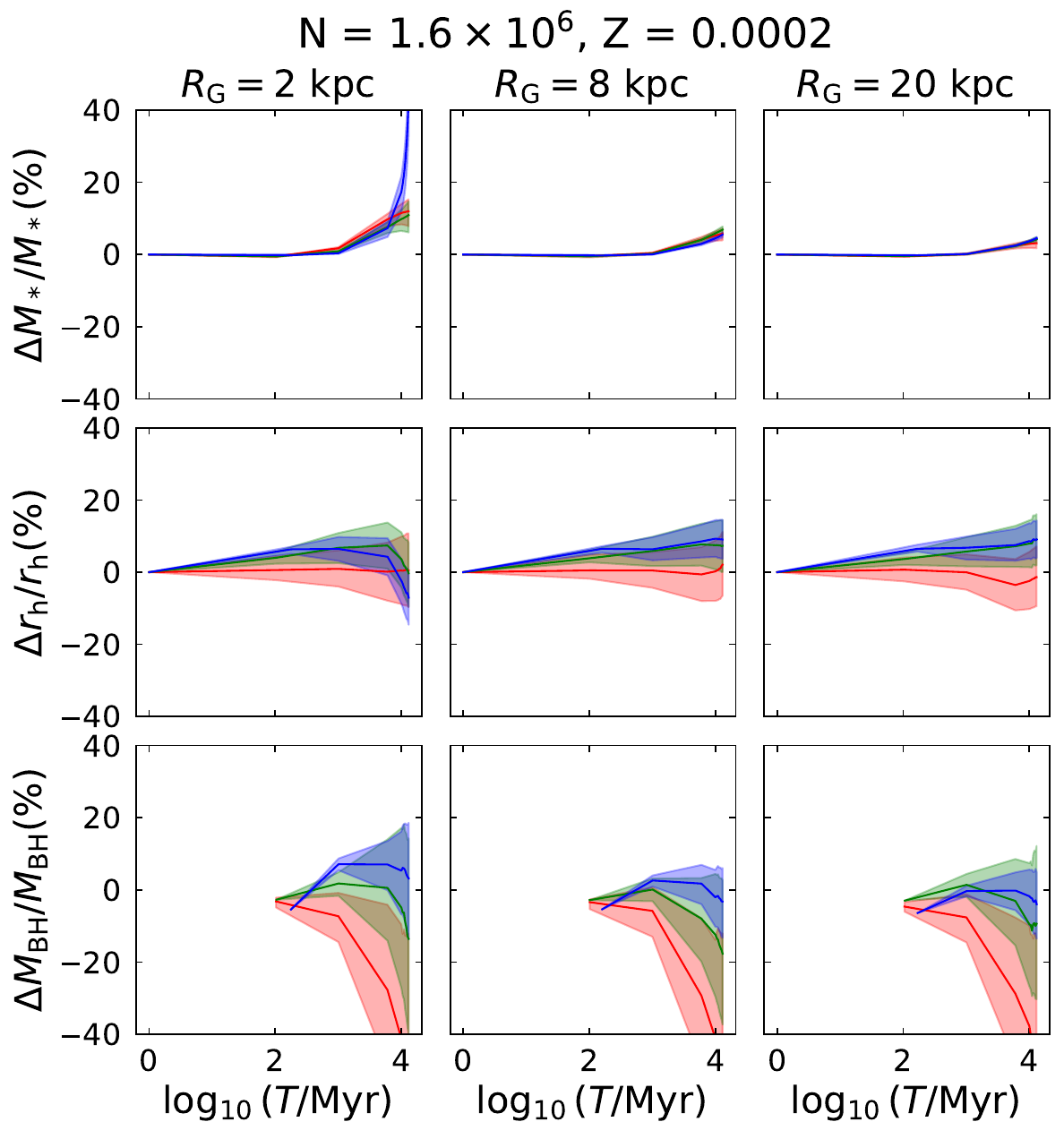}\\
\caption{Same as Figure~\ref{Fig_N8e5_Z0.0002}, but for models with $N=1.6\times10^6$ and $Z=0.0002$.}
\label{Fig_N1.6e6_Z0.0002}
\end{figure*}

\begin{figure*}
\includegraphics[width=\columnwidth]{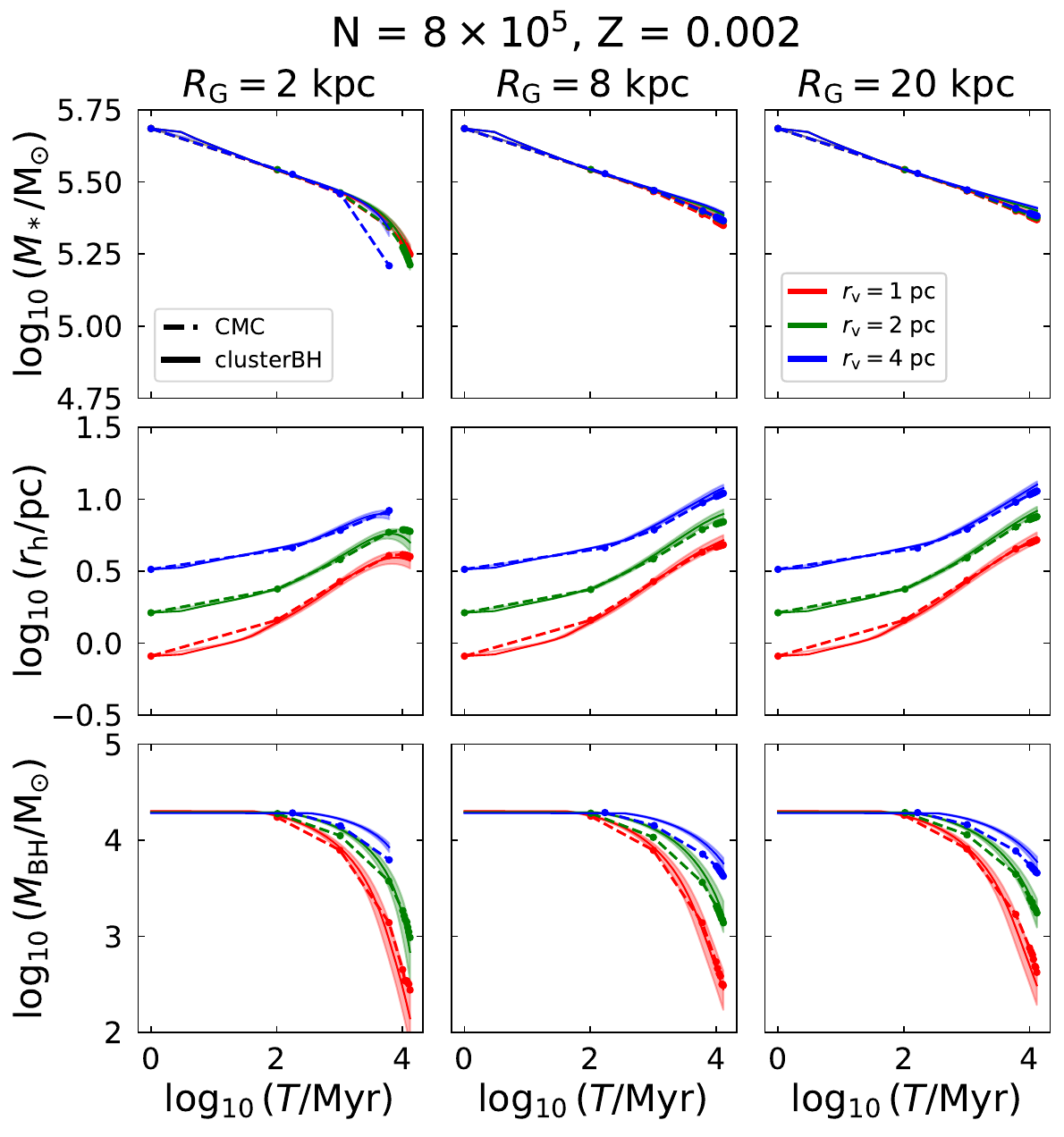}
\includegraphics[width=\columnwidth]{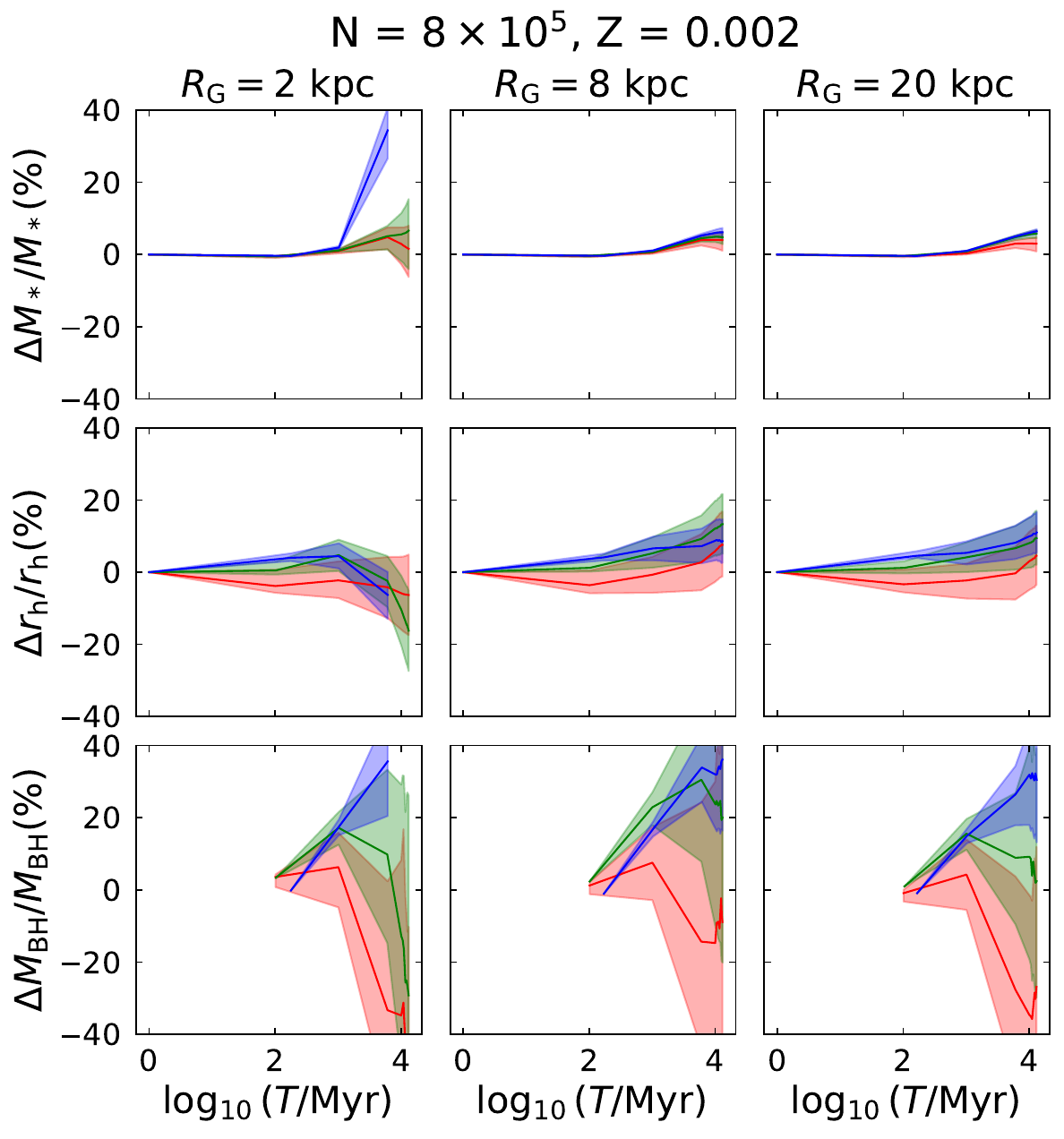}\\
\caption{Same as Figure~\ref{Fig_N8e5_Z0.0002}, but for models with $N=8\times10^5$ and $Z=0.002$.}
\label{Fig_N8e5_Z0.002}
\end{figure*}

\begin{figure*}
\includegraphics[width=\columnwidth]{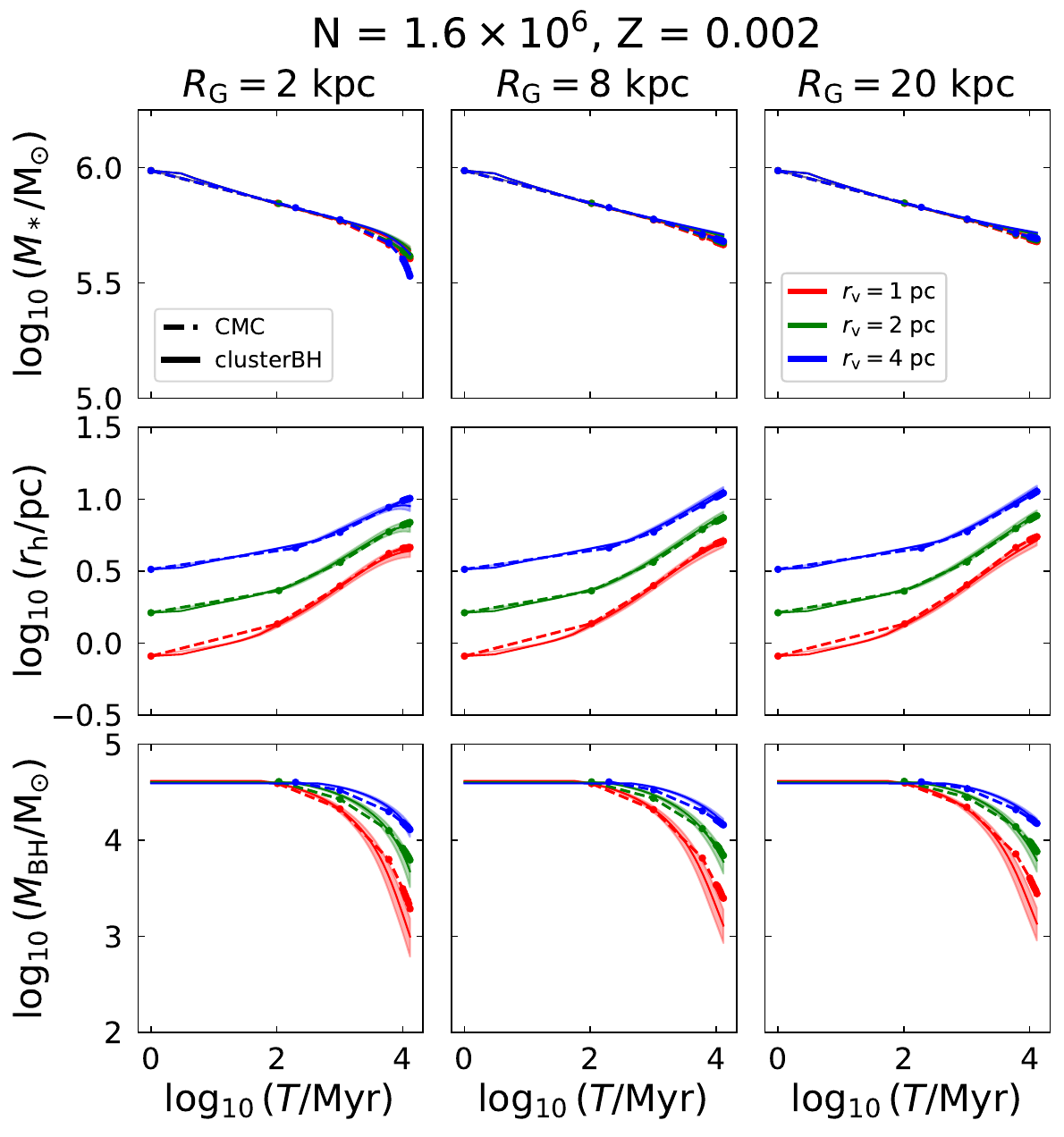}
\includegraphics[width=\columnwidth]{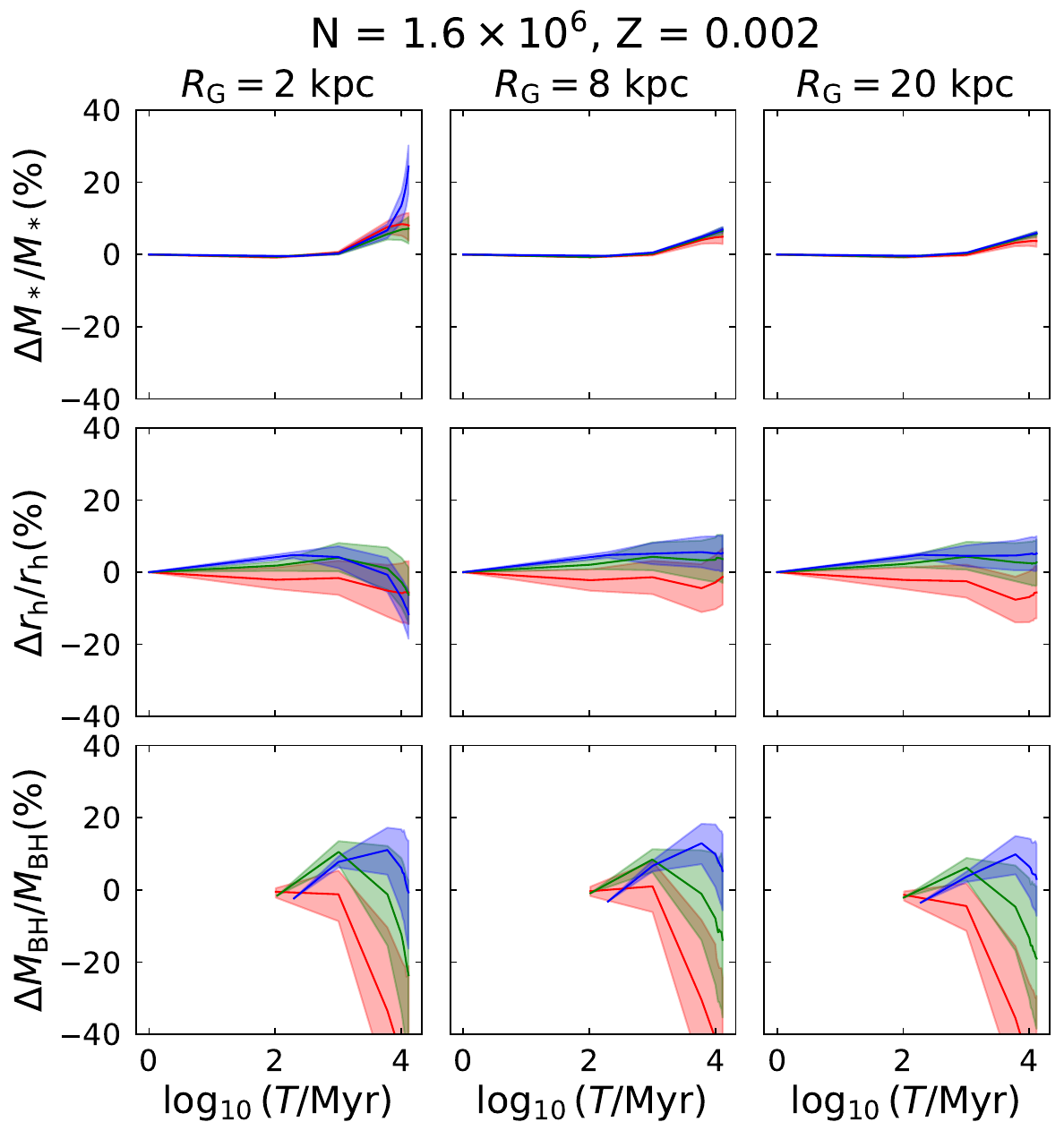}\\
\caption{Same as Figure~\ref{Fig_N8e5_Z0.0002}, but for models with $N=1.6\times10^6$ and $Z=0.002$.}
\label{Fig_N1.6e6_Z0.002}
\end{figure*}

\begin{figure*}
\includegraphics[width=\columnwidth]{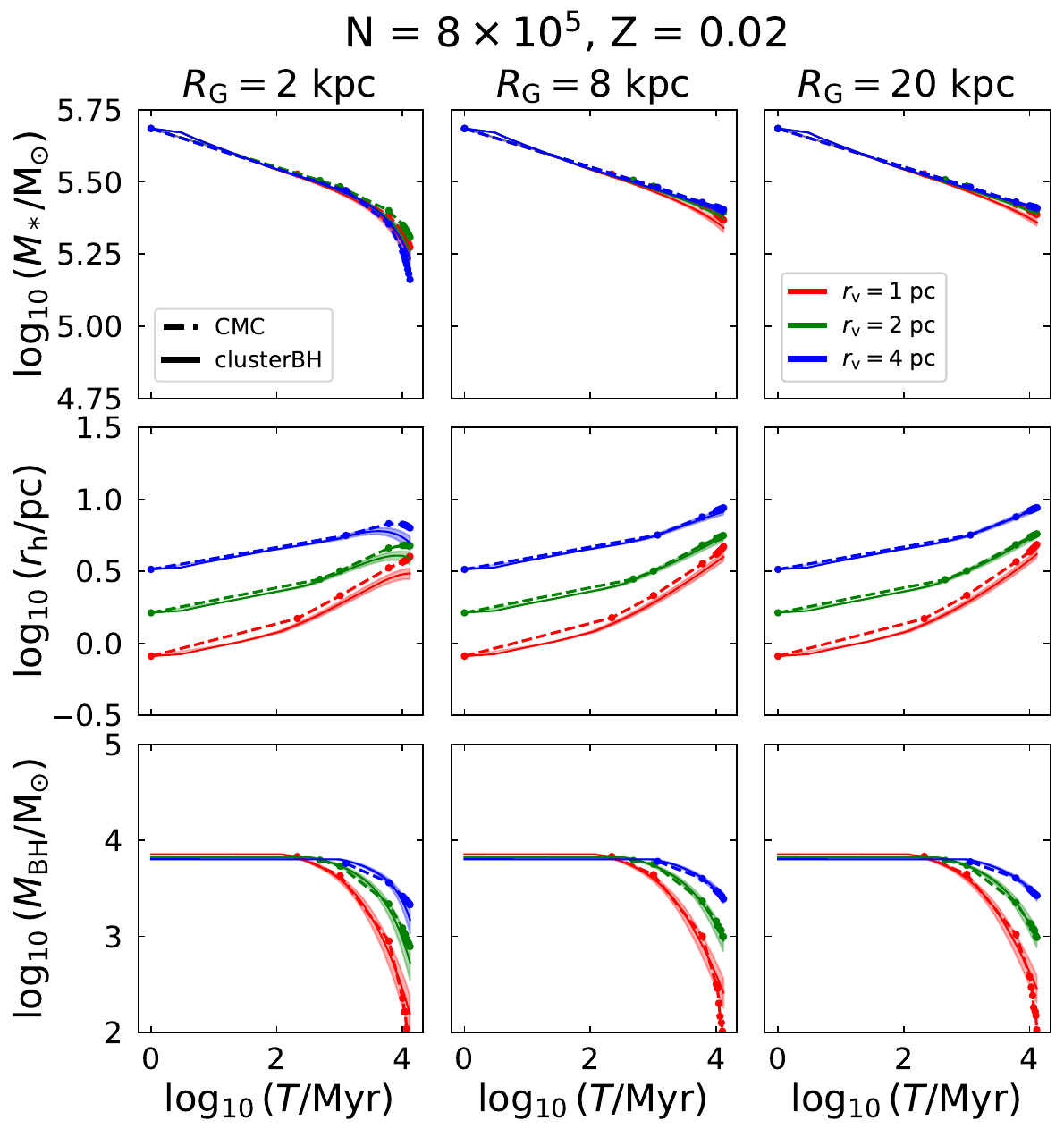}
\includegraphics[width=\columnwidth]{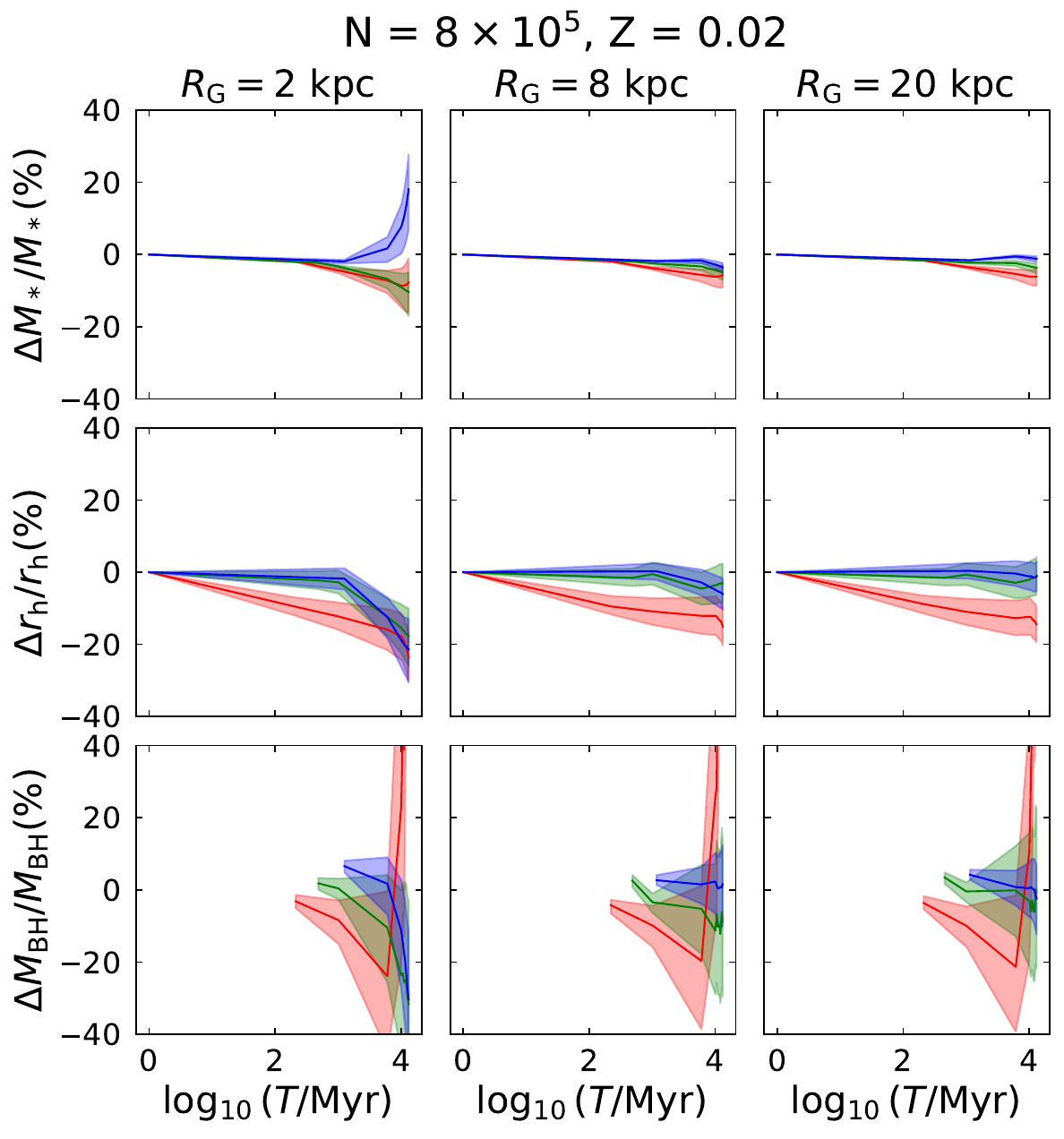}\\
\caption{Same as Figure~\ref{Fig_N8e5_Z0.0002}, but for models with $N=8\times10^5$ and $Z=0.02$.}
\label{Fig_N8e5_Z0.02}
\end{figure*}
\begin{figure*}
\includegraphics[width=\columnwidth]{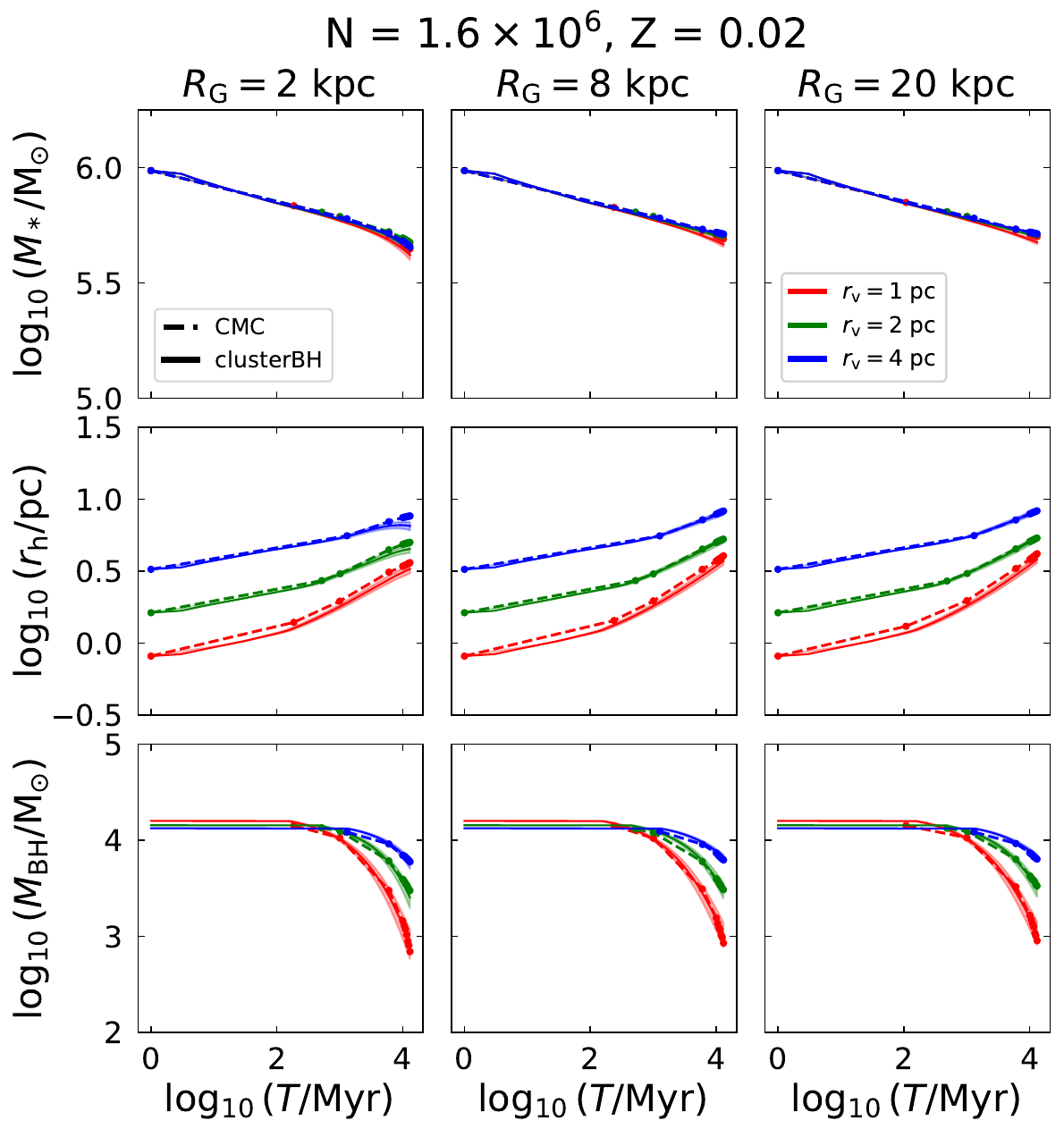}
\includegraphics[width=\columnwidth]{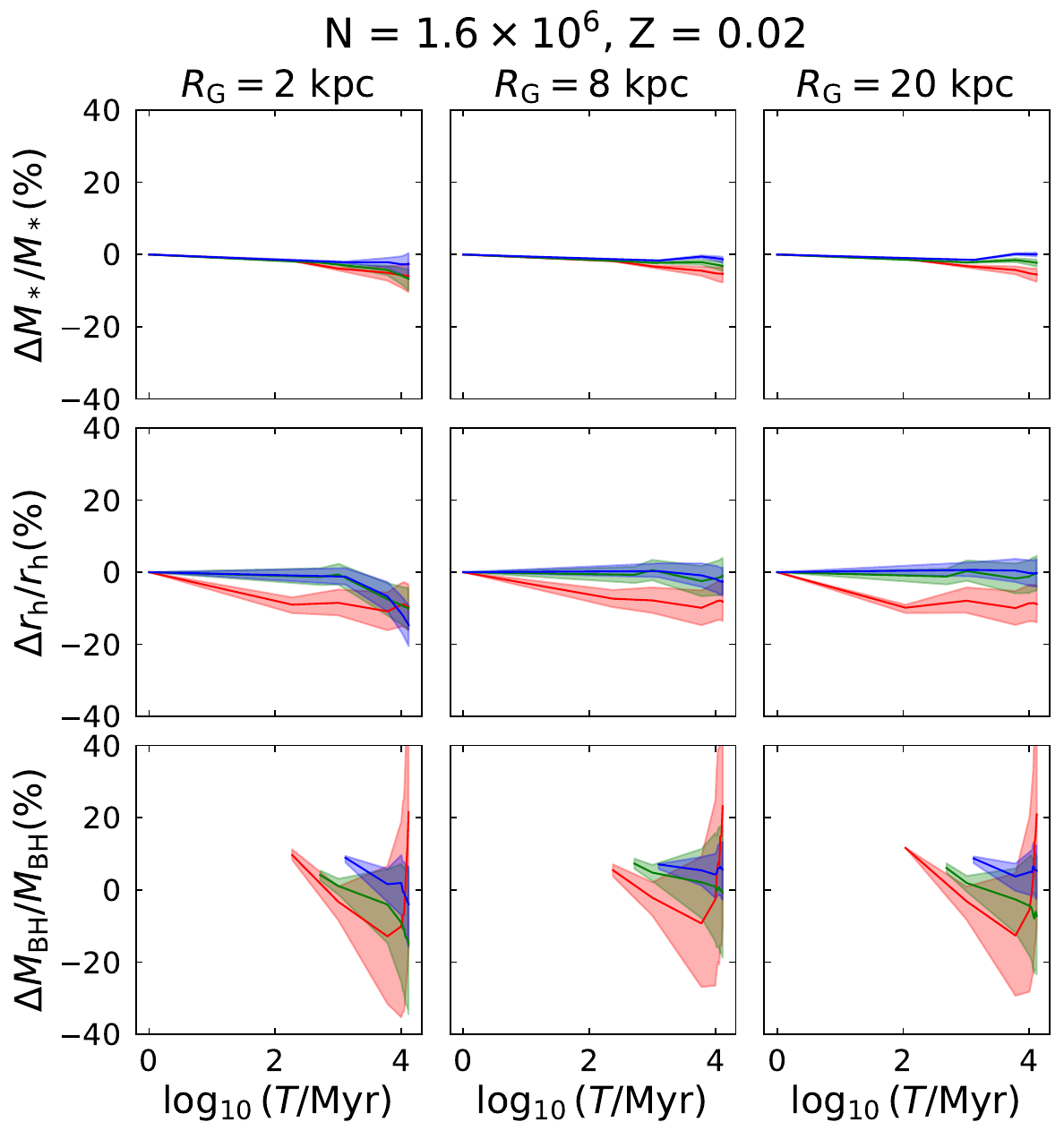}\\
\caption{Same as Figure~\ref{Fig_N8e5_Z0.0002}, but for models with $N=1.6\times10^6$ and $Z=0.02$.}
\label{Fig_N1.6e6_Z0.02}
\end{figure*}

\begin{figure*}
\includegraphics[width=\columnwidth]{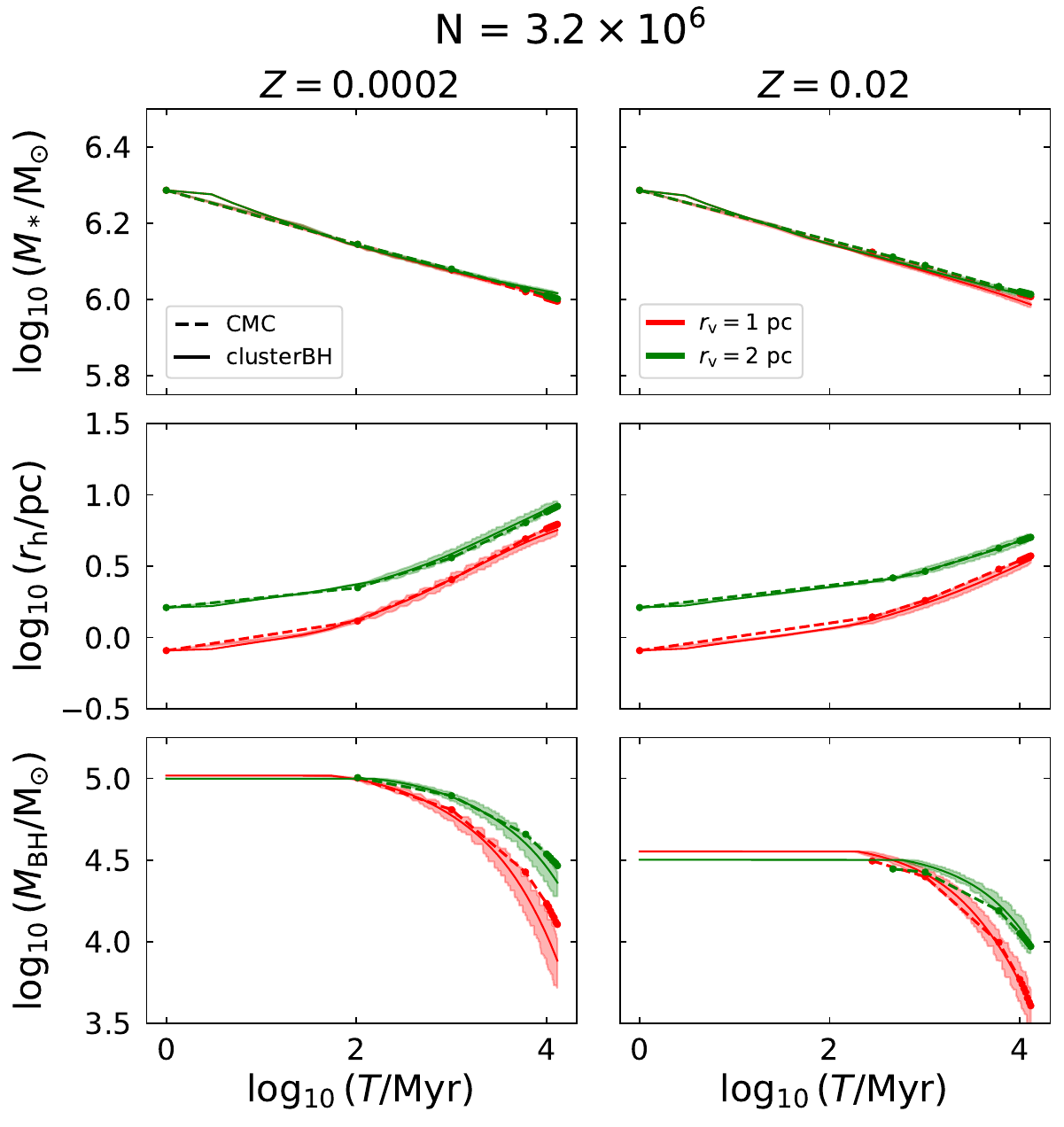}
\includegraphics[width=\columnwidth]{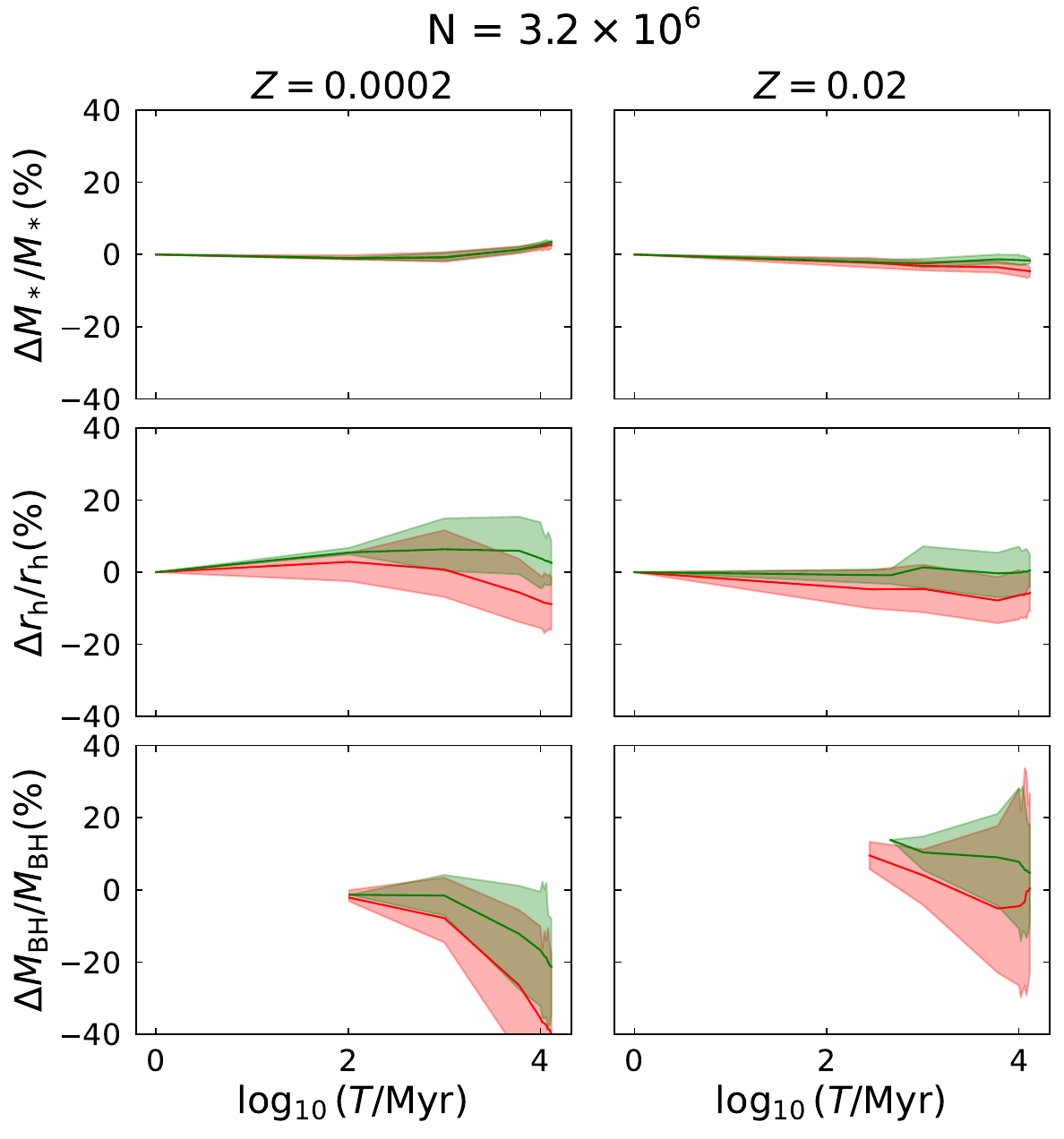}\\
\caption{Predictions of \cbh\ (solid lines) compared to \cmc\,(dashed lines) for clusters with initial $N=3.2\times10^6$ stars at $\RG=20\,\kpc$. Columns show the results for metallicities $Z=[0.0002, 0.02]$ from left to right, for $\rv=[1,2]\,\pc$ with colors red and green respectively.}
\label{Fig_N3.2e6}
\end{figure*}

\subsection{Likelihood}
To constrain the eight fitting parameters for \cbh, we use three quantities: $\Mst$, $\Mbh$, and $\rh$. We assume that the uncertainty of the \cmc\ models  for each quantity in each snapshot is proportional to the value itself. Specifically, let $X_{i}^{\cmc}$ with $i\in[1,3]$ represent $\Mst$, $\Mbh$, and $\rh$ from the \cmc\,models. Then for the mass of stars and the half-mass radius, the standard deviation is simply (AG20) $\sigma^{\cmc}_{i}=\delta X_{i}^{\cmc}$ where $\delta$ is the same for all quantities $X_{i}^{\cmc}$, and serves as a measure of the average fractional difference between \cbh\ and \cmc, to be determined in the fit. For the BH population, an additional Poisson error is included in $\sigma^{\cmc}_{i}$ such that
\begin{equation}
\label{eq:BHerror}
\sigma^{\cmc}_{\rm BH}=\sqrt{(\delta M^{\cmc}_{\rm BH})^2 + N_{\rm BH}\mbh^2}\, .
\end{equation}
This is because the BH population in certain models is close to depletion. The parameter $\delta$ accounts for both the stochastic nature of the \cmc\ models (different random seeds yield different outcomes) and the systematic approximations of \cbh. This approach follows the prescription of AG20 (their equation~18). By treating $\delta$ as a free parameter in the likelihood, we introduce an uncertainty in the \cmc\ data, preventing the fit from over‑constraining the physical parameters to a specific realisation of the Monte Carlo noise. Because $\delta$ is typically larger than the fractional uncertainties on any single model parameter derived from the fit, we adopt $\delta$ as the primary measure of the model’s predictive accuracy. Such a value serves as a measure of the upper limit of the systematic error of \cbh\ relative to higher‑fidelity models, demonstrating that the simplified model captures the essential physics without being fine‑tuned to the stochastic variations of the \cmc\ grid.

By considering the contribution of each snapshot, the log likelihood can be written as
\begin{equation}
\label{eq:likelihood}
\ln\mathcal{L}=-\sum_{ijk}\bigg[ \frac{1}{2}\bigg(\frac{X_{ijk}^{\cbh}-X_{ijk}^{\cmc}}{\sigma_{ijk}^{\cmc}}\bigg)^2 + \ln\bigg(\sqrt{2\pi}\sigma_{ijk}^{\cmc}\bigg) \bigg]\, ,
\end{equation}
where index $i$ sums over the three cluster quantities, $j$ sums over $58$ cluster models and $k$ sums over the $11$ snapshots, at times approximately $[0, 0.1, 1, 6, 10, 10.55, 11, 11.55, 12, 12.5, 13]\,\Gyr$. We use the same IMF as the \cmc\ models (see Section~\ref{ssec:cmcmodels}).

We use the Bayesian nested sampling tool \dynesty\,\citep{2020MNRAS.493.3132S} to constrain the eight parameters. We adopt a uniform prior probability distribution for each parameter across the range specified in Table~\ref{tab:intervals}. 
\begin{table}[h]
    \centering
    \renewcommand{\arraystretch}{1.2} 
    \begin{tabular}{l c c}
        \toprule
        \textbf{Parameter} & \textbf{Prior Range} & \textbf{Fitted Value} \\
        \midrule
        $\zeta$    & [0.01, 0.15]    & $0.0875^{+0.0030}_{-0.0027}$ \\
        $\beta_0$  & [0.0001, 0.5]   & $0.0679^{+0.0027}_{-0.0025}$ \\
        $\Rht$     & [0.001, 1]      & $0.0711^{+0.0022}_{-0.0021}$ \\
        $\xi_0$    & [0, 1]          & $0.147^{+0.011}_{-0.009}$ \\
        $N_{\trh}$ & [0.1, 5]        & $0.916^{+0.099}_{-0.088}$ \\
        $\eta_1$   & [3, 10]          & $4.17^{+0.12}_{-0.14}$ \\
        $S_0$      & [0.0001, 10]    & $0.629^{+0.011}_{-0.010}$ \\
        $\delta$   & [0.01, 1]       & $0.115^{+0.002}_{-0.003}$ \\
        \bottomrule
    \end{tabular}
    \caption{Fitted parameters for \cbh. Uniform prior ranges used for the fit are shown in the second column, and the third column reports the median posterior estimates with $1\sigma$ uncertainties, defined as the $68\%$ central credible intervals.}
    \label{tab:intervals}
\end{table}

\subsection{Results}
\label{sec:results}
In this section we compare the best-fit \cbh\ models to the \cmc\ models. 
The posterior distributions of the parameters are shown in Fig.~\ref{Fig_paramplots} and the inferred values are listed in Table~\ref{tab:intervals}. 
A direct comparison of the models is available in Figures~\ref{Fig_N8e5_Z0.0002}-\ref{Fig_rhM}. 
We discuss each parameter in turn, building a coherent picture of the physical processes that govern cluster evolution.

The best‑fit value of the dimensionless energy–production rate is $\zeta = 0.0875$, slightly smaller than the value $\zeta=0.1$ used in the isolated \cbh\ model of AG20. 
Interestingly, $\zeta$ is closer to the value derived by  \citet{1965AnAp...28...62H} for isolated clusters ($\zeta=0.0926$) than to the tidally limited cluster value ($\zeta=0.0725$; \citet{1961AnAp...24..369H}). 
The dimensionless BH ejection rate is $\beta_0\zeta = 0.0059$, which matches the theoretical expectation of \citet{breen_heggie_2013} ($0.006$) and is  $2.1$ times larger than the AG20 value\footnote{In AG20 the BH ejection rate differs from equation~(\ref{eq:Mbhdot}) because the energy emission rate $\zeta$ was absorbed into $\beta$.}. 
This increase is because here $\mst$ drops to about half its initial value within the first few $\Myrs$, making relaxation roughly twice as long for the same $M$, $\rh$, and $f_{\rm BH}$ compared to the constant‐$\mst$ assumption in AG20. The exponential model for tides that is used here prefers the set of values $\Rht=0.0711$ and $\xi_0=0.147$. The values obtained here facilitate mass loss for clusters in orbits at small galactocentric distances.

The relaxation factor is $N_{\rm trh}=0.916$, nearly $3$ times smaller than the value $3.21$ used in AG20. 
This difference likely arises because AG20 adopted Plummer models, which have a larger ratio $r_{\rm c}/\rh$ (where $r_{\rm c}$ is the core radius). 
The final energy‐production efficiency parameter $\eta_1$ is $4.17$ and controls the early expansion of the cluster due to mass segregation. It differs from AG20 as it was fixed to $3$, while it is also $5\%$ larger than the value of $4$ used in \emacss\,(even though a differential equation was used for describing the evolution of such parameter, instead of the linear in $t$ formula used here). 

The parameter that governs the suppression of BH ejections is $S_0\simeq0.629$ (Table~\ref{tab:intervals}). 
Table~\ref{tab:S_values} lists the values of $S$ for all models at the first snapshot containing BHs ($t=0.1\;\Gyr$). 
For metal‐rich clusters ($Z=0.02$), the initial BH fraction is low ($f_{\rm BH}\simeq2\%$) and the average BH mass is small ($\mbh\simeq9\;\msun$); consequently these clusters have $S\simeq0.9$, close to the fitted $S_0$. 
As defined in equation~\ref{eq:beta_form}, when $S$ is comparable to $S_0$ the BH ejection rate $\beta$ is suppressed relative to its maximum $\beta_0$, explaining why metal‐rich clusters retain their BHs longer. 

Overall, the fitting error of \cbh\ relative to the \cmc\ models is $\delta\simeq0.115$, i.e. $11.5\%$. Given the simplified nature of \cbh, we consider this an acceptable accuracy. Because this $\sim10\%$ systematic uncertainty dominates over the fractional uncertainties of any individual model parameter derived from the fit, we adopt $\delta$ as the primary measure of our model’s predictive accuracy, representing a balance between model simplicity and predictive power.

The \cbh\ model was calibrated on 58 \cmc\ models with $M\ge 8\times10^5,\msun$ and $\rv\ge 1,\pc$, yielding $\sim1900$ data points for 8 free parameters, making overfitting unlikely. It generalizes well to lower-mass clusters and to models outside the fitting range (Appendix~D), though discrepancies increase for very compact clusters ($\rv=0.5,\pc$). We therefore recommend applying \cbh\ for $\rv\gtrsim 1,\pc$, and use caution at smaller radii.

\begin{table*}[htbp]
\centering
\caption{Predicted value of $S$ from \cbh\, (left) versus numerical value from \cmc\,(right) at $t=0.1\,\Gyr$ for all models considered.}
\resizebox{\textwidth}{!}{%
\begin{tabular}{|c|ccc|ccc|ccc|}
\hline
\textbf{Virial Radius} 
& \multicolumn{3}{c|}{$Z = 0.0002$} 
& \multicolumn{3}{c|}{$Z = 0.002$} 
& \multicolumn{3}{c|}{$Z = 0.02$} \\
\cline{2-10}
& $8\times10^5$ & $1.6\times10^6$ & $3.2\times10^6$ & $8\times10^5$ & $1.6\times10^6$ & $3.2\times10^6$ & $8\times10^5$ & $1.6\times10^6$ & $3.2\times10^6$ \\
\hline
1 pc & 7.44 $\mid$ 6.87 & 8.36 $\mid$ 7.38 & 8.80 $\mid$ 7.86 & 5.16 $\mid$ 5.01 & 5.65 $\mid$ 5.33 & -- $\mid$ -- & 0.86 $\mid$ 0.80 & 0.87 $\mid$ 0.90 & 0.98 $\mid$ 1.02 \\
2 pc & 8.72 $\mid$ 8.09 & 9.22 $\mid$ 8.34 & 9.31 $\mid$ 8.50 & 6.05 $\mid$ 5.90 & 6.36 $\mid$ 5.92 & -- $\mid$ -- & 0.89 $\mid$ 0.91 & 0.97 $\mid$ 1.00 & 0.98 $\mid$ 1.09 \\
4 pc & 9.46 $\mid$ 8.40 & 10.08 $\mid$ 8.78 & -- $\mid$ -- & 6.85 $\mid$ 6.31 & 7.15 $\mid$ 6.43 & -- $\mid$ -- & 0.99 $\mid$ 1.03 & 1.05 $\mid$ 1.11 & -- $\mid$ -- \\
\hline
\end{tabular}%
}
\label{tab:S_values}
\end{table*}

\subsubsection{Dependence on $N$}

For a given $\rv$, $\RG$ and metallicity, more massive clusters have longer relaxation times and therefore retain more BHs over a Hubble time. In the \cmc\ models, clusters with $N=8\times10^{5}$ lose their BHs earlier than those with $N=1.6\times10^{6}$ or $3.2\times10^{6}$. \cbh\ reproduces this trend accurately. For example, at $Z=0.0002$ and $\RG=20\,\kpc$ (as seen from $M_{\rm BH}(t)$ in Figures~\ref{Fig_N8e5_Z0.0002}-\ref{Fig_N8e5_Z0.02}), the BH mass at $13\,\Gyr$ increases from $\sim 10^2\,\msun$ to $\sim 10^3\,\msun$ as the initial number of stars rises by a factor of four. This demonstrates that more massive clusters retain a larger fraction of their BH population over a Hubble time, in agreement with the \cmc\ models. The initial BH mass $M_{\mathrm{BH,0}}$ is within $\lesssim10\%$ of the \cmc\ value, and the subsequent evolution of $\Mst$, $\rh$ and $\Mbh$ agrees to within 20–30\% for most models (Figures.~\ref{Fig_N8e5_Z0.0002}-\ref{Fig_N3.2e6}). The agreement is best for the largest clusters ($N=3.2\times10^{6}$, Figure~\ref{Fig_N3.2e6}) because they are only available at $\RG=20\,\kpc$ where tides are weak and internal dynamics dominate. The small discrepancies that appear at low metallicity and large $\rv$ (e.g., $N=8\times10^{5}$, $\rv=4\,\pc$, $Z=0.0002$) are due to a  feedback loop: an overestimate of $\Mbh$ increases $\rh$, increases the relaxation time, reduces the BH ejection rate, and leads to a further overestimate of $\Mbh$ and $\rh$. Despite this, the overall performance for $N\gtrsim8\times10^{5}$ is satisfactory.

\subsubsection{Dependence on virial radius}

\begin{figure}
\includegraphics[width=\columnwidth]{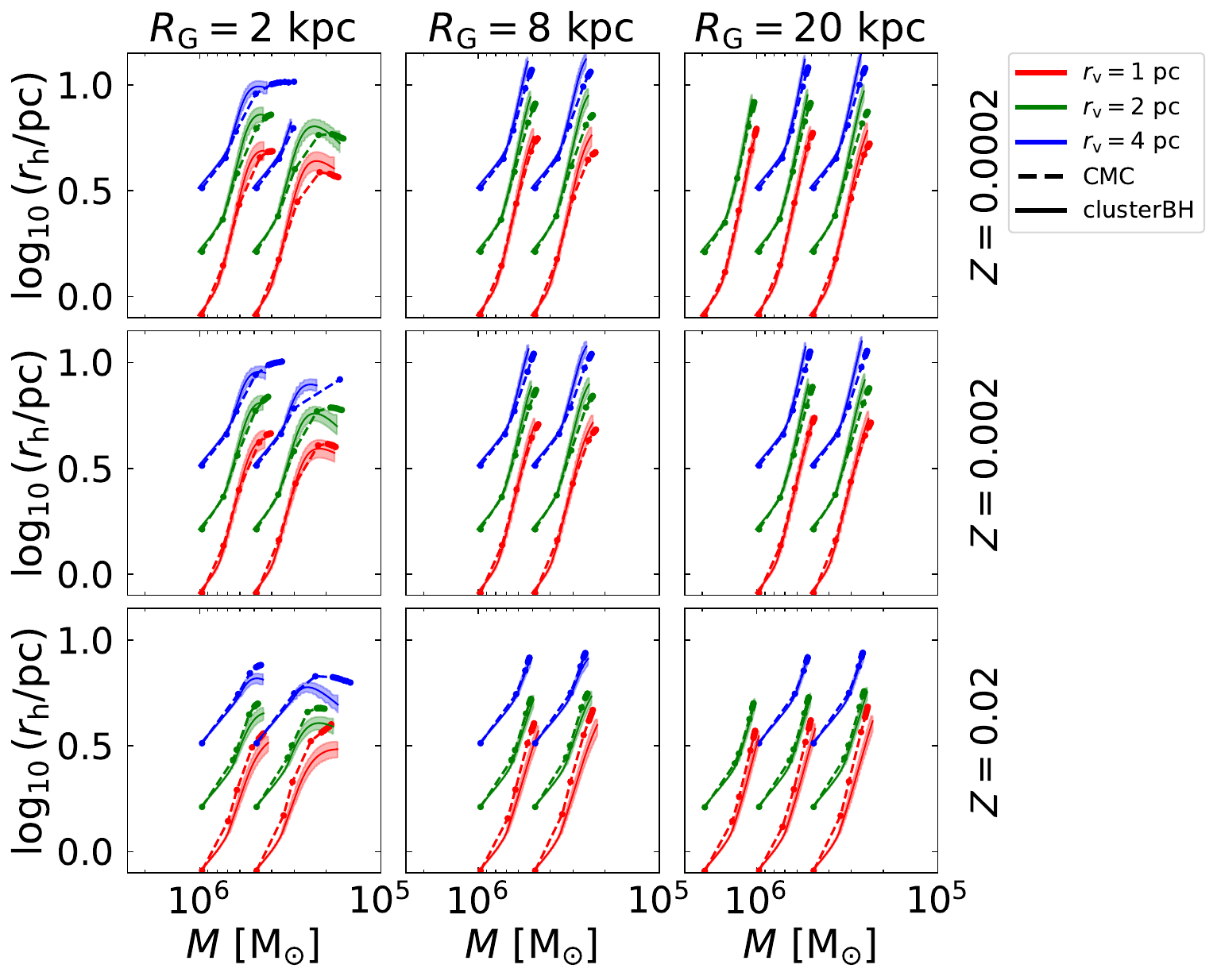}
\caption{Evolution of $\rh$ as a function of $M$ for all \cmc\, models used in the fitting. The \cbh\ prediction is shown for comparison.}
\label{Fig_rhM}
\end{figure}

The virial radius $\rv$ sets the initial density $\rhoh\propto M\rv^{-3}$. Denser clusters (smaller $\rv$) have shorter relaxation times and eject their BHs more rapidly. In the \cmc\ models, compact clusters ($\rv=1\,\pc$) lose most of their BHs within a few $\Gyr$, while extended clusters ($\rv=4\,\pc$) retain BHs for much longer, driving a strong expansion of $\rh$ (see \citealt{2020ApJS..247...48K}). \cbh\ captures this behaviour well, but with a systematic trend. For $\rv=1\,\pc$ it slightly underestimates $\Mbh$ at late ages (especially for metal‑poor clusters), whereas for $\rv=4\,\pc$ it tends to overestimate $\Mbh$ and $\rh$ (Figures~\ref{Fig_N8e5_Z0.0002}–\ref{Fig_N1.6e6_Z0.02}). The $\rv=2\,\pc$ models show the best agreement. This pattern is explained by the same feedback loop described above: an initial offset in $\Mbh$ is amplified because the relaxation time depends on the BH population. The fact that the most compact clusters are slightly under‑predicted in $\Mbh$ may also reflect that \cmc\ includes a small binary fraction ($5\%$), which can enhance BH ejection via binary–single interactions – an effect not  included in \cbh. Nevertheless, the overall behaviour as a function of initial density is correctly reproduced. Note that BHs break the self‑similar evolution in the $M-\rh$ plane of post‑collapse, equal‑mass clusters \citep{1961AnAp...24..369H,1965AnAp...28...62H,2012MNRAS.422.3415A,2014MNRAS.437..916G}, as shown in Figure~\ref{Fig_rhM}.

\subsubsection{Dependence on galactocentric distance}

The tidal field strength is set by $\RG$ (circular orbit, $\Vc=220\,\kms$. At large distances ($\RG=20\,\kpc$) the tidal radius is large, evaporation is negligible, and clusters evolve as effectively isolated systems. Here \cbh\ and \cmc\ agree exceptionally well for all quantities ($\RG=20\,\kpc$ panels in Figures.~\ref{Fig_N8e5_Z0.0002}–\ref{Fig_N3.2e6}), confirming that the internal dynamics (stellar evolution, relaxation, BH ejections) are correctly calibrated. At intermediate distances ($\RG=8\,\kpc$) tides start to matter for extended clusters ($\rv=4\,\pc$), and \cbh\ captures the increased mass loss with only minor deviations. The reason why sparse clusters are affected first is because the tidal mass-loss rate is sensitive to the fraction $\rh/\rt$ which is large for large $\rh$ and/or small $\RG$. At small galactocentric distances ($\RG=2\,\kpc$), tides are strong and drive rapid dissolution, especially for low‑density clusters. The exponential expression for the dimensionless evaporation rate (equation~\ref{eq:xi}) with the fitted parameters $\Rht=0.0711$ and $\xi_{0}=0.147$ successfully reproduces the evaporation mass loss seen in \cmc. The only notable discrepancy is that for the lowest metallicity and largest $\rv$ ($Z=0.0002$, $\rv=4\,\pc$), \cbh\ predicts slightly slower mass loss than \cmc\ (Figure ~\ref{Fig_N8e5_Z0.0002}, right column). This suggests that the most extended, metal‑poor clusters may require an even stronger tidal efficiency. Nevertheless, the same tidal parameters work across all metallicities and masses, validating the prescription.

\subsubsection{Dependence on metallicity}

Metallicity controls the initial BHMF: metal‑poor clusters ($Z=0.0002$) form many massive BHs that are retained ($f_{\rm BH}\sim7-8\%$, $\langle m_{\rm BH}\rangle\sim20\,\msun$), giving a large Spitzer parameter $S$; metal‑rich clusters ($Z=0.02$) form fewer and lighter BHs ($f_{\rm BH}\sim2\%$, $\langle m_{\rm BH}\rangle\sim9\,\msun$), resulting in a smaller $S$. At the time of core collapse, the initial BH mass predicted by \cbh\ agrees with the \cmc\ models to within $\sim10\%$ for metal‑poor clusters, and even better for $Z=0.02$.\footnote{The small discrepancy in the initial BHMFs between \cbh\ and the \cmc\ grid likely arises from differences in the underlying stellar and binary evolution prescriptions. While both use similar assumptions for supernovae kicks and remnant‑mass relations, details in the implementation may vary. The stellar evolution module employed by the \cmc\ public grid is an earlier version of \textsc{COSMIC} \citep{2020ApJ...898...71B}, while \ssp\ is based on \sse\ \citep{2020A&A...639A..41B}, and does not account for, e.g., binary systems. Moreover, \cmc\ adopts a $5\%$ binary fraction. While \ssp\ closely reproduces the results of the latest \textsc{COSMIC} version, it is not identical to it.} Adjusting the BH ejection rate with an $S$‑dependent 
$\beta$ is especially important for metal‑rich clusters, as fewer and on average lower mass BHs form inside them, which results in already sufficiently low $S$ and therefore inefficient ejection, as described in Section \ref{sect:bhejec}. In the \cmc\ models, metal‑poor clusters eject most BHs within a few $\Gyr$, while metal‑rich clusters retain BHs for many $\Gyr$ because the ejection rate $\beta$ is suppressed when $S$ is close to $S_{0}$ (equation~\ref{eq:beta_form}). \cbh\ agrees well with \cmc\ for $Z=0.0002$ over the first $\sim10\,\Gyr$; at later ages the fractional differences in $\Mbh$ can reach up to $40\%$. However, at those late times the absolute BH mass is very small (often $\lesssim100\,\msun$), and the large fractional error is dominated by Poisson noise in the \cmc\ models. For $Z=0.02$ the agreement is remarkably good (Figures~\ref{Fig_N8e5_Z0.02}–\ref{Fig_N1.6e6_Z0.02}), with all quantities within $20\%$ over $13\,\Gyr$. This success directly demonstrates the necessity of the $S$-dependent $\beta$: without it, \cbh\ would eject BHs far too quickly in metal‑rich clusters. The intermediate metallicity $Z=0.002$ shows a smooth transition, and the small remaining offsets are within the global $\sim10\%$ error $\delta$. A minor discrepancy persists for the most compact metal‑poor clusters ($\rv=1\,\pc$, $Z=0.0002$), where \cbh\ slightly underestimates $\Mbh$; this may also be related to the absence of primordial binaries in \cbh, as noted in the footnote.

\subsection{Performance of \cbh\ at 13 $\Gyr$}
\begin{figure}
\includegraphics[width=1\columnwidth]{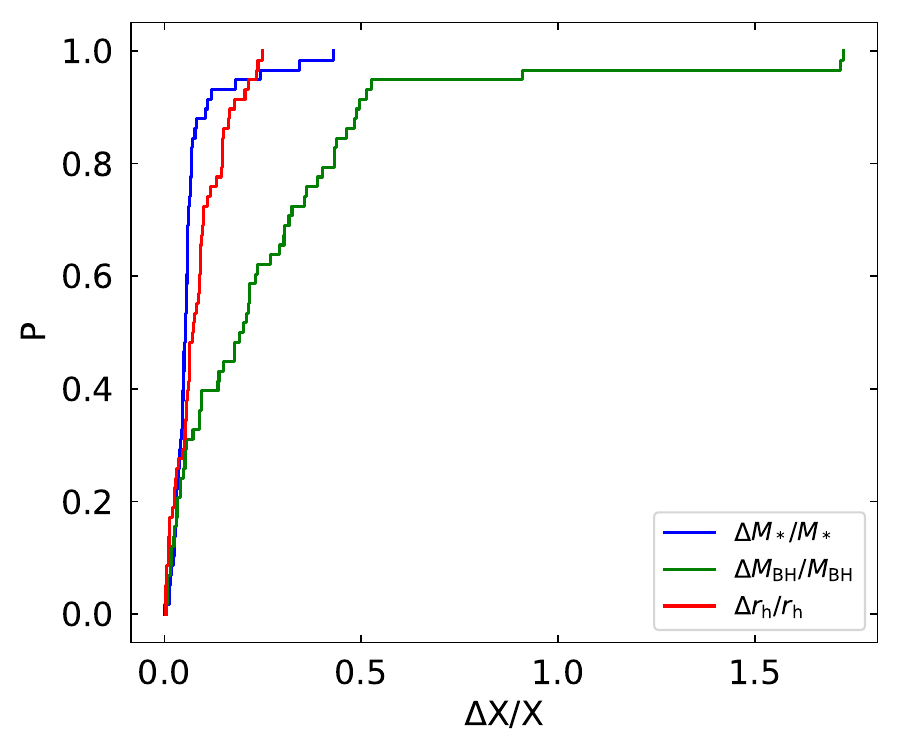}
\caption{Comparison between \cbh\,and \cmc\,at the final snapshot. The y-axis represents the cumulative probability distribution of the fractional differences. All models have a difference in $\rh$ below $30\%$, $96.5\%$ for $\Mst$ and $60\%$ for $\Mbh$ for all \cmc\, models fitted.}
\label{Fig_Final_Values}
\end{figure}

The new \cbh\ version generally reproduces most of the \cmc\ models accurately at different times, but for applications to GCs, it is most relevant how the performance is at old ages. Figure~\ref{Fig_Final_Values} shows the comparison of $\Mst$, $\Mbh$, and $\rh$ between \cbh\ and \cmc\ at the time of the final snapshots for all models at $t=13\,\Gyr$. If the clusters were not evolved up until $13\, \Gyr$, the final snapshot available was used. Considering a tolerance of $30\%$ in the fractional difference, $96.5\%$ of models agree for $\Mst$, $100\%$ for $\rh$, and $59\%$ for $\Mbh$. The agreement for $\Mbh$ improves to $83\%$ when a $50\%$ tolerance is used. A few \cbh\ models overestimate the BH fraction by more than $100\%$. These outliers correspond to clusters with metallicity $Z=0.02$, or to models having $N = 8\times10^{5}$, $\rv = 2\,\pc$ or $4\,\pc$, and $Z = 0.0002$. In these cases the absolute BH mass in the \cmc\ model is very small (typically $\lesssim100\,\msun$), while \cbh\ predicts a few hundred solar masses. Such large fractional differences arise from modest absolute discrepancies (a few $10^2\,\msun$) in the regime where the BH population is nearly depleted. Moreover, in this stochastic regime the \cmc\ results themselves are subject to Poisson fluctuations because they track individual BHs; rerunning the same \cmc\ model with a different random seed can change the outcome by a similar amount. Hence, the outliers do not indicate a systematic failure of \cbh\ but rather highlight the limitations of comparing deterministic models against Monte Carlo simulations when only a handful of BHs remain.

\subsection{Comparison with $N$-body models}
\label{sect:nbody}
\begin{figure*}
\includegraphics[width=\columnwidth]{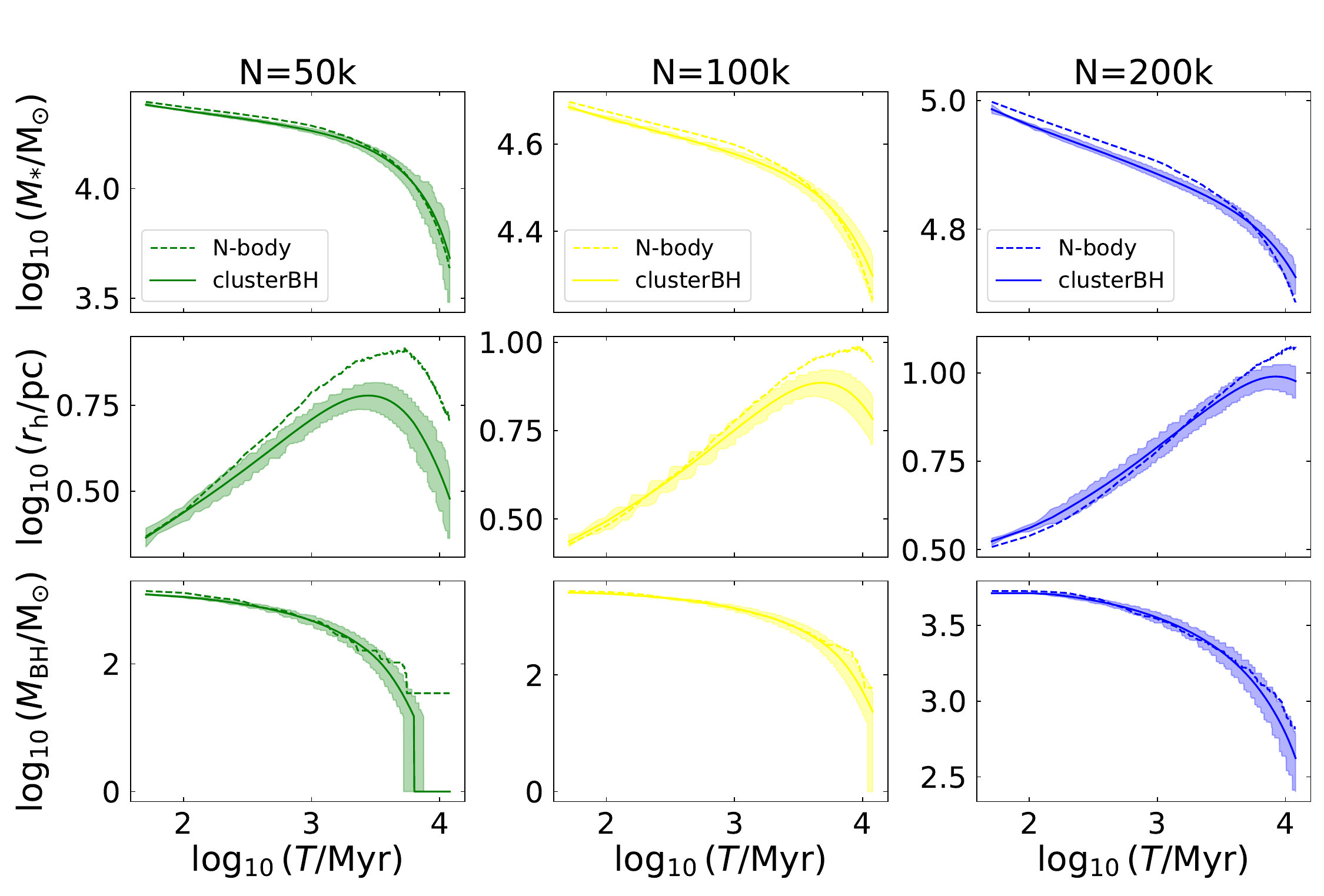}
\includegraphics[width=\columnwidth]{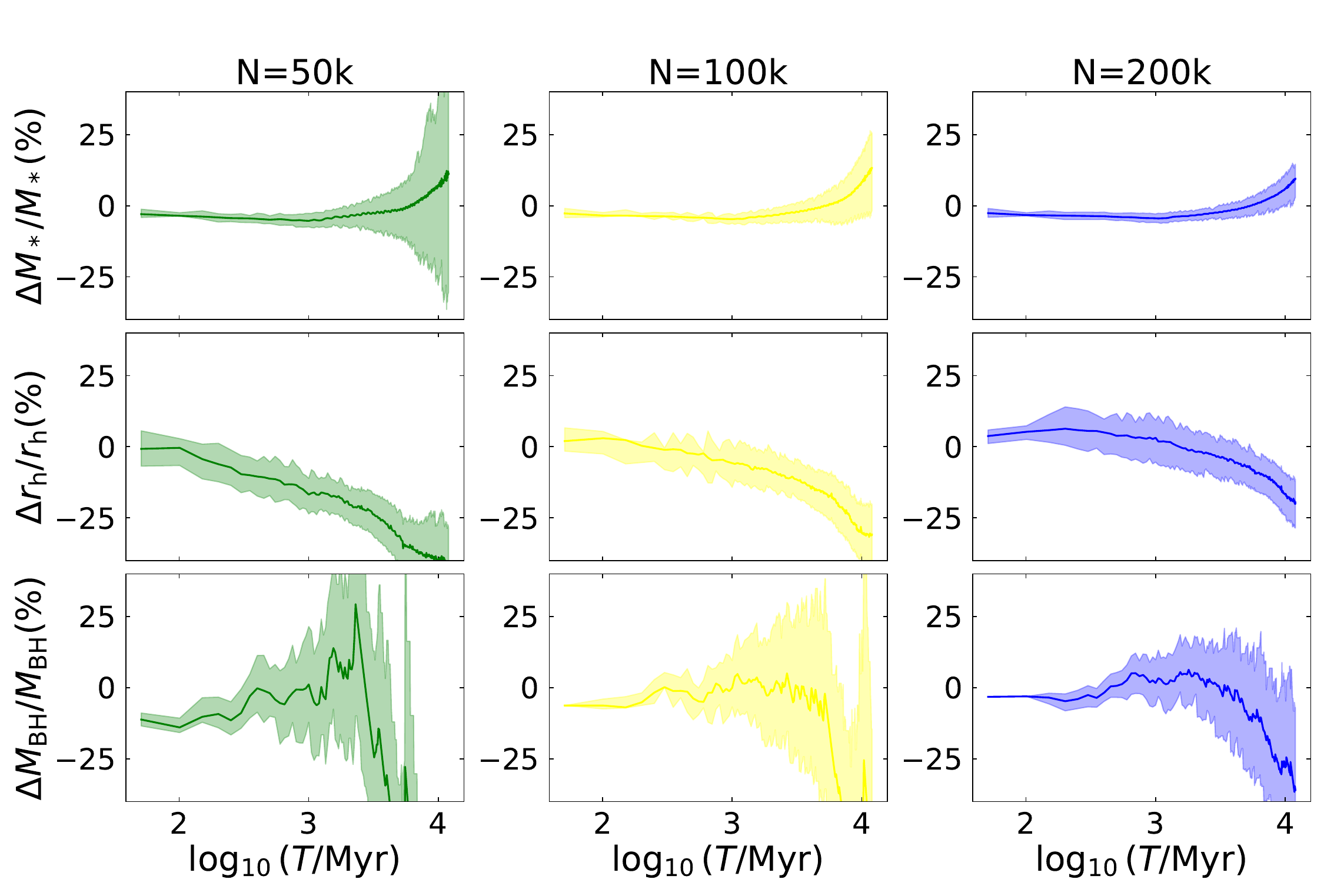}\\
\caption{Comparison between \cbh\,and $N$-body models with initial density $\rho_h=1000\,\msun\pc^{-3}$ and metallicity $[\mathrm{Fe/H}]=-1.5$.}
\label{Fig_Nbody}
\end{figure*}

We now make a brief comparison between \cbh\ and the $N$-body models studied in \cite{2021NatAs...5..957G}. We use the models with an initial density of $\rho_{\rm h} = 10^3 \, \msun\ \pc^{-3}$. The metallicity is $[\mathrm{Fe/H}]=-1.5$, and the models adopt a Kroupa IMF between $[0.1, 100]\,\msun$ resulting in an initial average mass $\m_0=0.638\,\msun$. The clusters are on eccentric orbits, and the tidal mass loss can be approximated by a circular orbit with galactocentric distance of $\RG=9.15\,\kpc$ \citep{2023MNRAS.522.5340G} with circular velocity $V_{\rm c}=220\,\kms$. In Fig.\ref{Fig_Nbody} the comparison with \cbh\, is available. Although these $N$-body models were not included in the fitting, the agreement is satisfactory. The code is therefore reliable across a wide range of cluster masses, from $\sim10^4\,M_\odot$ (tested against $N$‑body models) up to several $10^6\,M_\odot$.

\section{Predictions for BBH mergers}
\label{sect:bbhmergers}
In this Section, we compare the predicted number of BBH mergers in \cbhbd\ to the results from \cmc\ and the lower-mass $N$-body models of \cite{banerjee_2021}.

\begin{figure*}
    \centering
    \includegraphics[width=17cm]{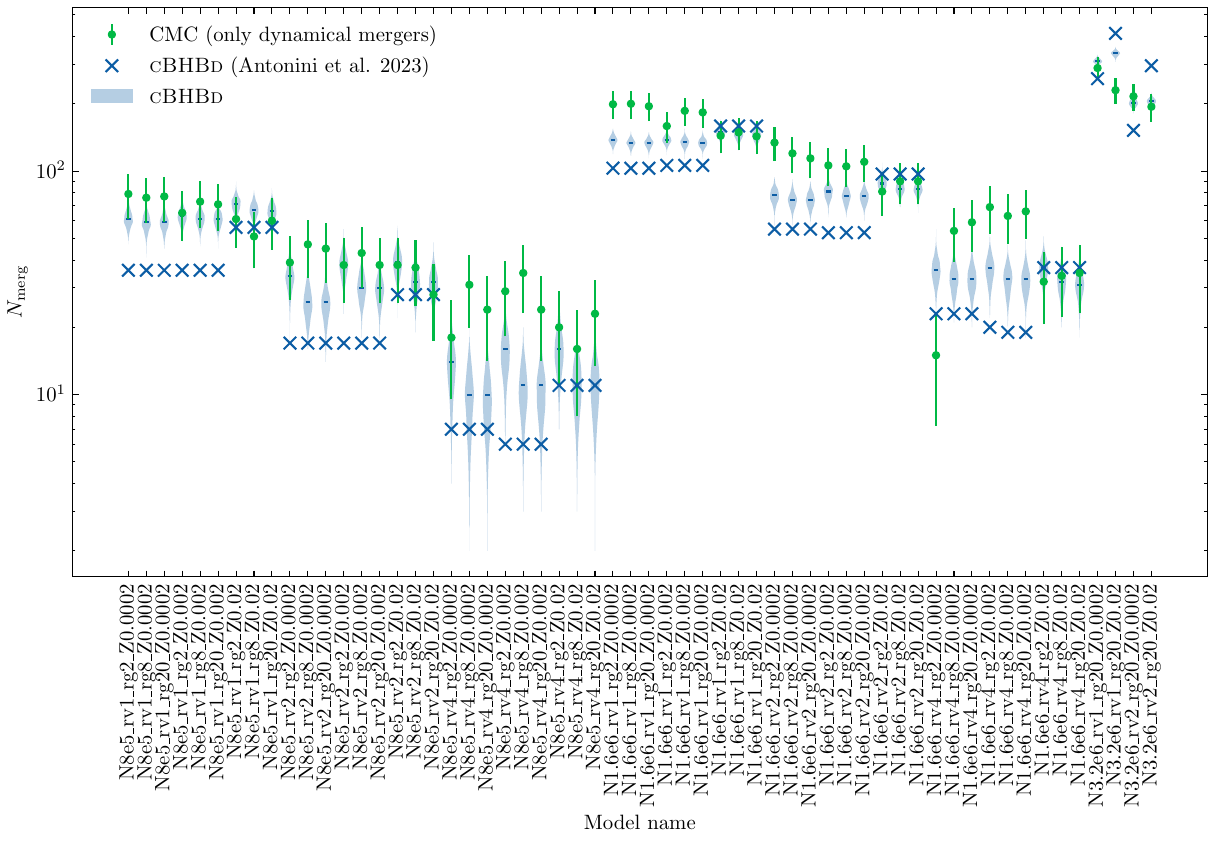}
    \caption{Detailed comparison of the number of dynamical BBH mergers in the new version of \cbhbd\ (light blue) and \cmc\ (green), separated by initial conditions. The gray area represents the distribution of the number of mergers in \cbhbd, found using 1000 runs for each set of initial conditions; the median is marked with a vertical bar. The green errorbars represent the 95\% confidence intervals in the \cmc\ models, estimated from a single run for each set of initial conditions. The blue crosses represent the median of the number of mergers when using the previous version of \cbhbd.}
    \label{fig:Nm_model}
\end{figure*}

To do this, we use the initial conditions of the \cmc\ models and run them using both the new and previous versions of \cbhbd. The results are shown in Fig.~\ref{fig:Nm_model}. We find that the updated \cbhbd\ can accurately reproduce the number of dynamical mergers for most of the parameter space of initial conditions. The typical factor-of-a-few discrepancies between \cbhbd\, and \cmc\, are comparable to the systematic uncertainties inherent in population synthesis predictions for the GC channel, which can exhibit variations of similar magnitude even when using the same physical model (see e.g., \cite{2022LRR....25....1M}, their Figure. 3). For models with lower densities, the number of predicted mergers shows a systematic deviation, with \cbhbd\ underpredicting them. This discrepancy likely arises because the \cmc\ models include a $5\%$ binary fraction, making mergers triggered by dynamical interactions involving primordial binaries non‑negligible in low‑density environments.

\begin{figure}
    \centering
    \includegraphics[width=\columnwidth]{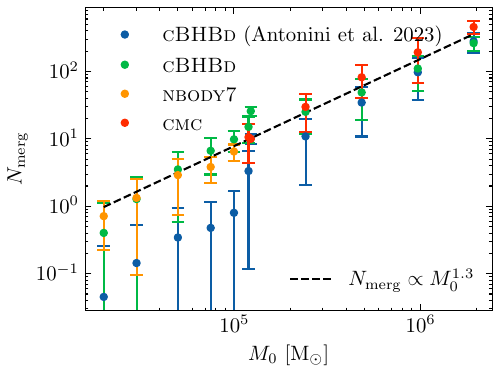}
    \caption{Comparison of the number of dynamical BBH mergers, binned by initial cluster mass, for the different codes. In red, the \cmc\ models; in yellow, the \nbodys\ models of \cite{banerjee_2021}; in green, the new version of \cbhbd; and in blue, the previous version of \cbhbd\ \citep{antonini_gieles_2023}.}
    \label{fig:Nm-M0}
\end{figure}

The discrepancies between the previous and updated versions of \cbhbd\ can be highlighted by binning in cluster masses, as shown in Figure~\ref{fig:Nm-M0}. The new version has an overall better agreement with the \cmc\ models, with a median percentual difference of $\lesssim 20\%$. The agreement with $N$-body models is especially improved at lower cluster masses ($\leq 10^5\,\msun$), where the previous version of \cbhbd\ underpredicts the number of mergers. The overall improvement is due to the inclusion of GW captures in direct interactions, which account for $\sim 20\%$ of all mergers, while the improvement in the low-mass regime is due to the inclusion of GW captures during BBH-BBH interactions. 

While it was previously reported that the mass dependence of BBH mergers was superlinear, $N_\mathrm{merg}\propto M_0^{1.6}$ \citep{2020MNRAS.492.2936A}, by performing a least-squares fit, we find a relationship close to linear, $N_\mathrm{merg}\propto M_0^{1.3}$. We note that this fit is performed over our family of models, which might obfuscate some degeneracies in the initial conditions. In particular, the \cmc\ models have higher initial densities than the $N$-body models. If we run all models using \cbhbd\ but with the same initial density (fixed at $\rhoh = 10^4\,\msun\,\pc^{-3}$) and for the same amount of time ($t=13\,\Gyr$), we find $N_\mathrm{merg}\propto M_0^{1.1}$. 
The almost-linear scaling with $M_0$ implies that the merger efficiency defined as the number of mergers per unit of initial cluster mass, is roughly constant across all masses. For an initial cluster mass function that scales as $M_0^{-2}$ \citep*[for example][]{2010ARA&A..48..431P} (and assuming that the initial $\rho_{\rm h}$ does not depend on mass), low-mass clusters contribute roughly the same to the merger rate as massive clusters.

\section{Conclusion}
\label{sect:conclusion}
In this study we present the computational capabilities and advantages of the fast code \cbhbd\ for modeling the evolution of star clusters and their BBH mergers under tidal field effects on circular orbits, with different metallicities and IMFs. By focusing on macroscopic properties such as mass-loss mechanisms and energy production, \cbh\ provides accurate descriptions of clusters with varying initial conditions. The study analyzed clusters with $N$ from $8\times10^5$ to $3.2\times10^6$
stars initially using \cmc\, models as comparison data, considering different initial half-mass radii, galactocentric distances, and metallicities. Results confirm that the current version of \cbhbd\,effectively models cluster evolution and resulting BBH mergers across diverse scenarios at low computational costs.

While the present study focuses on validating the total BBH merger rate, we recognize that the astrophysical interpretation of GW events requires accurate predictions of the full demographic distributions, including BH masses, mass ratios, and eccentricities. A systematic comparison of these properties with high‑fidelity models such as \cmc\ and direct 
$N$‑body simulations will be the subject of a forthcoming study. For reference, a detailed comparison of BBH properties within the \cbh\ framework can be found in AG20. Such a validation is essential to ensure that \cbhbd\ reliably captures rare but scientifically important populations, such as hierarchical mergers in the mass gap or eccentric mergers that retain a dynamical signature. In addition, we emphasize that \cbh\ is best suited for clusters with virial radii $\gtrsim 1\, \pc$, i.e., within the range used for calibration. For more compact clusters, the model uncertainties increase, as shown in Appendix~\ref{app:LowN}. Future work should further test the code against higher‑fidelity simulations.

We envision that in future work \cbhbd\ can be used to address various science cases. A particularly important application will be the inference of the initial conditions, structure, stellar content and BH populations of nearby GCs by constraining their dynamical histories using fast evolutionary models like \cbh. This would help bridge the gap between present-day observables and theoretical predictions of initial conditions, allowing for the improved modelling of cluster formation and evolution. A forthcoming study by Dickson et al. (in prep) will combine \cbh\ with detailed cluster mass models to explore implications on the initial conditions and evolution of a large sample of Milky Way GCs based on present-day observations \citep{2023MNRAS.522.5320D,2024MNRAS.529..331D}. 

Tidal streams originating from GCs also provide valuable insights into their disruption processes and the dynamics of the host galaxy \citep{2008MNRAS.390.1437C,2011MNRAS.415.1569S,2018ApJ...862..114S,2019MNRAS.482.1525V,2024ApJ...976...54P,2025ApJ...980...71V,2025ApJ...980L..18C}. Observations of these streams reveal their kinematic and structural properties, helping to trace the gravitational potential of the host galaxy. The BH population plays a key role in tidal tail formation, as the rapid loss of stars increases the BH fraction, leading to cluster expansion and accelerated evaporation \citep{2021NatAs...5..957G}. In mass modeling of GC streams, efficient evolutionary prescriptions allow for more reliable inference of their present-day mass distribution, particularly in the presence of BHs \citep{2025ApJ...980L..18C,2025ApJS..276...32C,2025MNRAS.tmp..409R}. While a dedicated fast evolution code for modeling clusters and their resulting streams is not yet available to our knowledge, developing such an approach building upon a fast cluster evolution code like \cbhbd\ holds great promise for accurately capturing the formation and evolution of streams. We aim to address this issue in the future when tidally dissolved clusters can be adequately described by \cbhbd.

The improved GW recipes in \cbhbd\ combined with its speed make \cbhbd\ an ideal tool for population synthesis modelling of BBH mergers in star clusters, which is very relevant in the current era of GW astronomy, marked by a rapidly growing number of detections (e.g., the fourth gravitational‑wave transient catalogue, GWTC-4.0, \citealt{2025arXiv250818082T}) and an increasing demand for detailed population modelling. The added description of GW captures in BBH-BBH interactions have improved the code particularly for low-mass clusters ($\lesssim10^5\,\msun$). Combined with the new ability to model clusters with different metallicities, \cbhbd\ is now ideally suited to model low-mass, metal-rich clusters that are forming across cosmic time. This would be an interesting addition to the \cbhbd\ population models for old, massive (globular) clusters of \citet{2023MNRAS.522..466A}.

%________________________________________

\section*{Acknowledgments}
FFP acknowledges the  “la Caixa” Foundation (ID100010434) for financial support in the form of a Doctoral INPhINIT fellowship (fellowship code LCF/BQ/DI23/11990067). ND is grateful for the support of the Durland Scholarship in Graduate Research. DMP and MG acknowledge financial support from the grants PRE2020-091801, PID2021-125485NB-C22, PID2024-155720NB-I00, CEX2019-000918-M, CEX2024-001451-M funded by MCIN/AEI/10.13039/501100011033 (State Agency for Research of the Spanish Ministry of Science and Innovation). VHB acknowledges the support of the Natural Sciences and Engineering Research Council of Canada (NSERC) through grant RGPIN-2020-05990. 
FA acknowledges the support of STFC grants ST/V005618/1.
We thank Peter J. Smith for helpful discussions and suggestions during the development of this work.

%______________________________________

\bibliographystyle{aa} 
\bibliography{total} 

%______________________________________
\begin{appendix}

\section{Differences between {\cbh\ }and \emacss}
\label{app:cbh_emacss_diff}
It is worth briefly discussing the capabilities of \cbh\,in comparison to another available fast evolution code, \emacss, as detailed in \cite{2012MNRAS.422.3415A,2014MNRAS.437..916G,2014MNRAS.442.1265A}. In summary, while both codes can be used to study stellar clusters efficiently and with sufficient accuracy, there are a few main differences. First, \emacss\,includes additional physical phenomena, such as mass segregation and its impact on the average stellar mass, induced mass loss mechanisms as well as the time evolution of the ratio $\rh / \rv$. As a result, \emacss\,requires additional parameters to properly describe these effects. In contrast, the current version of \cbh\,does not include such effects yet. The second difference lies in the treatment of particle populations. \cbh\,is specifically designed so that the BH population is created and evolves until it is  depleted, whereas \emacss\, does not have this specific focus. The balanced phase is also treated differently in \cbh. Here, each effect contributing to mass loss and energy production is accounted for at $t>\tcc$. On the other hand, \emacss\,makes a smooth connection prior core collapse to post core collapse while in the balanced phase only Hénon's contribution is considered. Regarding stellar evolution, \emacss\ makes use of a similar expression as equation~(\ref{eq:Edotsev}) however unlike \cbh, the ratio $\mst/\m$ is used on the right hand side, with $\mst$ not being influenced by tides. Finally \cbh\, makes no predictions for the evolution of the core.

\section{Energy change following mass loss}
\label{app:dEdM}
The parameter $\eta$, first introduced in equation (\ref{eq:Edotsev}), quantifies the level of segregation and is related to the efficiency of energy transfer to the cluster's outer layers. Mass loss occurs at the cluster's center, and it is assumed that the time scale is longer than the local crossing time. This applies to both BH ejections during the balanced phase and stellar mass loss from main sequence stars. The heaviest stars, which lose more mass, segregate faster. 

Given that the cluster is in virial equilibrium, the relation $U = 2E$ holds where $U$ is its potential energy. Let $\mvsq$ be the mean square velocity such that $T = \frac{1}{2} M \mvsq$. Mass loss affects both kinetic and potential energy, as a removed particle carries away a fraction of these energies. The variations in potential and kinetic energy can be expressed as,

\begin{equation}
\label{eq:variations}
\frac{\mathrm{d}U}{U} = 2\frac{\langle\phi_{\Delta}\rangle}{\langle \phi \rangle}\frac{\mathrm{d}M}{M}, \quad \frac{\mathrm{d}T}{T} = \frac{\langle\vsqd\rangle}{\mvsq}\frac{\mathrm{d}M}{M},
\end{equation}
where $\phi$ is the specific potential and $\langle \phi \rangle = 2U/M$. Subscript $\Delta$ denotes removal. The total rate of energy change, derived from the virial theorem, is connected to mass loss through,

\begin{equation}
\label{eq:dEdM}
\frac{\mathrm{d}E}{E} = \left[ 4 \frac{\langle\phi_\Delta\rangle}{\langle \phi \rangle} - \frac{\langle\vsqd\rangle}{\mvsq} \right] \frac{\mathrm{d}M}{M}.
\end{equation}
This expression defines $\eta$ in equation (\ref{eq:Edotsev}), implying that the level of segregation depends on the fractions of potential and squared velocity carried away by escaping particles, regardless of the escape method. Thus,

\begin{equation}
\label{eq:Mval}
\eta = 4 \frac{\langle\phi_\Delta\rangle}{\langle \phi \rangle} - \frac{\langle\vsqd\rangle}{\mvsq}.
\end{equation}
This expression generally increases over time, indicating a growing concentration in the central regions of the cluster. For a homologous distribution, the ratios $\langle\phi_\Delta\rangle/\langle \phi \rangle$ and $\langle\vsqd\rangle/\mvsq$ are both equal to unity, giving $\eta = 3$.

\section{Computation of $\psi$}
\label{app:comp_h}
Here we compare properties within \(\rh\) to global cluster values, since the parameter \(\psi\) in equation (\ref{eq:psi}) is meant to reflect conditions within \(\rh\). The subscript \(\mathrm{h}\) denotes values within \(\rh\). The key quantity is \(\langle m \rangle_{\mathrm{h}} \psi\), because it appears in the denominator of the relaxation time (equation~\ref{eq:trh}), which sets the timescale of BH ejection, evaporation  and expansion.

To understand the behavior of $\psi$, all the \cmc\,models are considered. Since the level of segregation is different, clusters will be separated based on the initial relaxation. Since no information of $\psi(t=0)$ is available, the formula for relaxation for the one component model, that is equation~(\ref{eq:trh}) with $\psi=1$, at $t=0$ is considered to separate clusters based on relaxation. The stellar distribution initially is random, therefore the same value of $\psi$ is expected for all models. The same assumption was made in  \cite{2002ApJ...576..899P} where core collapse was investigated. 

\begin{figure}
\includegraphics[width=1.\columnwidth]{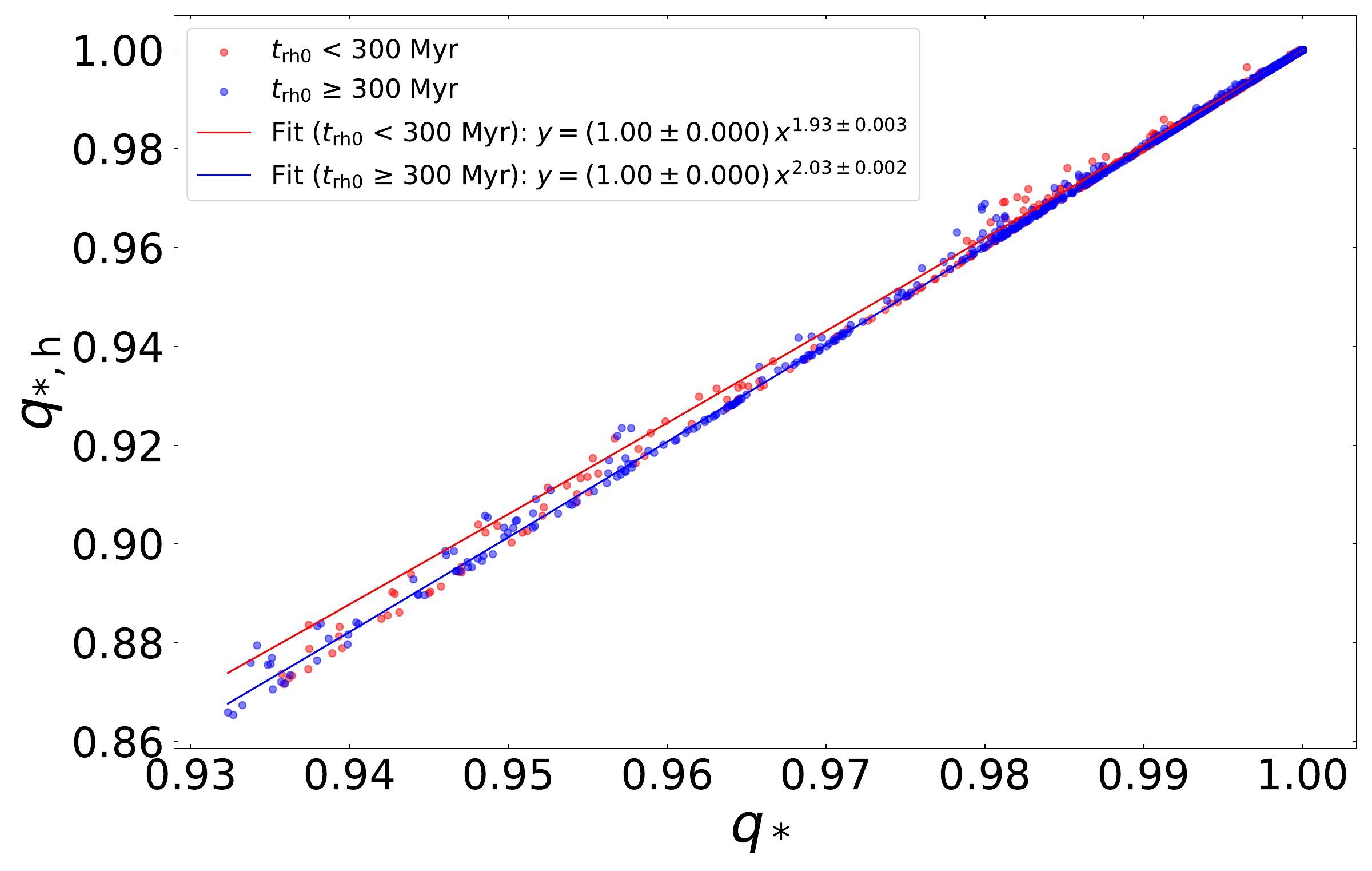}
\caption{The relation between $q_*$ within the half-mass radius and the global $q_*$ is well-described by a quadratic power-law. Since $q_*$ values are close to unity, this contribution in equation~(\ref{eq:psi_approx}) can be approximated as constant.}
\label{Fig_qst}
\end{figure}

\begin{figure}
\includegraphics[width=1.\columnwidth]{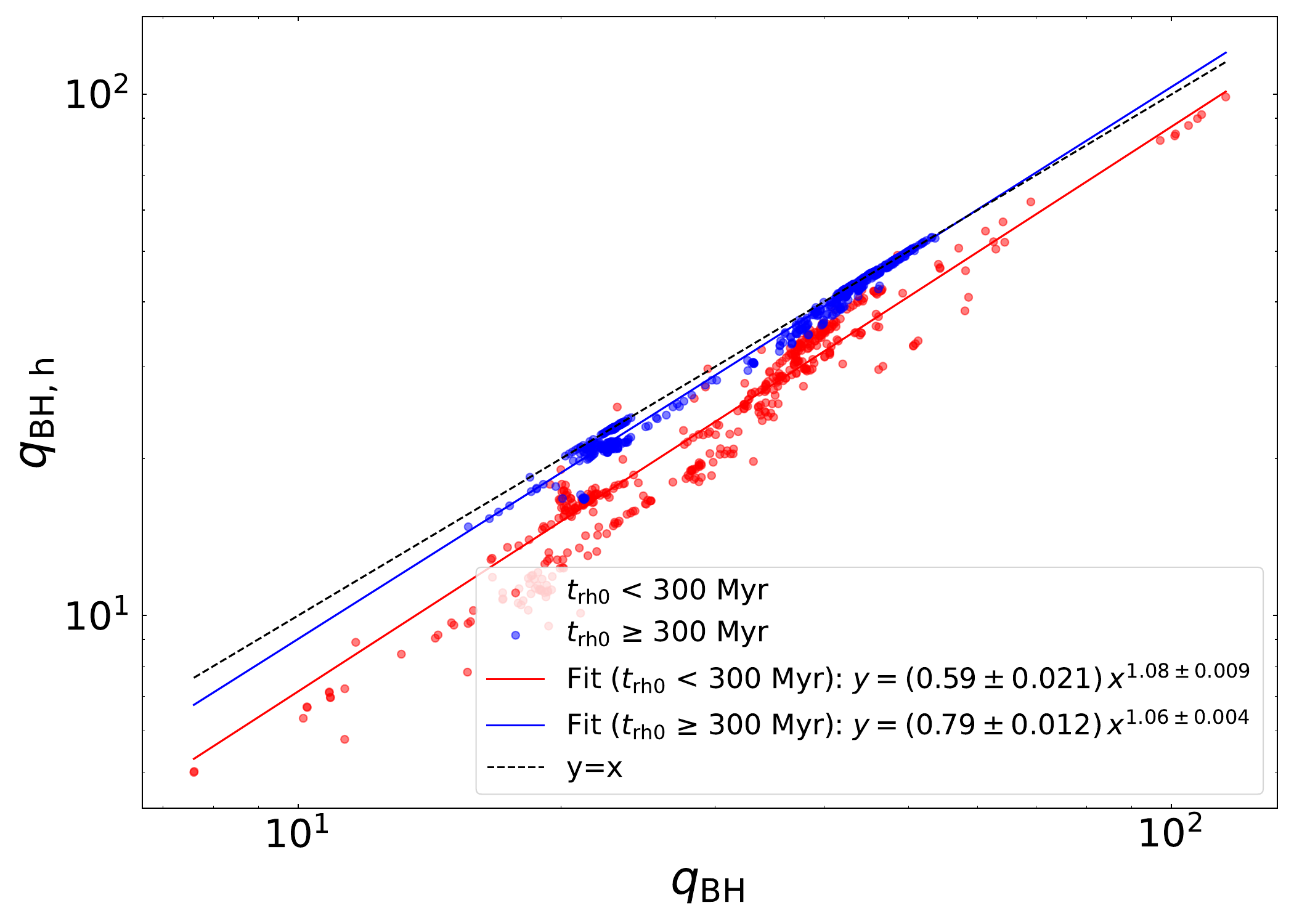}
\caption{The relation between the fraction $\mbh/\m$ within $\rh$ with the global property. Points with large initial relaxation align closely to the $y=x$ line, indicating non-segregated clusters.}
\label{Fig_qmbh}
\end{figure}

\begin{figure}
\includegraphics[width=1.\columnwidth]{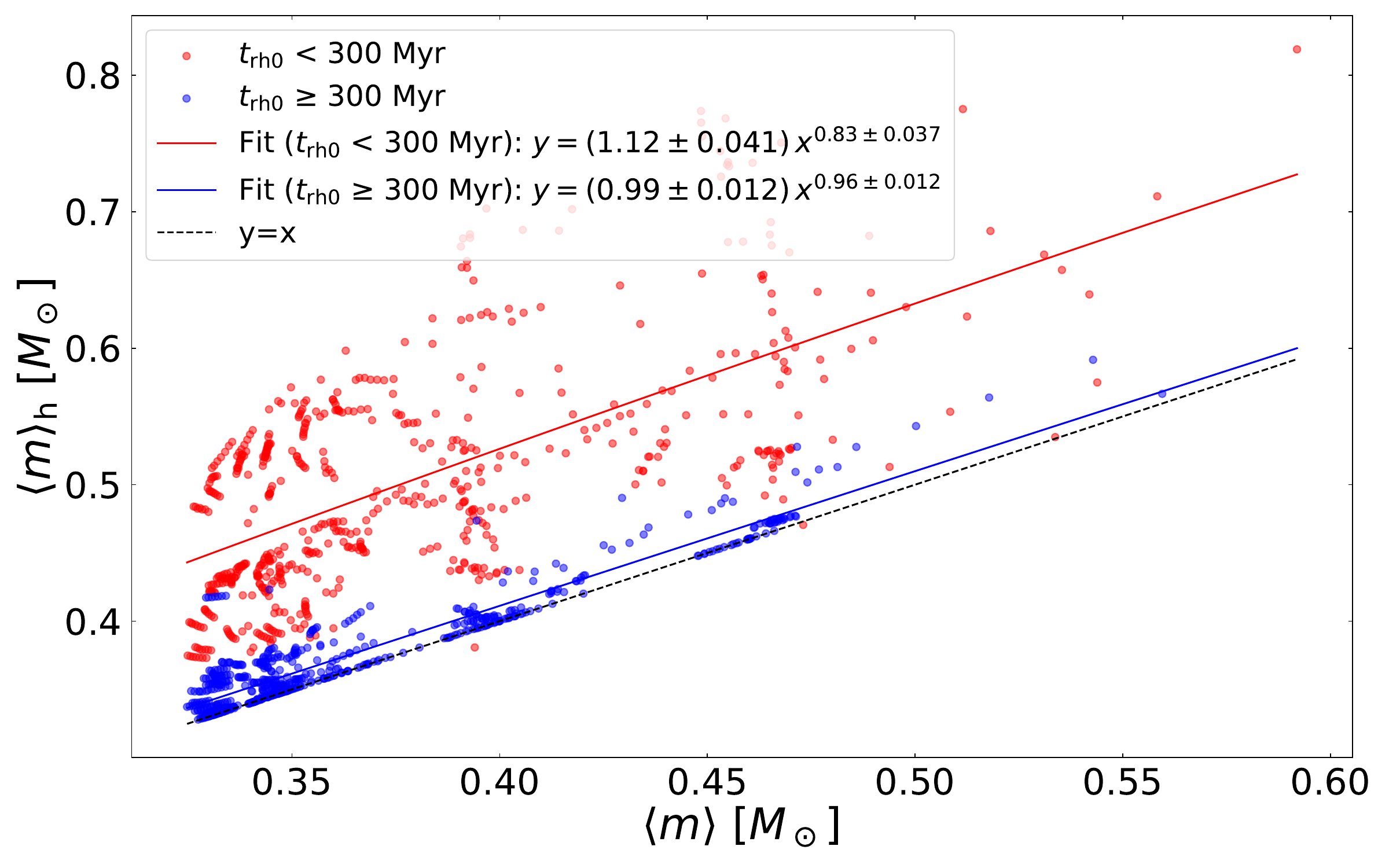}
\caption{Relation of the average mass within $\rh$ and the global by separating clusters based on initial relaxation. Points at $t=0$ are dropped for consistency with Fig.\ref{Fig_qmbh}. The parallel lines suggest that a single set of parameters can describe both categories of clusters.}
\label{Fig_m}
\end{figure}

Consider the ratio \(q_{*} = \mst / \m\). Both \(\mst\) and \(\m\) within \(\rh\) are larger than $\mst$ and $\m$ of the cluster respectively, so we assume 
$q_{*,\mathrm{h}} = q_{*}^{a}$, 
with prefactor 1 because \(q_{*,\mathrm{h}}=q_{*}\) when no BHs are present. In the stellar mass, we include everything that is not a BH, even remnants such as white dwarfs or neutron stars. Figure~\ref{Fig_qst} shows that \(\alpha = 2\) is preferred, but since global $q_{*}$ is close to unity, we simply set this factor is $\psi$ equal to 1 in equation~(\ref{eq:psi}) (as in AG20). Consequently, the BH population determines \(\psi\) at early times, and once all BHs are ejected, \(\psi = 1\). This substitution is valid for \(t > \tcc\), since BHs already dominate \(\psi\) during the relaxation‐driven expansion.

Similarly the BH fraction $f_{\rm BH,h}$ is studied with a power-law. Because the majority of BHs reside within $\rh$, the relation between $f_{\rm BH, rh}$ and $f_{\rm BH}$ should be fairly trivial. It is expected that the prefactor of the power-law is quite close to $2$ while the exponent is close to $1$. 

The final ingredient in computing \(\psi\) is the BH mass fraction within \(\rh\), defined as \(q_{\mathrm{BH}} = \mbh / \m\). A power-law fit for this quantity is shown in Fig.~\ref{Fig_qmbh}. Clusters with initially long relaxation times tend to lie close to the identity line (\(y = x\)), while those with short relaxation times fall on a distinct second branch parallel to the first. This distinction arises because clusters with long initial relaxation do not segregate rapidly. However, this does not necessitate treating the two populations separately, since the relaxation timescale is governed by the product \(\langle m \rangle_{\rm h} \psi\), which varies by less than \(20\%\) across the two categories.

\begin{figure}
\includegraphics[width=1.\columnwidth]{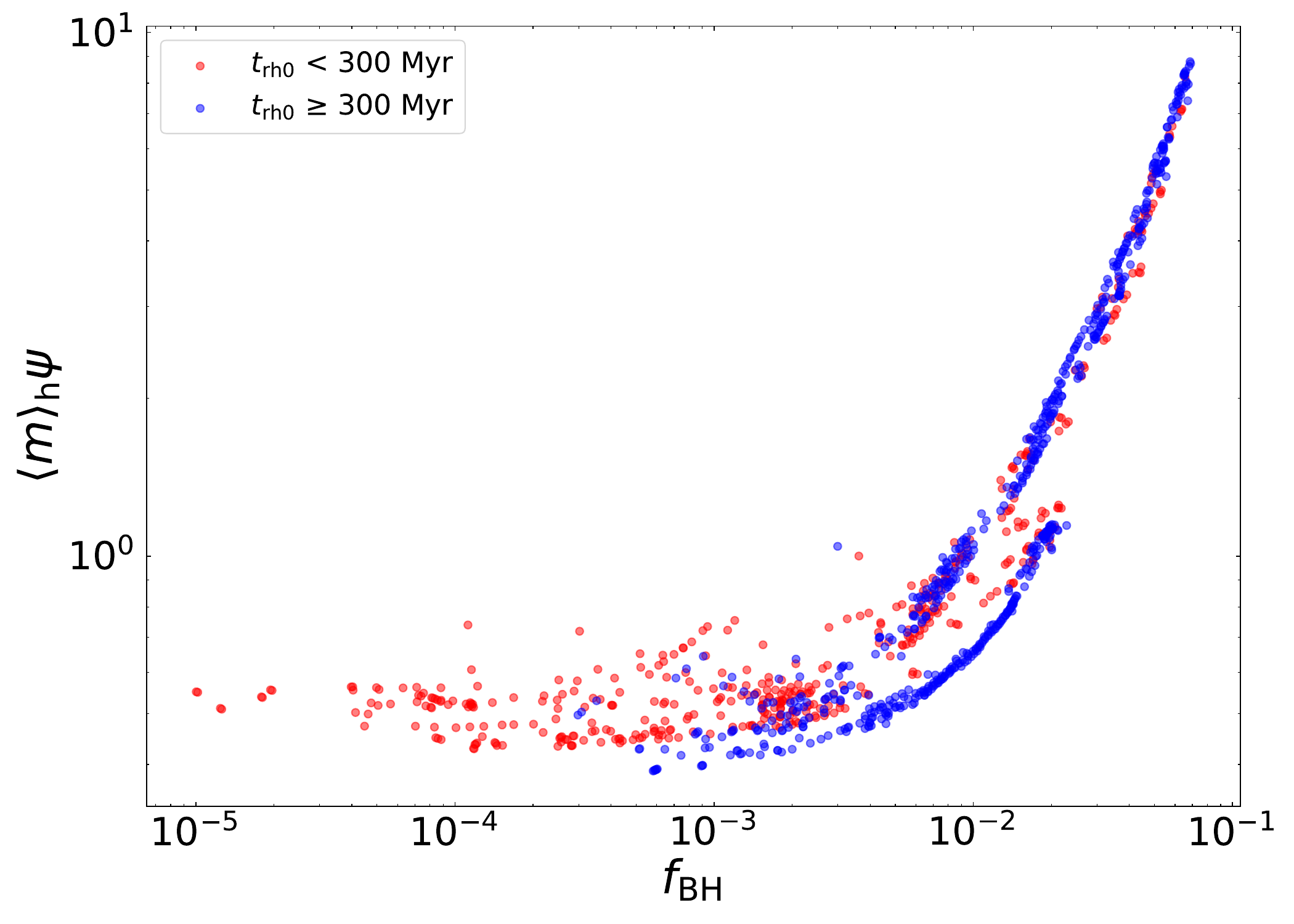}
\caption{The product $\langle m\rangle_{\rm h}\psi$.  Each point represents a snapshot for a given model. Regardless of initial relaxation, the behavior is the same therefore a single set of values suffices. The figure is made using equation~(\ref{eq:psi_approx}). Second branch on the left plot belongs to $Z=0.02$ clusters which for the same BH fraction as the rest clusters, they have a different BH mass.}
\label{Fig_mpsi}
\end{figure}

In Fig.~\ref{Fig_m}, we show how the average mass within \(\rh\) compares to the cluster average for both branches, while Fig.~\ref{Fig_mpsi} shows that the product \(\langle m \rangle_{\mathrm{h}} \psi\) behaves similarly regardless of the initial relaxation state. This justifies the use of a unified set of parameters to model clusters with different initial conditions. Parameters tuned to match the high relaxation group are more convenient and therefore selected.

\begin{figure}
\includegraphics[width=1.\columnwidth]{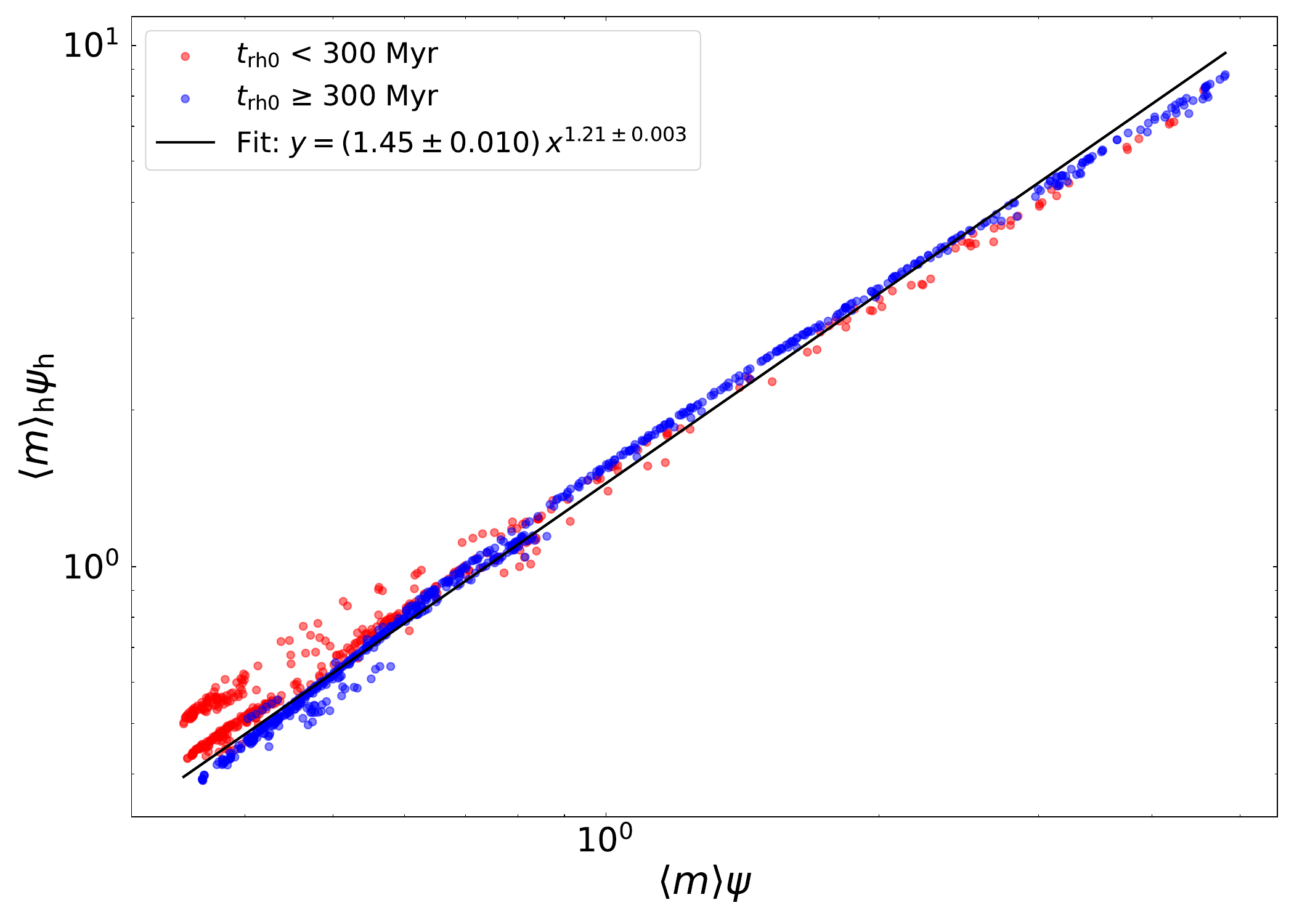}
\caption{ Comparison between $\m\psi$ within $\rh$ and the approximated form.}
\label{Fig_mpsi2}
\end{figure}
Based on the prefactor obtained from Fig.~\ref{Fig_qmbh} for large relaxation clusters, adopting \(\lambda = 0.25\) yields \(\psi \simeq 1 + 1.47S\), identical to the form used in AG20. While this relation can be employed, a simpler alternative is to use the approximation \(\psi \simeq 1 + S\), as shown in Fig.~\ref{Fig_mpsi}. This results in a simple power-law formula as shown in Fig.~~\ref{Fig_mpsi2}. Using a different prefactor in \(\psi\) effectively rescales the model parameters fitted previously.

Compared to the fit presented in the main text, the values of \(\zeta\), \(\Rht\), \(N_{\trh}\), and \(S_0\) change if $S'$ is used in $\psi$. Specifically, using $\psi(S')$ shortens the relaxation timescale, necessitating a higher value of \(N_{\trh}\) for consistency. Consequently, the energy emission rate \(\zeta\) decreases to maintain the same product \(\zeta \psi\), which governs the BH ejection rate. Since lower \(\zeta\) could lead to overestimated BH retention at low BH fractions, the parameter \(S_0\) is increased to compensate. To preserve the evaporation rate, \(\xi_0\) is also increased. In summary, although modifying the scaling in \(\psi\) alters the individual parameters, the overall model behavior remains robust due to compensating shifts. In the main text, we adopt the linear approximation \(\psi = 1 + S\) for simplicity.

\section{Comparison to {\cmc\ }models not included in the fits}
\label{app:LowN}

\begin{figure*}
\includegraphics[width=\columnwidth]{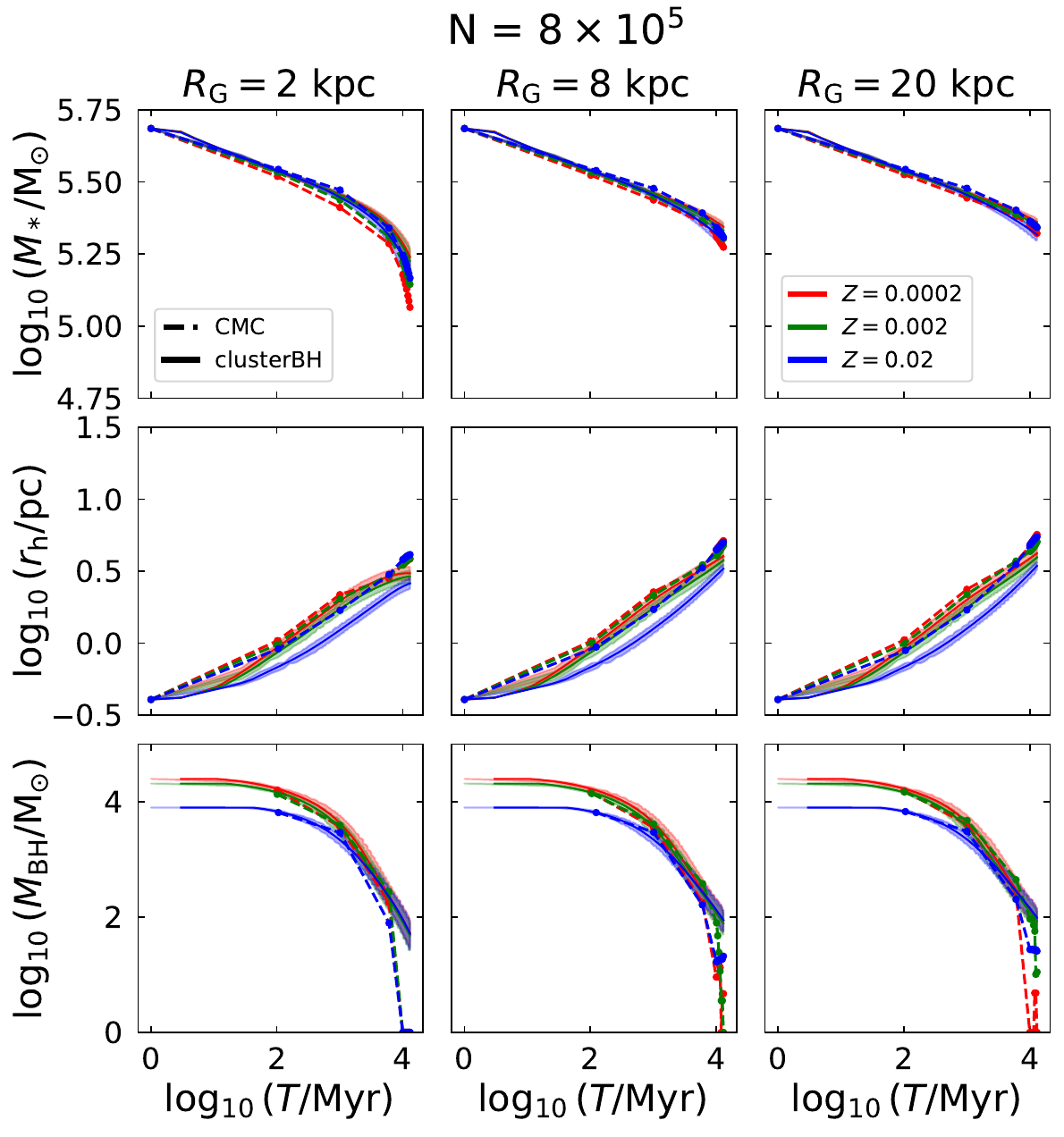}
\includegraphics[width=\columnwidth]{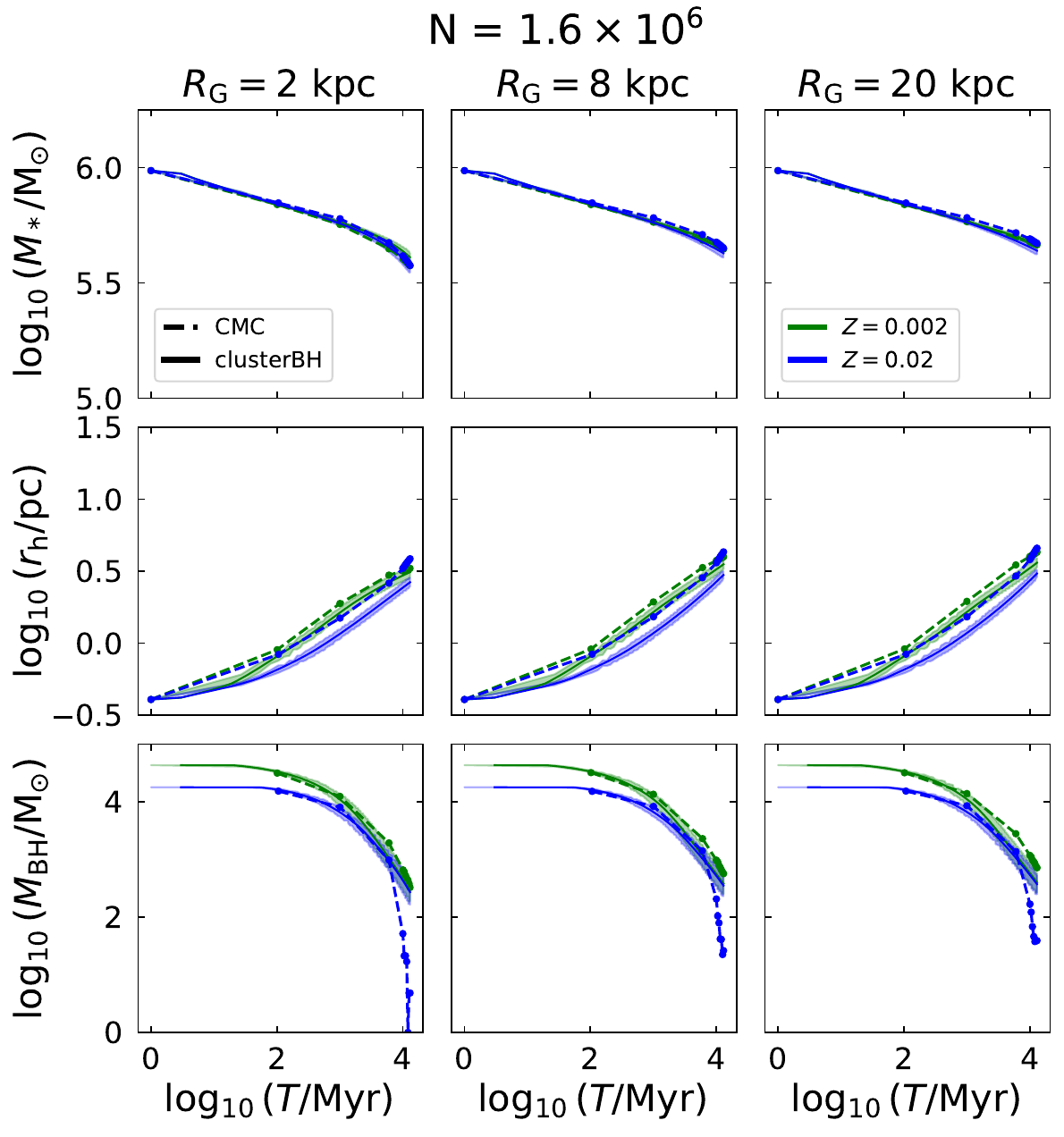}\\
\caption{Comparison between \cbh\, and \cmc\, for models with initial $\rv=0.5\,\pc$ and $N=8\times10^5$ stars (left) and $N=1.6\times10^6$ (right).}
\label{Fig_0.5}
\end{figure*}

\begin{figure*}
\includegraphics[width=\columnwidth]{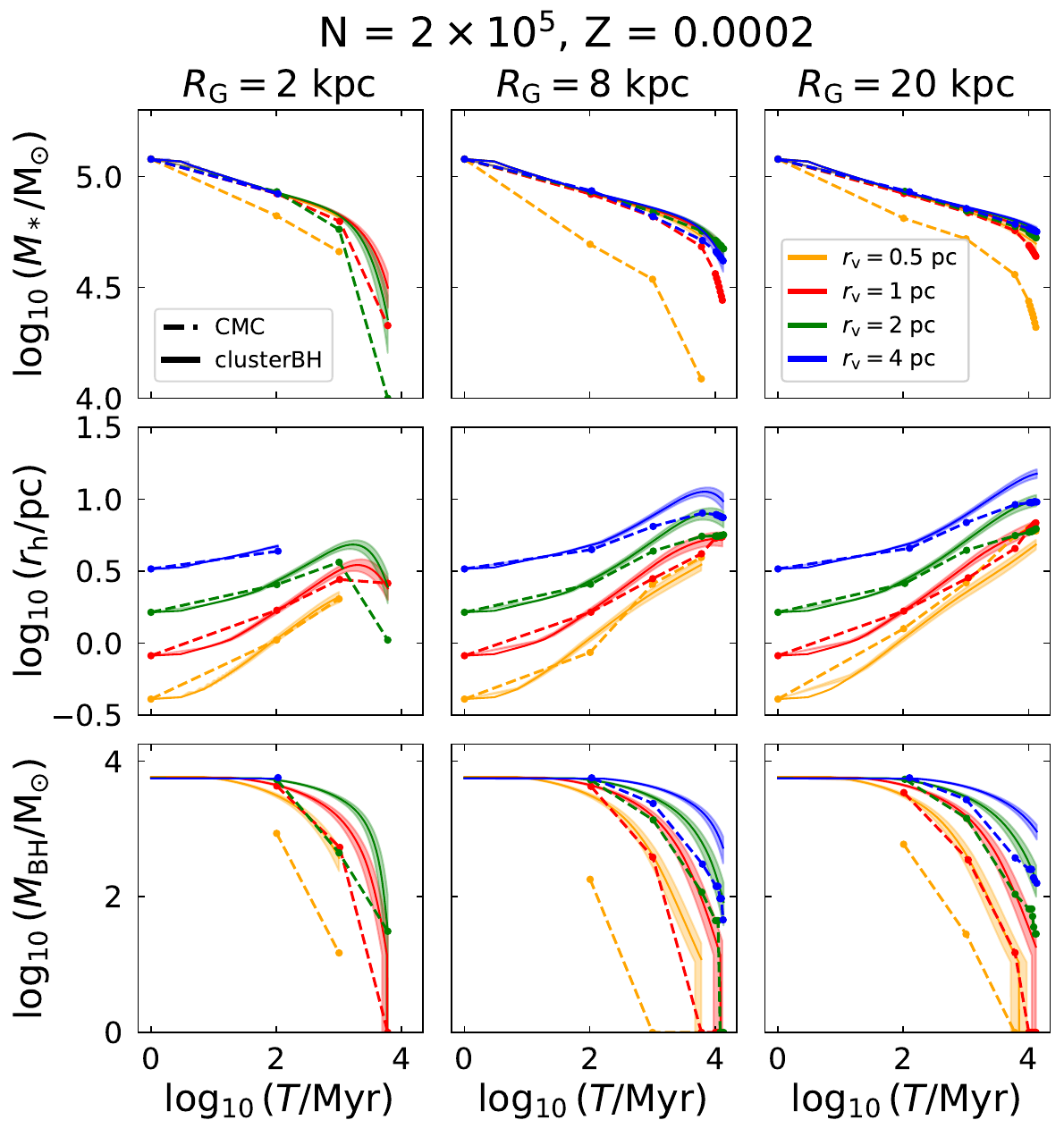}
\includegraphics[width=\columnwidth]{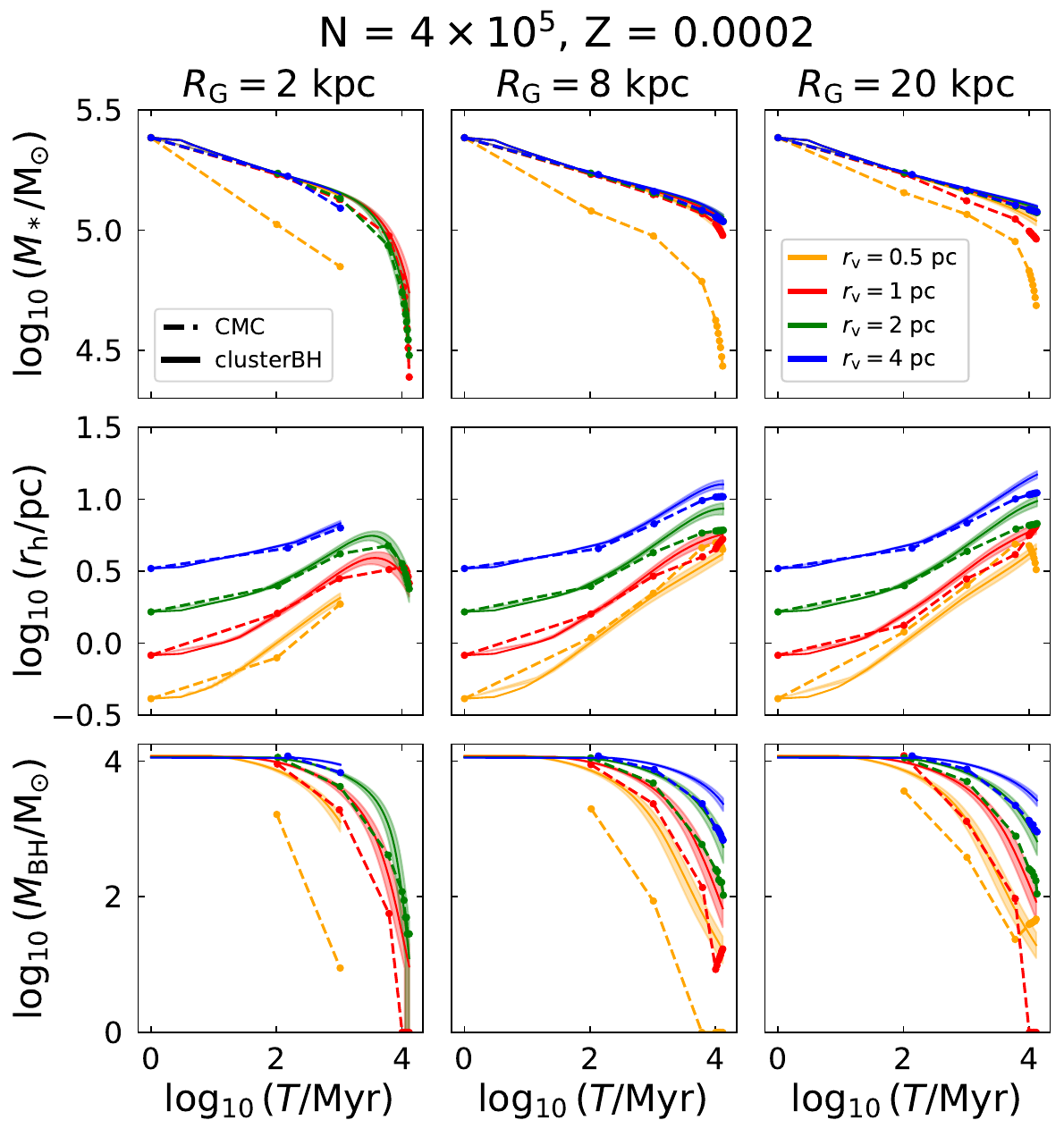}
\caption{Clusters with $N=2\times10^5$ (left) and $N=4\times10^5$ (right) stars for Z=0.0002.}
\label{Fig_low_Z0.0002}
\end{figure*}
Our model parameters were calibrated using massive clusters with $N=[8\times10^5,1.6\times10^6,3.2\times10^6]$ and  $\rv=[1,2,4]\,\pc$. Here we compare \cbh\, to the remaining \cmc\, models from the public grid that were left out of the fit. The estimated error to these models is extracted from the root mean square

\begin{equation}
\label{eq:delta_nofit}
\tilde{\delta}=\sqrt{\frac{1}{N_{\rm tot}}\sum_{ijk}\bigg(\frac{X_{ijk}^{\cbh}-X_{ijk}^{\cmc}}{X_{ijk}^{\cmc}}\bigg)^2}
\end{equation}
where $N_{\rm tot}$ is the total number of snapshots and is approximately $\tilde{\delta}\simeq0.401$. \cbh\ is capable of describing $\rh$ within $30\%$, $\Mst$ close to $50\%$ mainly because of $\rv=0.5\,\pc$ models that lose mass faster in \cmc. $\Mbh$ is overestimated in \cbh\ mainly for metal-poor clusters. The stellar masses of clusters with $\rv=0.5\,\pc$ and $Z=0.0002$ and the BH masses of clusters with $N\leq4\times10^5$ and $Z=[0.0002, 0.002]$ are the ones that drive the value of $\tilde{\delta}$ up.

In Figure~\ref{Fig_0.5}, the models with $\rv=0.5\,\pc$ for $N=[8\times10^5, 1.6\times10^6]$ are shown. The figure shows that \cbh\ underestimates the stellar mass loss rate, with the difference in $\Mst$ being within $10\%$ and increases with age for models at $\RG=2\,\kpc$ with $Z=0.0002$. It is worth noting that it is the metal-poor clusters that experience an increased mass loss in \cmc, which cannot be captured by \cbh\ at the moment. This was one of the reasons that these clusters were not included in the fit. $\rh$ is also underestimated in \cbh\ with the difference being within $20-40\%$. The deviation occurs before core collapse so it is not due to BHs. In the end, the BH population is described reasonably well by \cbh\, as the ejections are overestimated slightly but primarily when the BH population has dropped to about $\Mbh=100\,\msun$.

\begin{figure*}
\includegraphics[width=\columnwidth]{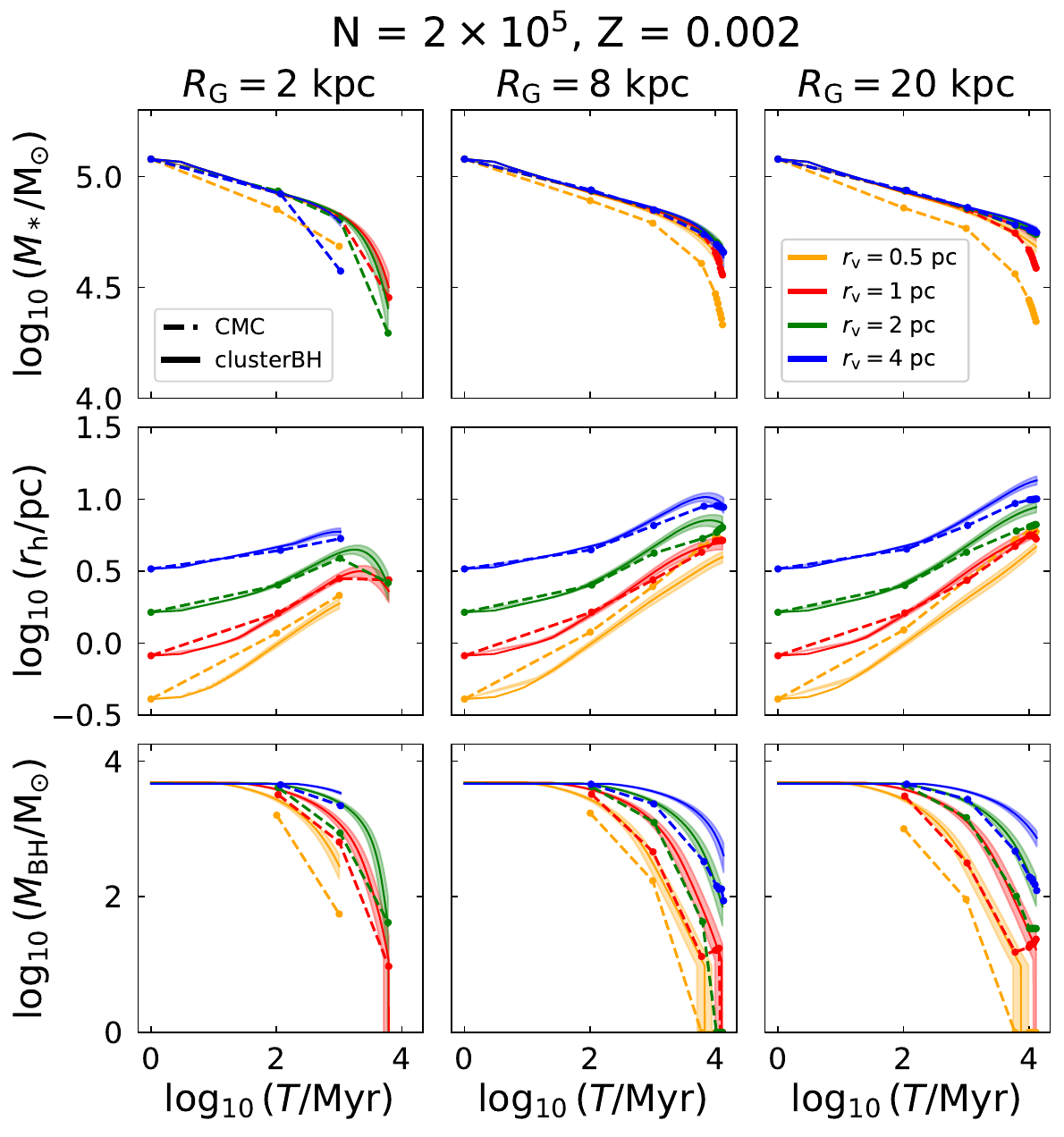}
\includegraphics[width=\columnwidth]{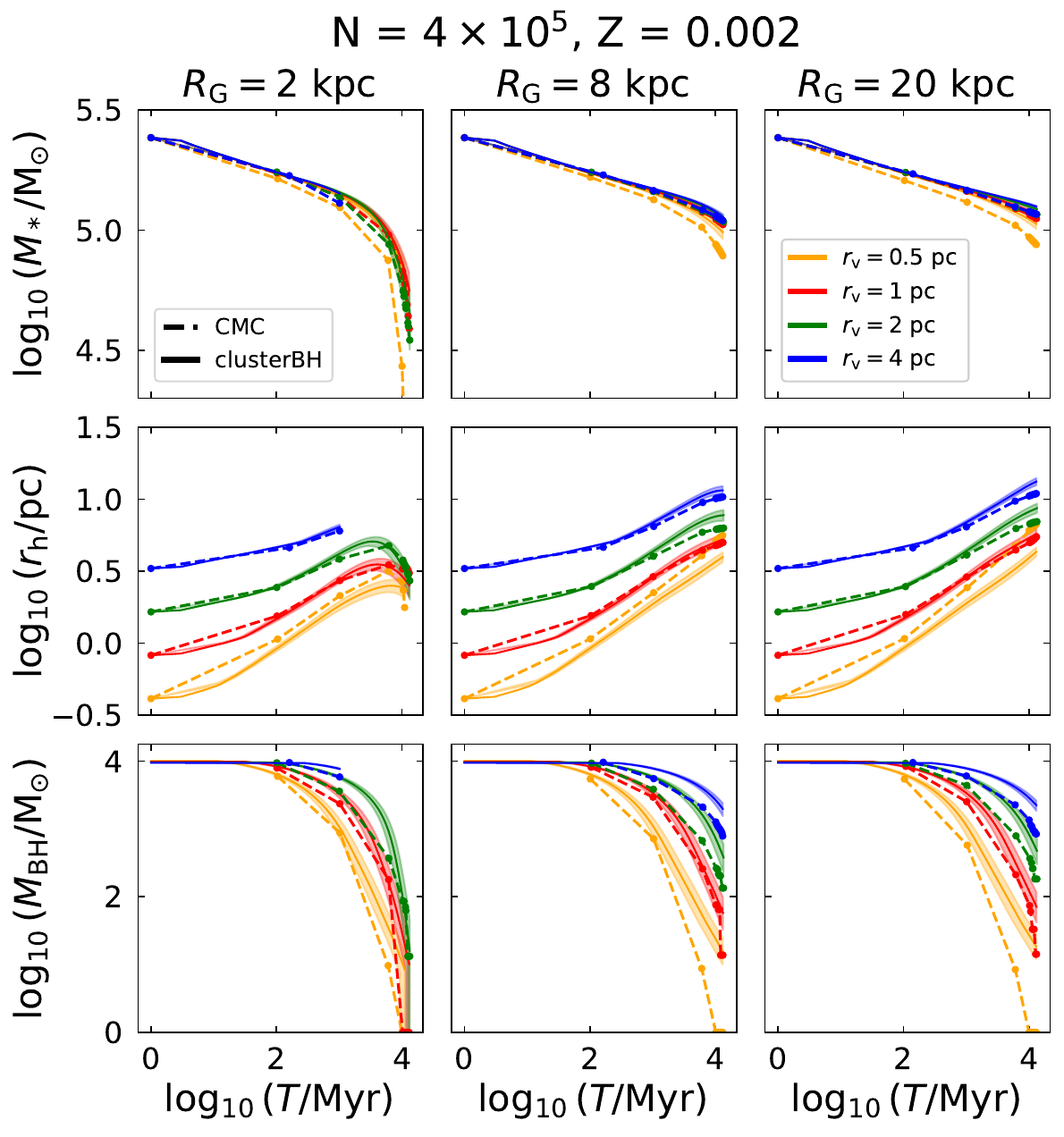}
\caption{Cluster models for $Z=0.002$ with $N=2\times10^5$ (left) and $N=4\times10^5$ (right).}
\label{Fig_low_Z0.002}
\end{figure*}
\begin{figure*}
\includegraphics[width=\columnwidth]{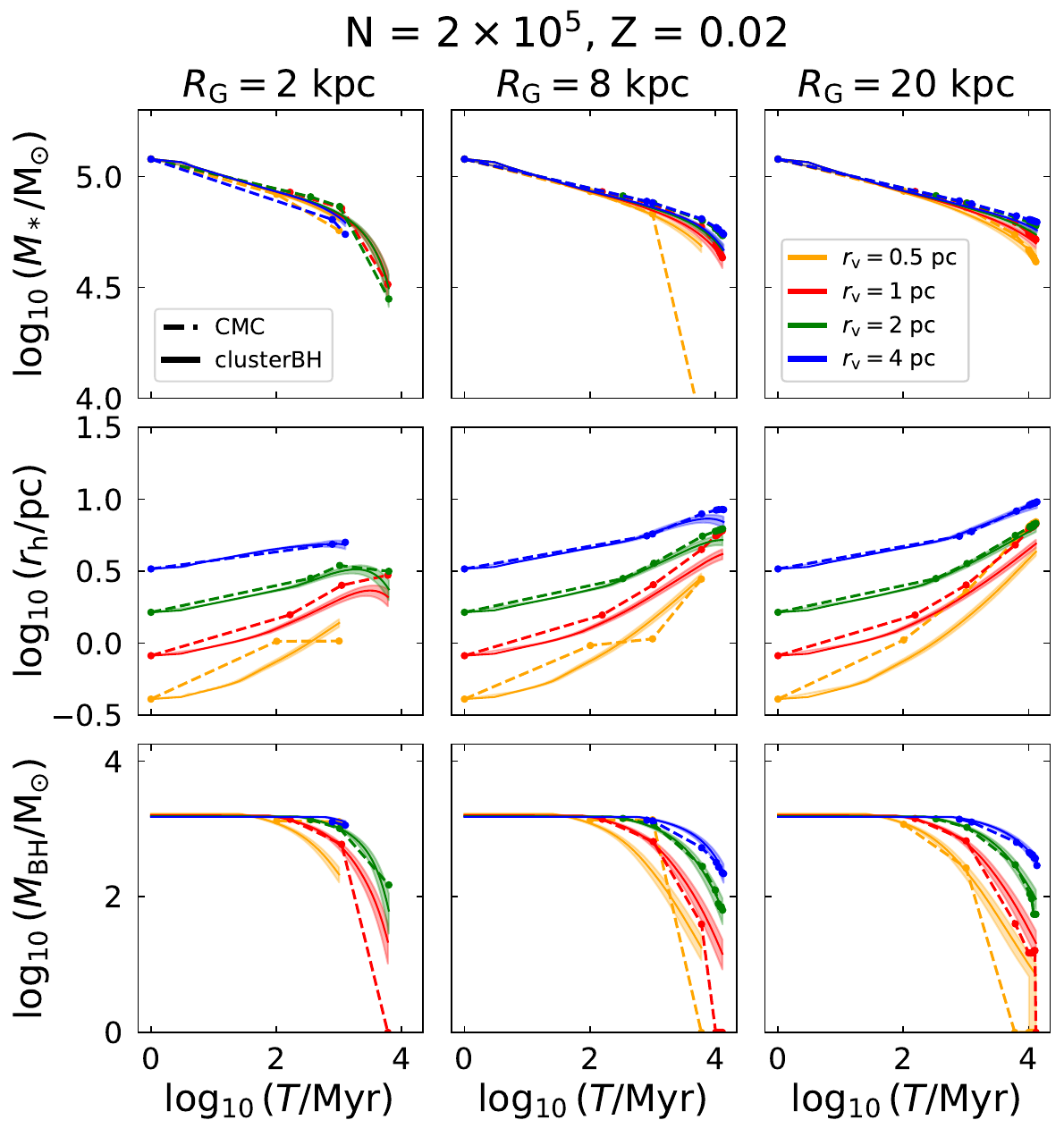}
\includegraphics[width=\columnwidth]{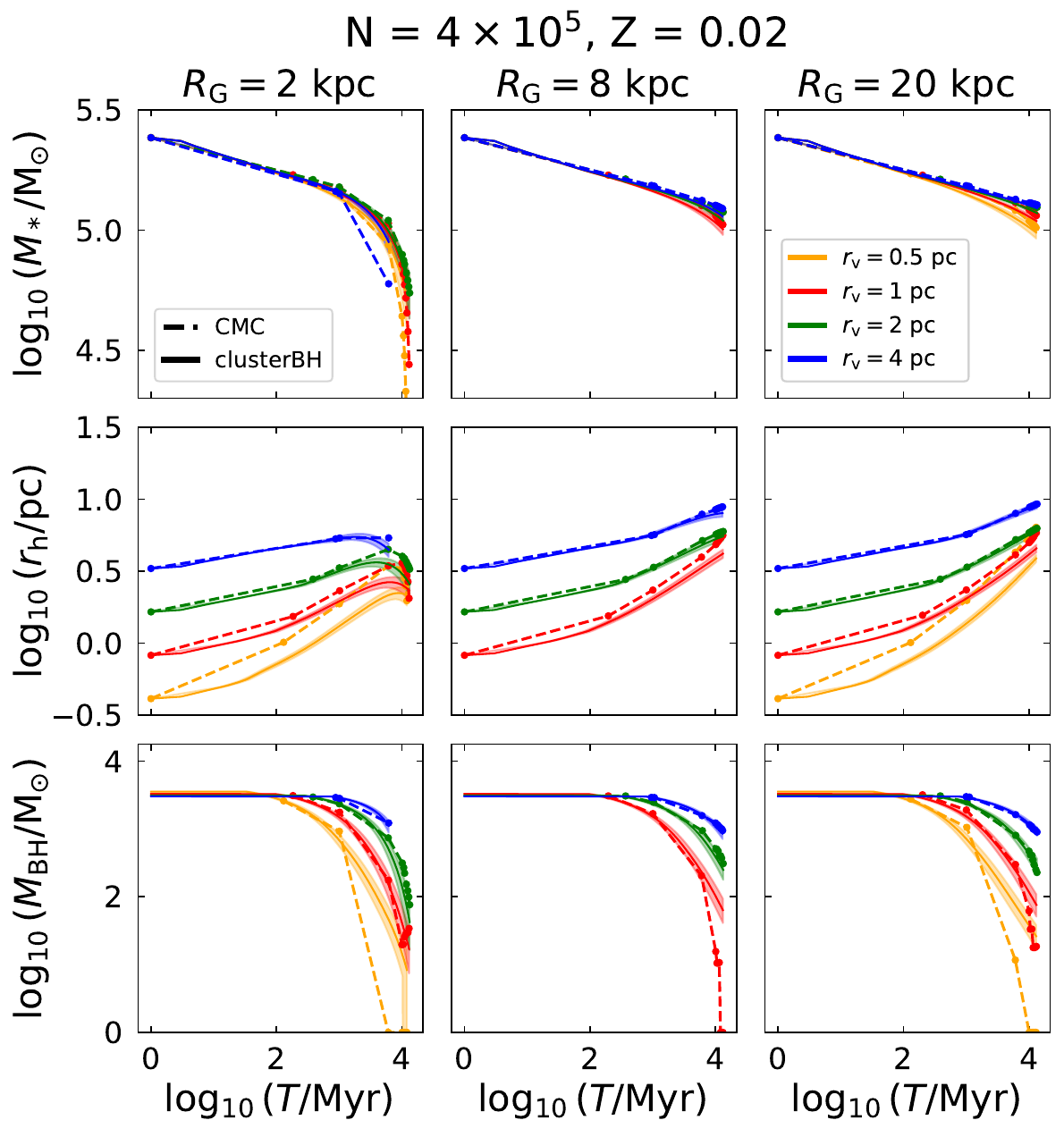}
\caption{Same as Figure \ref{Fig_low_Z0.0002} but for $Z=0.02$.}
\label{Fig_low_Z0.02}
\end{figure*}

Figures~\ref{Fig_low_Z0.0002}, \ref{Fig_low_Z0.002}, and \ref{Fig_low_Z0.02} show the results for all three metallicities for $N=[2\times10^5,4\times10^5]$. By plotting the low \(N\) models, we find that most clusters have stellar masses and half-mass radii within a 30\% error margin. Clusters with metallicity $Z=0.0002$ and $\rv=[2, 4]\,\pc$ tend to expand more according to \cbh\ while those with $\rv=[0.5,1]\,\pc$ expand less. This follows the pattern observed in the models that were fitted. However, the BH masses, especially for $Z=0.0002$, are overestimated in \cbh. Increasing the BH ejection rate solely for such models is difficult to deal with since, to our knowledge, no scaling with the stellar population should be observed. At \(\RG = 2\,\kpc\), clusters with \(\rv = 0.5\,\pc\) and \(\rv = 4\,\pc\) also exhibit excess mass loss, indicating that our current tidal prescription cannot capture the rapid dissolution of these systems. Such models are expected to have a large ratio $\rh/\rt$ so the contribution of $\dot E_{\rm ev}$ previously neglected should become important.

\begin{figure}
\includegraphics[width=1\columnwidth]{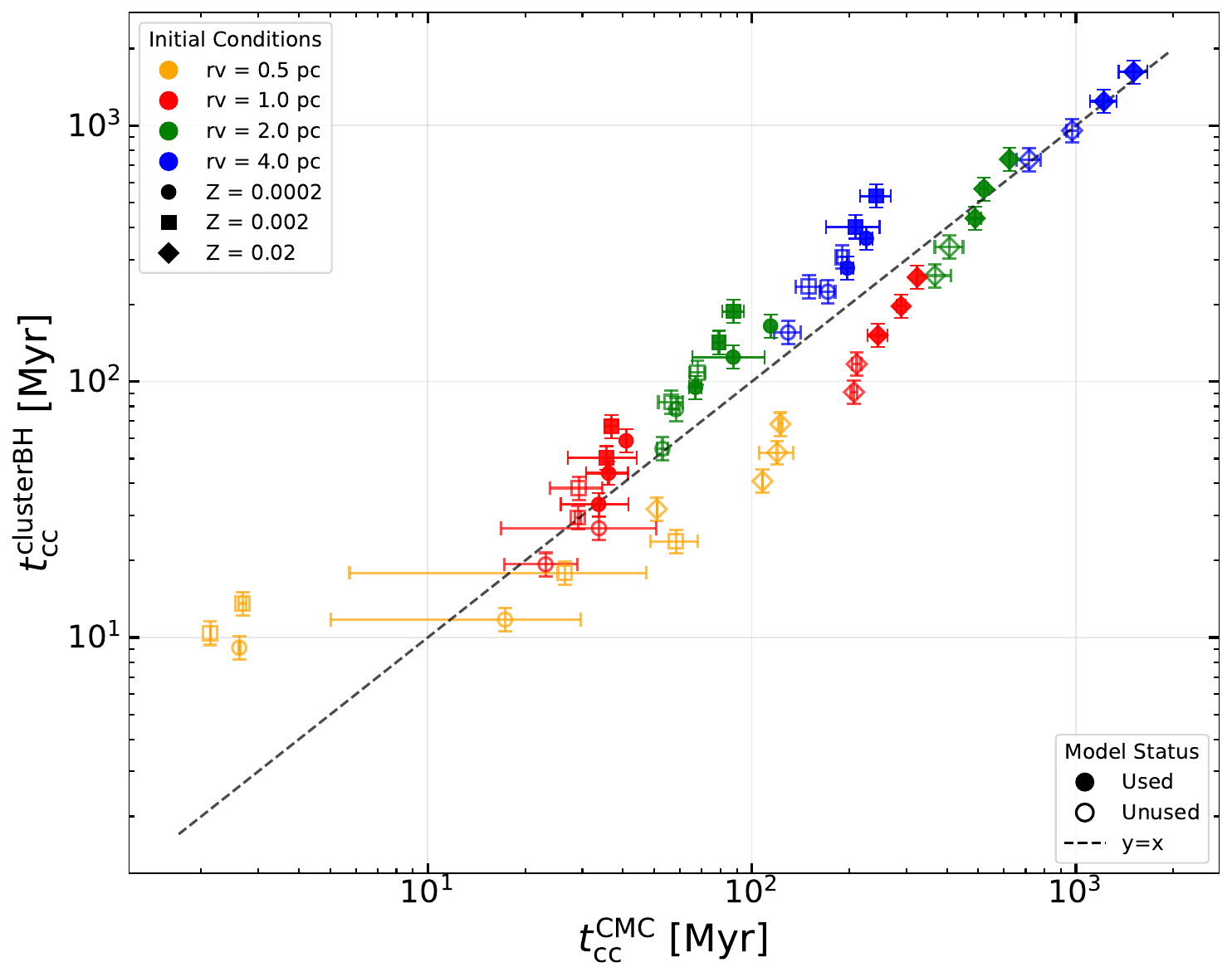}
\caption{Core collapse in $\cbh\,$versus $\cmc\,$. Error in x axis is due to the standard deviation from combining different galactocentric distances. Error in the y axis is due to the $1\sigma$ uncertainty in the number of initial relaxations $N_{\rm{trh}}$.}
\label{Fig_tcc}
\end{figure}

As a final comment, we showcase the predicted core--collapse times \(t_{\mathrm{cc}}\) in \cbh\ compared to those obtained with \cmc\ in Fig.~\ref{Fig_tcc}. The \cmc\ values are determined by analyzing the evolution of the core radius and identifying the first local minimum as the time of core collapse. 

The figure shows a noticeable spread along both axes. This spread arises because, for \cbh, the \(1\sigma\) uncertainty in the parameter \(N_{\mathrm{trh}}\) is considered, while for the \cmc\ models, results from all galactocentric distances are combined, and the spread represents the corresponding standard deviation. Models are also distinguished based on whether they were included in the likelihood analysis.

It becomes evident that, for most clusters with short initial relaxation, the \cbh\ predictions of \(t_{\mathrm{cc}}\) deviate by roughly 20\% from the \cmc\ values. This suggests that even though, for these clusters, the initial relaxation time \(\tau_{\mathrm{rh0}}\) depends primarily on the stellar IMF rather than on the BH fraction through \(\psi\), using the expression 
$t_{\mathrm{cc}} = N_{\mathrm{trh}} \, \tau_{\mathrm{rh0}},$
where \(\tau_{\mathrm{rh0}}\) depends on \(\psi\), still provides satisfactory results. As a comment, using the elapsed relaxation

\begin{equation}
\label{eq:Nrlx}
N_{\rm{trh}}=\int_0^{\tcc}\frac{dt}{\trh}
\end{equation}
produces similar results. Therefore, to keep the model simple, we selected to use $\tcc=N_{\rm{trh}}\tau_{\rm{rh0}}$.

Examining the models with \(Z = 0.02\), it becomes clear that \(t_{\mathrm{cc}}\) is significantly larger. This is expected, since metal-rich clusters produce lighter BHs, which take longer to segregate. Rather than introducing an explicit metallicity dependence directly on \(t_{\mathrm{cc}}\), it is more physically motivated—though not entirely accurate—to incorporate a \(Z\)--dependence through \(\psi\), thereby accounting for variations in the BHMF.

\end{appendix}

\end{document}